\documentclass[a4paper,fleqn,usenatbib,useAMS]{mnras}
\pdfoutput=1 

\usepackage{graphicx}  
\usepackage{amsmath}
\usepackage{amssymb}
\usepackage{multicol}
\usepackage{bm}
\usepackage{pdflscape}
\usepackage[T1]{fontenc}
\usepackage{ae,aecompl}
\usepackage{newtxtext,newtxmath}

\title[BH-LMXBs from BH retaining GCs]{Low-mass X-ray binaries from black-hole retaining globular clusters}

\author[Giesler, Clausen, \& Ott]{Matthew Giesler,$^{1}$\thanks{Email: \href{mailto:mgiesler@tapir.caltech.edu}{mgiesler@tapir.caltech.edu}} Drew Clausen,$^{1}$  Christian D Ott$^{1}$ \\
  $^{1}$TAPIR, Walter Burke Institute for Theoretical Physics,
  California Institute of Technology, Pasadena, CA 91125, USA}

\date{\today}
\pubyear{2017}

\begin{document}
\label{firstpage}
\pagerange{\pageref{firstpage}--\pageref{lastpage}}
\maketitle

\begin{abstract}
  Recent studies suggest that globular clusters (GCs) may retain a substantial
  population of stellar-mass black holes (BHs), in contrast to the long-held belief
  of a few to zero BHs. We model the population of BH low-mass
  X-ray binaries (BH-LMXBs), an ideal observable proxy for elusive single BHs, produced from a
  representative group of Milky Way GCs with variable BH populations. We simulate the formation of
  BH-binaries in GCs through exchange interactions between binary and single stars in the company
  of tens to hundreds of BHs.
  Additionally, we consider the impact of the BH population on the
  rate of compact binaries undergoing gravitational wave driven mergers.
  The characteristics of the BH-LMXB population and binary
  properties are sensitive to the GCs structural parameters
  as well as its unobservable BH population. We find that GCs retaining $\sim 1000$ BHs
  produce a galactic population of $\sim 150$ ejected BH-LMXBs whereas GCs
  retaining only $\sim20$ BHs produce zero ejected BH-LMXBs.
  Moreover, we explore the possibility that some of the
  presently known BH-LMXBs might have originated in GCs and identify five candidate systems.
\end{abstract}

\begin{nokeywords}
\end{nokeywords}

\section{Introduction}\label{sec:intro}
The fate of the population of stellar-mass black holes (BH) in globular clusters (GCs)
is still widely uncertain. It is expected that tens to hundreds and possibly thousands
of BHs are formed in GCs,
of which some fraction might be ejected
early due to a kick at formation~\citep{Belczynski:2006}.
In the standard GC evolution picture, the remainder
of the BHs should rapidly sink to the core due to mass segregation. There they are
subject to a high rate of dynamical interactions that are likely to eject the BHs
as singles or in binaries. It was long accepted that this process would lead to
repeated ejections from the GC leaving a few to zero BHs~(e.g.,~\citealt{Kulkarni:1993};~\citealt{Sigurdsson:1993}).
Historically, this was supported by the lack
of observational evidence for a BH in a GC; however, BHs are difficult to observe unless they
are actively accreting from a stellar companion.

In order to explore the population of BHs within and outside of GCs, black-hole low-mass
X-ray binaries (BH-LMXBs) can serve as an ideal proxy. In an evolved cluster, a main-sequence star
(MS) will necessarily be less than the MS turnoff mass,
yielding an abundance of potential low-mass companions. This, coupled with a
high rate of encounters due to the high-density environment of GCs, makes GCs
ideal BH-LMXB factories. However, this assumes that a significant number of BHs are
retained by GCs and that the BHs avoid segregating completely from the lower-mass stars.

The discovery of two BH-LMXB systems in the Milky Way GC M22~\citep{Strader:2012}
has led to a renewed interest in GC BH retention. 
This observation coupled with an estimate for the fraction of the BH population
expected to be in accreting binaries~\citep{Ivanova:2010}
suggests that M22 may contain between $5-100$ BHs~\citep{Strader:2012}.
Recent theoretical studies, including some detailed \textit{N}-body
simulations~(e.g.,~\citealt{Aarseth:2012};~\citealt{Wang:2016}), support the idea that GCs are capable of
retaining from a few to hundreds of
BHs~(e.g.,~\citealt{Breen:2013};~\citealt{Morscher:2013};~\citealt{Sippel:2013};~\citealt{Rodriguez:2016b}).

There is an increasing number of BH-LMXB candidates identified in the
Milky Way galaxy. BlackCAT~\citep{Corral-Santana:2016}, a catalog of BH-LMXBs,
has to date identified $59$ candidate Milky Way BH-LMXBs. An LMXB is
identified as a candidate BH-LMXB if the X-ray spectrum
rules out a neutron star (NS) as the compact accretor~\citep{McClintock:2006}.
Of the $59$ candidate BH-LMXBs in BlackCAT, $22$ are currently considered to
be `confirmed' BH-LMXBs. A BH-LMXB labeled as `confirmed' has
a dynamical measurement of the primary mass or
mass-function $f(M_\mathrm{BH})$~(see, e.g.,~\citealt{Casares:2014}).

Roughly one-fifth of the observed BH-LMXBs reside at an absolute distance $|z|$
perpendicular to the galactic plane greater than $1 \, \mathrm{kpc}$
(e.g.,~\citealt{Jonker:2004};~\citealt{Corral-Santana:2016}).
The distribution of the candidate and confirmed BH-LMXBs within the Milky Way
gives rise to the idea that BHs might be subject to high-velocity kicks at formation
(e.g., \citealt{Gualandris:2005};~\citealt{Fragos:2009};~\citealt{Repetto:2012};~\citealt{Repetto:2015}).
In some cases, the velocity needed
for the binary to reach large $|z|$ 
exceeds the contribution from a Blaauw kick~\citep{Blaauw:1961}.
This is the velocity imparted to a binary in the case of sudden mass loss,
i.e. in the BH progenitor's supernova explosion.
The exceptional high-velocity BH-LMXB cases have led to the idea of
high-velocity formation kicks, also known as `natal' kicks, where 
the binary receives a large kick through an
asymmetric explosion launched prior to BH formation~(\citealt{Janka:2013};~\citealt{Janka:2017}).
Due to the long-held assumption that GCs maintain a near-zero population of
BHs, the possibility that some of these systems originated in GCs
has been largely ignored.
BH-LMXBs sourced by BH-retaining GCs might help to explain some of the peculiar
properties of the observed Milky Way BH-LMXB population.
Although GCs are not likely to describe the entire
population of BH-LMXBs, the halo-orbits of GCs in the Milky Way
make GCs ideal candidate sources for the high-$|z|$ systems.
In light of the recent studies that suggest GCs might harbor
a large number of BHs, we revisit in this paper the possibility of
GCs as a potential origination point for a subset of the observed
BH-LMXB systems.

In addition to using BH-LMXBs as probes of BH retention in
GCs, the BH-BH merger rates might also serve
to place some constraints on GC BH retention.
The recent success in observing merging BH-BH binaries by advanced LIGO (aLIGO) makes this a
realistic possibility~(\citealt{LIGO:2016_main};~\citealt{Abbott:2016a};~\citealt{Abbott:2016b}).
Furthermore, binary BH mergers occurring
in GCs may be characteristically eccentric due to dynamical formation channels.
Although these eccentric systems are likely to have
circularized by the time they are visible in the aLIGO
frequency band, the eccentricity is potentially detectable
at lower frequencies. 
The addition of a space-based gravitational
wave observatory (e.g., LISA) in the future, designed for
sensitivity at lower frequencies, further improves the
prospect of using BH-BH mergers to probe GC dynamics.

In this study, we explicitly evolve `test' binaries in a
fixed cluster background subject to dynamical friction and
single-binary interactions. Additionally, we include 
an updated prescription for allowing single BHs to exchange
into existing binaries.
The GCs are chosen to represent a realistic subset
of Milky Way GCs with varying BH populations in order to investigate the
effects of BH retention in clusters. Each GC background is
described by an isotropic multi-mass King model.
We produce a large number of realizations for each set of initial parameters
to obtain statistical distributions of the number of ejected binaries and
their relevant properties.
Using the statistics from the GC simulations, we then perform
Monte Carlo simulations to obtain a population of BH-LMXBs
produced by GCs. The GCs and the ejected binaries are evolved in time
through the Milky Way potential while simultaneously accounting for
the stellar evolution of the ejected binaries. 
The resulting
mass-transferring systems make up a previously unexplored
galactic population of BH-LMXBs from GCs. We investigate the
distribution and properties of the resulting population and
its dependence on BH retention in GCs.
Specifically, we find that
in the case of minimal BH retention ($N_\mathrm{BH} = 20$)
no observable BH-LMXBs are produced, while the
$N_\mathrm{BH} = 200$ and $N_\mathrm{BH} = 1000$
cases respectively yield galactic populations of
$25^{+10}_{-6}$ and $156^{+26}_{-24}$ BH-LMXBs.
Furthermore,
we use the resulting population to
determine the most likely candidates for a GC origin in the
population of observed Milky Way BH-LMXBs:
the five systems that are compatible with our simulated
population of BH-LMXBs from GCs are SWIFT J1357.2-0933,
SWIFT J1753.5-0127, XTE J1118+480, and GRO J0422+32.
Future measurements will be necessary to increase support
for a GC origin theory, but if we can confidently attribute a
BH-LMXB to a GC, this would provide strong evidence for
significant BH retention in GCs.

The remainder of this paper is structured as follows.
In section~\ref{sec:methods}, we describe our model
for the GCs and the evolution of a test-binary in a static cluster background.
In section~\ref{sec:simulations}, we lay out how we 
generate the present-day BH-LMXB population from our simulations of Milky Way
GCs. In section~\ref{sec:results}, we review the properties of the
ejected BH binaries along with the distribution and properties of the present-day BH-LMXBs
from GCs. Additionally, we explore the effects of BH retention on the BH-BH merger rate in GCs.
We conclude the section by comparing our results with observations and previous work.
Finally, in section~\ref{sec:conclusion}, we provide concluding remarks.

\section{Methods}\label{sec:methods}
GCs typically contain $\sim10^5-10^6$ stars, which makes them accessible to
modern \textit{N}-body simulations~(e.g.,~\citealt{Zonoozi:2011};~\citealt{Wang:2016})
that can track GC evolution. However,
full \textit{N}-body cluster evolution simulations are
still very computationally expensive, making this method poorly suited
for studying many realizations of different GCs necessary for building
statistics on the evolution of BH binaries inside clusters.
Fokker-Planck methods are more approximate and describe GCs with a
phase-space distribution function for its constituent stars that
evolves via the Fokker-Planck equation, a Boltzmann equation with a
small local collision term that modifies only velocities~(see, e.g.,
~\citealt{Spitzer:1987}).
The Fokker-Planck equation can be numerically
integrated directly~(e.g.,~\citealt{Cohn:1979,Chernoff:1990})
or,
more commonly, integrated with Monte Carlo
methods~(see, e.g.,~\citealt{Henon:1971,Spitzer:1971} and~\citealt{Rodriguez:2016b} for a
comparison between \textit{N}-body and the Monte Carlo approaches).
However, here we are concerned with the evolution of BH binaries in
GCs and not with the GC evolution itself. Hence, we adopt the approach
of \cite{Sigurdsson:1995} and model the evolution of binaries in a
fixed cluster background. We approximate the collision term in the
Fokker-Planck equation analytically to model the effects of distant
encounters as the binary evolves
through the GC. Near encounters are accounted for by explicitly integrating
the three-body equations of motion.
We build up statistics by carrying out simulations of many random realizations of
binaries for a given GC background model. In the following sections,
we describe our method in detail.

\subsection{Model}
\label{sec:model}
Our model, based on~\cite{Sigurdsson:1995},
incorporates a number of assumptions that simplify
the simulations and allow us to perform $\sim10^4$ realizations
for a given cluster model with relatively minimal computational needs.
The three key assumptions are: (i) GCs are well described by a `lowered Maxwellian' distribution function,
(ii) the gravitational potential and distribution functions are stationary, and (iii)
the effect of distant interactions is well described by the leading order
terms in the Fokker-Planck equation.
The `lowered Maxwellian' distribution function,
which eliminates the tail of the Maxwellian velocity distribution,
introduces a maximum energy for stars within the cluster to remain bound. This maximum
energy $\phi(r_\mathrm{t})$ implies a finite mass and a
maximum radius $r_\mathrm{t}$, commonly referred to as the `tidal' radius,
as stars beyond this distance are pulled from the cluster by the
galactic tidal field. Models based on a `lowered Maxwellian',
commonly referred to as King models, 
readily describe many observed clusters~(\citealt{Peterson:1975};~\citealt{Bahcall:1977};~\citealt{Spitzer:1987}).

We evolve a single `test binary', initialized according to section~\ref{sec:bininit},
in a static cluster background described by an isotropic
multi-mass King model~\citep{King:1966} defined by single particle distribution functions
$f_{\alpha}(\bm{r},\bm{v},m_{\alpha})$ for a discrete set of mass groups.
Here, $\bm{r}$ and $\bm{v}$ are the radius and velocity in the cluster
center-of-mass frame and $m_{\alpha}$ is the representative mass of group $\alpha$.
The distribution function
for a given mass group is given by
\begin{equation}
  f_{\alpha}(\varepsilon) =
  \begin{cases}
    \frac{n_{0_{\alpha}}}{(2\pi\sigma_{\alpha}^2)^{3/2}} (\mathrm{e}^{-\varepsilon/\sigma_{\alpha}^2} - 1 )
    & \varepsilon < 0 \\
    0 & \varepsilon \ge 0 \,\,. \\
  \end{cases}
  \label{eq:distf}
\end{equation}
Here, $\varepsilon$ is the energy per unit mass,
$\varepsilon = v^2/2 - \Psi(r)$, and 
$\Psi(r) \equiv \phi(r_\mathrm{t}) - \phi(r)$ is the gravitational
potential relative to that at the tidal radius $r_\mathrm{t}$.
Additionally, $\sigma_{\alpha}$ is the group's velocity dispersion at the core of the cluster
and $n_{0_{\alpha}}$ is a normalization factor.
For an isotropic cluster, the velocity
dispersion reduces to the one-dimensional mean-square velocity, such that 
$3\sigma^2_\alpha = \bar{v}_\alpha^2$. The normalization factor in its full form
is 
\begin{equation}
  n_{0_{\alpha}} = \eta_{\alpha} \frac{n_\mathrm{o}}{\mathrm{e}^{\Psi(0)/\sigma^2_{\alpha}}\mathrm{erf}
    \bigg( \sqrt{\frac{\Psi(0)}{\sigma^2_{\alpha}}} \bigg)-\sqrt{\frac{4\Psi(0)}{\pi \sigma^2_{\alpha}}}
    \bigg(1 + \frac{2\Psi(0)}{3\sigma^2_{\alpha}} \bigg)} \,\,,
  \label{eq:norm}
\end{equation}
where $\eta_{\alpha} = N_{\alpha}/N$ is the number fraction for mass group $\alpha$
and $n_\mathrm{o} = n(0)$ is the central density.

The free structural parameters necessary to specify a 
model cluster, with specified mass groups,
are the mean core velocity dispersion $\bar{\sigma}$, the core number density
$n_\mathrm{o}$, and the potential depth, which is specified
by the dimensionless King parameter
 $W_\mathrm{o} = \Psi(0)/\bar{\sigma}^2$.
The remaining structural parameters, which are fully determined
by the free parameters, are: total mass $M_\mathrm{c}$,
core radius $r_\mathrm{c}$, tidal radius $r_\mathrm{t}$,
and concentration $c=$log$_{10}(r_\mathrm{t}/r_\mathrm{c})$.
The core radius $r_\mathrm{c}$ is defined as the radius at which
the surface brightness has dropped to half the value at the core.

For a given set of masses with corresponding distribution functions, the cluster
satisfies Poisson's equation for the relative potential 
$\nabla^2 \Psi(r) = -4\pi G \sum_{\alpha} \rho_{\alpha}$.
Here, $\rho_{\alpha} = m_{\alpha} n_{\alpha}$, where $n_{\alpha}$ is the number
density of mass group $\alpha$ given by
\begin{equation}
  n_{\alpha} = \int_{0}^{v(r_\mathrm{t})} f_{\alpha}(\bm{r},\bm{v},m_{\alpha}) \, 4 \pi v^2 dv \,\,.
  \label{eq:dens}
\end{equation}
The upper limit of the integral is the maximum allowed
velocity $v(r_\mathrm{t}) = \sqrt{2\Psi(r_\mathrm{t})}$, i.e. the escape velocity.
The object masses $m_\mathrm{\alpha}$ and number fraction $\eta_{0_{\alpha}}$
are determined by the evolved mass function, discussed in section~\ref{sec:emf}. We generate a
model cluster that satisfies Poisson's equation for the specified masses and number
fractions in an iterative fashion. We begin by integrating Poisson's equation out to a radius
$r_\mathrm{t}$, implicitly determined by $\Psi(r_\mathrm{t}) = 0$,
with boundary conditions $\Psi(0) = W_\mathrm{o}$ and $\nabla \Psi(0) = 0$,
and take $\eta_{\alpha} = \eta_{0_{\alpha}}$ as our initial guess.
The actual number fraction of each mass group,
$\eta_{\alpha} = N_{\alpha}/N$, is then calculated using
\begin{equation}
  N_{\alpha} = \int_{0}^{r_\mathrm{t}} n_{\alpha}(r) \, 4 \pi r^2 dr \,\,,
  \label{eq:num}
\end{equation}
along with $N = \sum_{\alpha} N_{\alpha}$. We then update our guess to
$\eta_{\alpha} = (\eta_{\alpha_\mathrm{new}} + \eta_{\alpha_\mathrm{old}})/2$,
where 
$\eta_{\alpha_\mathrm{new}} \rightarrow \eta_{\alpha_\mathrm{old}} \times (\eta_{0_{\alpha}}/\eta_{\alpha})$.
We repeat the above steps until $(\eta_{0_{\alpha}} - \eta_{\alpha})/\eta_{0_{\alpha}} < \delta$
is satisfied for all mass groups, where we have made the
somewhat arbitrary choice of $\delta = 6.25 \times 10^{-3}$ for our
convergence threshold.
This iterative procedure determines the normalization constant $n_{0_{\alpha}}$ and $r_\mathrm{t}$.
Once $r_\mathrm{t}$ is found, the concentration $c=\mathrm{log}_{10}(r_\mathrm{t}/r_\mathrm{c})$ is determined
and the total mass of the cluster $M_\mathrm{c}$ is obtained from
\begin{equation}
  -\nabla \Psi(r_\mathrm{t}) = \frac{GM}{r_\mathrm{t}^2} \,\,.
  \label{eq:cmass}
\end{equation}

The evolution of our `test binary' in the cluster background is affected
by long-range and short-range interactions,
which modify the magnitude and direction of the binary's velocity.
The short-range encounters are accounted for by fully resolving
the three-body interactions, detailed in section~\ref{sec:encres}.
We account for the velocity fluctuations due to long-range
interactions with `field stars', distant cluster stars,
through the diffusion coefficients
$D(\Delta v_i)$ and $D(\Delta v_i \Delta v_j)$
in the Fokker-Planck equation,
\begin{equation}
  \frac{Df}{Dt} = \Bigg( \frac{\partial f}{\partial t} \Bigg)_{\mathrm{enc}} =
  \sum_{i,j} \bigg\{ -\frac{\partial }{\partial v_i}(D(\Delta v_i) f) + \frac{1}{2}\frac{\partial^2 }{\partial v_i \partial v_j}
  (D(\Delta v_i \Delta v_j) f) \bigg\}
  \,\,.
\end{equation}
In this context, a diffusion coefficient $D(X)$ for a variable $X$, 
corresponds to the average change in $X$ per unit time.
Here, we focus on velocity changes 
per unit time  as experienced by the binary due to interactions with the `field stars'.
The form of the coefficients can be derived, for a simple case,
by first considering the change in
velocity of a mass $m_1$, initially at rest, due to an encounter with a second
mass $m_2$ at a relative velocity $v$ with impact parameter $p,$
\begin{equation}
  (\Delta v)^2 = \frac{4 m_1^2}{(m_1+m_2)^2}\frac{v^2}{(1 + (\frac{p}{p_\mathrm{o}})^2)} \,\,,
  \label{eq:dvdt}
\end{equation}
where $p_\mathrm{o} \equiv G(m_1+m_2)/v^2 $ is a reference impact parameter which
causes a deflection of $\pi/2$, consistent with close encounters~(e.g.,~\citealt{Spitzer:1987}).
The average rate of change of the quantity in Equation~\ref{eq:dvdt},
per unit time, due to encounters is then obtained by integrating over the possible
impact parameters for a
given density of field stars $n$, 
\begin{equation}
  D(\Delta v^2)= 2 \pi \int_0^{p_\mathrm{max}} \Delta v^2 p n v dp \,\,,
\end{equation}
up to a maximum allowable impact parameter $p_\mathrm{max}$.
The maximum impact parameter is required to suppress the divergence of the
integral and essentially determines the maximum distance of long-range encounters that
contribute to the velocity perturbations. This maximum value, $p_\mathrm{max}$,
is not explicitly specified, but finds its way into the coefficient calculations
through the so-called Coulomb logarithm,
$\mathrm{ln}\, \Lambda \equiv \mathrm{ln} (p_\mathrm{max}/p_\mathrm{o})$,
which appears as a result of the integration.

We work out the details for the case of an isotropic velocity dispersion
with a density of field stars given by Equation~\ref{eq:dens} and restate the relevant
coefficients we use in our model (cf.~\citealt{Binney:1987}).
These coefficients, which describe the
average rate of change in the velocity of the binary due to long-range encounters,
are used to update the velocity of the binary at each time step. 
The implementation is described further in section~\ref{sec:clustev}.
A detailed derivation and a more general form of the
coefficients can be found in~\cite{Spitzer:1987}.
 
By choosing a coordinate system in which one axis is aligned with the velocity
of the binary, we can decompose $D(\Delta v_i)$
into a coefficient parallel to the binary's velocity
$D(\Delta v_{\parallel})$ and two mutually orthogonal coefficients
perpendicular to the velocity, $D( \Delta v_{\perp})_1$ and
$D(\Delta v_{\perp})_2$.
In an isotropic cluster, there is no preferred direction with regard to
the two perpendicular components, so the contributions from
$D( \Delta v_{\perp})_1$ and $D(\Delta v_{\perp})_2$ tend to cancel each other out.
Their squares, $D(\Delta v_{\perp}^2)_1$ and $D(\Delta v_{\perp}^2 )_2$,
on the other hand, do not and are non-vanishing.
Additionally, we include a quadratic term for the parallel
component $D(\Delta v_{\parallel}^2)$ and in consideration of the symmetry we
retain only the sum of the perpendicular components
$D(\Delta v_{\perp}^2) = D(\Delta v_{\perp}^2)_1 + D(\Delta v_{\perp}^2)_2$.

The diffusion coefficient $D(\Delta v_{\parallel})$ parallel to the binary's motion
is by analogy often referred to
as the coefficient of dynamical friction as it opposes the binary's direction
of motion,
\begin{equation}
  D(\Delta v_{\parallel}) = -\sum_{\alpha} \gamma_{\alpha} \bigg(1 + \frac{m_\mathrm{b}}{m_{\alpha}}\bigg)
  \int_0^v \bigg(\frac{v_{\alpha}}{v}\bigg)^2 f_{\alpha}(v_{\alpha}) dv_{\alpha} \,\,.
\end{equation}
Here, $m_\mathrm{b}$ is the mass of the binary and
$\gamma_{\alpha}\equiv (4\pi G m_{\alpha})^2 \, \text{ln} \, \Lambda$,
where we have chosen to set $\mathrm{ln}\, \Lambda = 10$, a value typical
for GCs~\citep{Spitzer:1987}.
The two remaining coefficients,
\begin{equation}
  \begin{split}
  D(\Delta v_{\parallel}^2) &= \sum_{\alpha} \frac{2}{3}v\gamma_{\alpha}
  \Bigg\{ \int_0^v \bigg(\frac{v_{\alpha}}{v}\bigg)^4 
  + \int_v^\infty \bigg(\frac{v_{\alpha}}{v}\bigg)
  \Bigg\}f_{\alpha}(v_{\alpha}) dv_{\alpha}
  \end{split}
\end{equation}
and
\begin{equation}
  \begin{split}
  D(\Delta v_{\perp}^2) &= \sum_{\alpha} \frac{2}{3}v\gamma_{\alpha} \\
  & \times \Bigg\{ \int_0^v \bigg[3\bigg(\frac{v_{\alpha}}{v}\bigg)^2 -
    \bigg(\frac{v_{\alpha}}{v}\bigg)^4\bigg]
  + 2\int_v^\infty \bigg(\frac{v_{\alpha}}{v}\bigg)
  \Bigg\}f_{\alpha}(v_{\alpha}) dv_{\alpha} \,\,,
  \end{split}
\end{equation}
are strictly positive. These coefficients are responsible for the
stochastic perturbations to the parallel and perpendicular components of the
velocity, which take the binary on a random walk through velocity space
and compete with the slowing due to dynamical friction. We implement
these `random kicks' as discrete changes to the binary's velocity by sampling
from a normalized distribution of the velocity perturbations,
described in section~\ref{sec:clustev}.

\subsection{Initial conditions}
\subsubsection{Evolved mass function}\label{sec:emf}

We obtain an initial distribution of masses in the range $0.08\,M_\odot<m<120\,M_\odot$
from the broken-power-law initial mass function (IMF)
\begin{equation}
  \xi(m) \propto
  \begin{cases}
    m^{-1.3} m_{\mathrm{x}}^{0.3-x_{*}} & m < m_\mathrm{x} \\
    m^{-1.0-x_{*}} & m \ge m_\mathrm{x} \,\,, \\
    \end{cases}
\end{equation}
with $x_{*} = 1.35$ and $m_\mathrm{x} = 0.55\,M_\odot$
chosen to incorporate a Salpeter IMF~\citep{Salpeter:1955} for masses above
$m_\mathrm{x}$ and a Kroupa `correction'~\citep{Kroupa:2001} to masses below $m_\mathrm{x}$
along with a normalization factor
for continuity. Stars with masses below the
main-sequence turn-off,
which we set to $m_{\mathrm{to}} = 0.85\,M_\odot$ \citep{Meylan:1997},
are assumed
not to evolve significantly on the timescale
of the simulations, while masses above $m_{\mathrm{to}}$
are assumed to be completely evolved according to a specified evolved mass function (EMF).
The evolved mass $m_\mathrm{e}$ is determined by the EMF
\begin{equation}
  m_\mathrm{e} = 
  \begin{cases}
    m_{\mathrm{MS}} = m & 0.08\,M_\odot < m \le m_{\mathrm{to}} \\
    m_{\mathrm{WD}} = 0.45 + 0.12(m - 1) &  m_{\mathrm{to}} < m < 8\,M_\odot \\
    m_{\mathrm{NS}} = 1.4 &  8\,M_\odot \le m < 20\,M_\odot \\
    m_{\mathrm{BH}}  = m_{\mathrm{BH}}(m,f_{\mathrm{s_{BH}}}) & 20\,M_\odot < m < 120\,M_\odot \,\,,
    \end{cases}
\end{equation}
where the mass subscripts label the object type and
refer to main sequence (MS), white dwarf (WD), neutron star (NS),
and black hole (BH). We occasionally refer to the set of MS and WD objects
as the non-compact (NC) population.
The MS stars below the turnoff mass
are set to their zero-age main-sequence (ZAMS) mass, the
WD stars are a linear function of their ZAMS mass~\citep{Catalan:2008}, NS
are simply set to $1.4\,M_{\odot}$.
Following the work of~\cite{Sana:2012},
the BHs are assumed to
have formed from two possible channels: stars with companions
that significantly affect the evolution of the star and those stars
that are `effectively single.' Effectively single is used to describe
stars that evolve in isolation as well as those stars
that evolve in wide binaries with minimal interaction. \cite{Sana:2012} estimate that
$\sim70$ per cent of massive stars will have their final state impacted by a companion, which
motivates setting $f_{\mathrm{s_{BH}}} = 0.3$ for the fraction of BHs
that formed in isolation.
This fraction of BHs that evolve from `effectively single' stars are void of the complexities of
binary stellar evolution and are assumed to lose a significant fraction of
their hydrogen shells
to stellar winds before collapsing to a BH.
For the low metallicities typical of GCs, we approximate the mass loss, 
as $\sim 10$ per cent of the initial mass and set $m_\mathrm{e} = 0.9m$.
The remaining $70$ per cent of BHs formed will have evolved with a companion and likely passed
though a common envelope phase, stripping the stars down to their helium (He) cores \citep{Sana:2012,de-Mink:2014}.
Using \verb#MESA#~\citep{Paxton:2011} 
to evolve masses in the range $20\,M_\odot<m<120\,M_\odot$, we obtained the He
core mass as a function of the ZAMS mass in order 
to determine the remnant mass for the remaining $(1-f_{\mathrm{s_{BH}}})$ fraction of BHs
\begin{equation}
 m_\mathrm{e} = m_{\mathrm{He}} = 0.2312 \big(m_{\mathrm{ZAMS}} \big)^{1.1797} \, M_\odot \,\,.
 \label{eq:mzams}
\end{equation}
The stellar evolution performed using \verb#MESA# version 6794, follows
the procedure laid out in ~\cite{Morozova:2015}.
\begin{figure}
  \includegraphics[width=0.99\columnwidth]{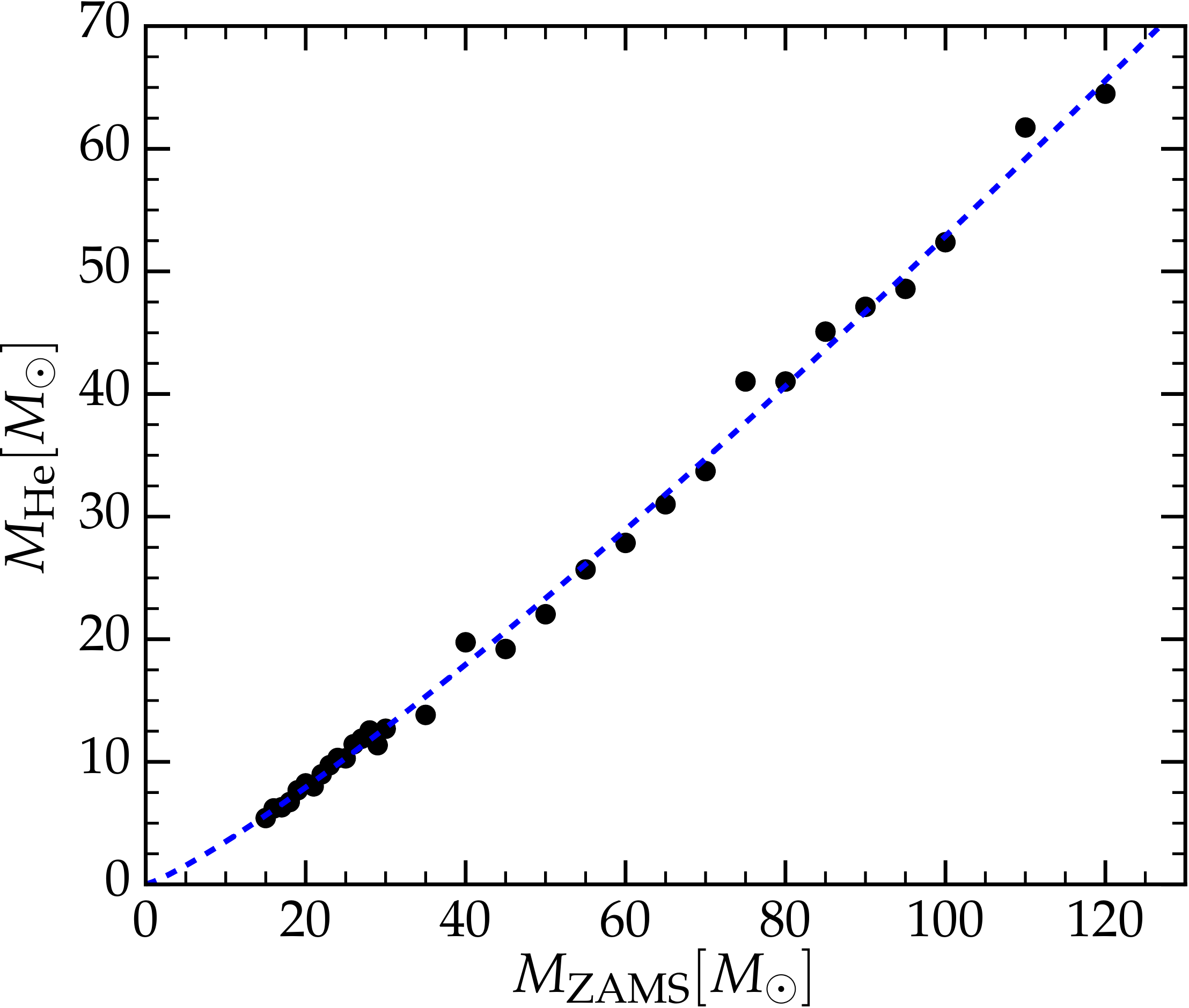}
  \caption{The He core mass (marked by circles) as a function of zero-age main-sequence mass
    from the \texttt{MESA}~\citep{Paxton:2011} runs, along with the fit (blue, dashed line)
    given by Equation~\ref{eq:mzams}.
    For the $\sim 70$ per cent of BHs formed in binaries,
    we approximate the remnant BH mass with the He core mass of the progenitor.
    The remnant mass for the remaining $\sim 30$ per cent of BHs is
    approximated by $0.9 M_\mathrm{ZAMS}$,
    which accounts for the hydrogen mass lost to stellar winds at low metallicity.
    \label{fig:mzamsFit}
    }
\end{figure}
Figure~\ref{fig:mzamsFit} displays the resulting He core mass as a
function of the ZAMS mass from the \verb#MESA# runs with metallicity $Z = 5\times10^{-4}$,
along with
the power-law fit of Equation~\ref{eq:mzams}. This metallicity corresponds to the higher
peak in the 
bimodal, GC metallicity distribution \citep{Harris:1996}.
We use Equation~\ref{eq:mzams} for all clusters,
since the He core masses from models with $Z = 5\times10^{-3}$, corresponding to 
the secondary peak, differ by $\la 10$ per cent.

\begin{table*}
  \begin{tabular}{c|c|c|c|c|c|c}
    Mass group & $m_\mathrm{min} ~[M_{\odot}]$ & $m_\mathrm{max} ~[M_{\odot}]$ & $\bar{m} ~[M_{\odot}]$ & $f_\mathrm{m}$ & $f_\mathrm{n}$ & $f_\mathrm{L}$ \\
\hline \hline
0  & 0.08 & 0.200  & 0.12827 & 0.17531 & 0.42853 & 1.0000 \\
1  & 0.20 & 0.350  & 0.26596 & 0.17757 & 0.20933 & 1.0000 \\
2  & 0.35 & 0.450  & 0.40704 & 0.13954 & 0.10748 & 0.7552 \\
3  & 0.45 & 0.600  & 0.51190 & 0.24921 & 0.15264 & 0.5763 \\
4  & 0.60 & 0.700  & 0.64624 & 0.10020 & 0.04861 & 0.7644 \\
5  & 0.70 & 0.850  & 0.76855 & 0.11027 & 0.04499 & 0.8233 \\
6  & 0.85 & 1.000  & 0.91758 & 0.01161 & 0.00397 & 0.0000 \\
7  & 1.00 & 1.200  & 1.08980 & 0.01005 & 0.00289 & 0.0000 \\
8  & 1.20 & 1.500  & 1.29547 & 0.00527 & 0.00128 & 0.0000 \\
9  & 1.50 & 10.00  & 8.87443 & 0.00143 & 0.00005 & 0.0000 \\
10 & 10.0 & 40.00  & 20.4808 & 0.01261 & 0.00019 & 0.0000 \\
11 & 40.0 & 120.0 & 57.1851 & 0.00693 & 0.00004 & 0.0000 \\

  \end{tabular}
  \caption{Evolved mass groups for NGC 6121 ($N_\mathrm{BH}=200$)
    with corresponding mass index, the lower boundary bin mass $m_\mathrm{min}$,
    the upper boundary bin mass $m_\mathrm{max}$, the average mass of the group $\bar{m}$,
    the fraction of the total mass in the cluster
    $f_\mathrm{m}$, the number fraction with respect to the total number of objects
    in the cluster $f_\mathrm{n}$, and the fraction of luminous
    objects in the group $f_\mathrm{L}$.
    For reference,
    the BH masses occupy the top three mass groups with mean masses of
    $8.87 \, M_\mathrm{\odot}$, $20.48 \, M_\mathrm{\odot}$, and $57.18 \, M_\mathrm{\odot}$.
    }
  \label{table:remnts}
\end{table*}
In addition to specifying the evolved masses, it is also necessary to
specify the number of NS and BH objects retained by the cluster
in its static state.
We specify the retained population of compact objects, comprised of NSs and BHs,
through the retention fractions $f_{\mathrm{r_{NS}}}$ and
$f_{\mathrm{r_{BH}}}$, respectively.
This is necessary since we are modeling the cluster in its
evolved state, a time at which many of the NS and BHs formed within
the cluster have already been ejected due to formation kicks.
Studies of the proper motion of
pulsars suggest that NSs receive kicks in the range of
$200 - 450 \, \mathrm{km} \, \mathrm{s}^{-1}$~\citep{Lyne:1994},
easily exceeding the typical escape velocity of clusters, which is
on the order of tens of $\mathrm{km} \, \mathrm{s}^{-1}$.
However, the observations of pulsars in GCs implies a `retention problem,'
since the observed fraction retained is inconsistent with the average
natal kick velocities being significantly greater than GC escape velocities.
This issue is somewhat reconciled by assuming some NSs form in binaries, which
dampen the kick and allow the GC to maintain a hold on the NS and companion~\citep{Pfahl:2002}.
In consideration of these observations, for the case of NSs,
we retain a constant fraction, $f_{\mathrm{r_{NS}}} = 0.1$, of those produced
by the IMF~(\citealt{Sigurdsson:1995};~\citealt{Pfahl:2002};~\citealt{Ivanova:2008b}).
In the BH case, the distribution of natal kicks is highly uncertain.
Rather than take the retention fraction $f_{\mathrm{r_{BH}}}$
to be a constant across clusters, as in the NS case, we utilize this fraction 
as a free parameter in our models to control the number of retained BHs in each modeled GC.

Once we have determined the evolved masses from the IMF, the masses
are binned into 12 groups. The small number of bins allows for a
proper representation of the true distribution while keeping the
computational costs to a minimum. Poisson's equation is then
integrated to determine the final structural parameters as discussed
in section~\ref{sec:model}. For illustrative purposes,
the evolved mass distribution for NGC 6121 with 200 retained BHs is
given in Table~\ref{table:remnts}.
The bins for each mass group, the mean mass in each bin, and the fraction of luminous objects
are constant across simulations, however the mass fraction and number fraction depend on the
structure of the cluster and the number of BHs.

\subsubsection{Core density}\label{sec:coredens}
As discussed in section~\ref{sec:model}, one of the free parameters in our model
when specifying a cluster's structure is the core number density $n_\mathrm{o}$.
However, because this parameter is not easily observable,
a GC's density is often reported in terms of a central luminosity density $\rho_\mathrm{L}$.
For each mass group we determine a central luminous
number density $n_{\mathrm{L}_{\alpha}}=f_{\mathrm{L}_{\alpha}} \bar{n}_{\alpha}$,
where $f_{L_{\alpha}}$ and $\bar{n}_{\alpha}$ are the fraction of luminous objects
and the core density, respectively, of mass group $\alpha$. The central luminosity
density is then given by $\rho_\mathrm{L} = \sum_{\alpha} L_{\alpha} n_{\mathrm{L}_{\alpha}}$.
In order to account for the variability in the mass-luminosity
relation with stellar mass, we use a parameterized luminosity for each group of
the form $L_{\alpha}=a(m_{\alpha})^b$,
with luminosity coefficients $a=0.23,b=2.3$ for $m_{\alpha}<0.43\,M_\odot$
and $a=1.0,b=4.0$ for the remaining luminous objects~\citep{Duric:2004}.
To ensure that our clusters appropriately model the Milky Way GCs of interest, we compute
$\rho_\mathrm{L}$ for each integrated cluster and adjust $n_\mathrm{o}$ accordingly
to match the observed quantity.

\subsubsection{Binary fraction}
In order to account for the uncertainty in the size of the binary
population within a cluster, we allow for a specifiable binary fraction.
The fraction of objects that are binaries is
\begin{equation}\label{eq:binfrac}
  f_\mathrm{b} = \frac{N_\mathrm{b}}{N_\mathrm{s} + N_\mathrm{b}} \,\,,
\end{equation}
where $N_\mathrm{s}$ and $N_\mathrm{b}$ are the number of single objects and binary
objects, respectively, and the total number of objects in our model clusters
is then $N = N_\mathrm{s} + 2N_\mathrm{b}$. Observations of the binary fraction are limited
to the luminous objects within the cluster. Due to this restriction, we take
the observed fraction to be determined solely by the MS star binary fraction
$f_{\mathrm{obs}} = N_{\mathrm{MS_b}}/(N_{\mathrm{MS_s}} + N_{\mathrm{MS_b}})$,
where, as above,
we respectively refer to $N_{\mathrm{MS_s}}$ and $N_{\mathrm{MS_b}}$ as the number of single and
binary MS stars.
Using the above definitions along with the fraction of all binaries that are MS-MS binaries,
$f_\mathrm{MS_b} = N_\mathrm{MS_b}/N_\mathrm{b}$, and the fraction of objects that are MS stars,
$f_\mathrm{MS} = N_\mathrm{MS}/N$, we convert the observed binary fraction
into a uniform total binary fraction for use in our models through the relation
\begin{equation}\label{eq:binf}
  f_\mathrm{b} = \Bigg(\frac{f_\mathrm{MS_b}}{f_\mathrm{MS}}\frac{(f_\mathrm{obs}+1)}{f_\mathrm{obs}}
  - 1\Bigg)^{-1} \,\,.
\end{equation}

The number
of MS stars $N_\mathrm{MS}$ is determined solely by the IMF and
for the simulations in this study
we use $f_\mathrm{MS_b} = 0.23$~\citep{Fregeau:2009}.
We perform our simulation with
$f_\mathrm{obs}$ covering a range of values, consistent with
observational constraints, between $5 - 20$ per cent~\citep{Milone:2012b}.
We complete an approximately equal number of simulations
for $f_\mathrm{obs}$ taking values from the set $\{0.05,0.10,0.20\}$.
However, we find that this parameter has a negligible effect on
the quantities of interest, so for conciseness, it is not specified in the
simulation parameters.

\subsubsection{Modified black-hole velocity dispersion}\label{sec:equip}
Recent studies of BH retention in GCs have shown clusters initially retain
between $65-90$ per cent of the BHs formed in cluster, with the remainder being
lost due to formation kicks~\citep{Morscher:2015}. This is in contrast
to the long-standing belief that present-day GCs should be nearly void
of BHs.
In addition to the
increase in retention,~\cite{Morscher:2015} also found that the retained
BHs remain well-mixed with the non-BH population.
Follow-up studies
support the idea of a large population of BHs that are spread throughout
the cluster and are consistent with a recent 
$10^6$ \textit{N}-body simulation~\citep{Rodriguez:2016b}.

In the standard King model, it is common to assume that the mass groups satisfy
an equipartition of energy. Specifically, 
\begin{equation}
m_{\alpha} \sigma_{\alpha}^{2} = \bar{m} \bar{\sigma}^{2} \,\,,
\label{eq:equip}
\end{equation}
where $m_\alpha$ and $\sigma_\alpha$ are the mass and velocity dispersion
of mass group $\alpha$, $\bar{m}$ is the mean mass of all objects in the cluster,
and $\bar{\sigma}$ is the mean velocity dispersion.
However, with this equipartition of kinetic energy amongst all mass groups,
the heavier objects then necessarily have lower random velocities compared
to the lighter objects and become trapped deep in the gravitational potential
at the core of the cluster.
With an equipartition of kinetic energy in place, the much more massive BHs
densely populate the central region of the cluster, driving the core radius
to a small fraction of the tidal radius. This disparity between the core radius
and tidal radius leads to concentrations that deviate from observations, limiting
the modeled clusters to supporting only a small number of BHs.
In order to generate clusters with a significant BH population
that are still representative of observed GCs, motivated by~\cite{Morscher:2015},
we implement a velocity dispersion for the BHs away from energy equipartition, 
We maintain an equipartition of energy among the lower-mass objects and use a modified
energy partitioning for the BHs of the form
\begin{equation}
  m_{\beta} \sigma_{\beta}^2 = \frac{\sum m_{\beta}}{\sum m_{\alpha}} \frac{1}{f_\mathrm{s}}
  \bar{m}\bar{\sigma}^2
  \,\,,
\label{eq:bhvel}
\end{equation}
where the indices $\beta$ and $\alpha$
label the mass groups corresponding to BHs and non-BHs, respectively.
Here, $f_\mathrm{s}$ is a specifiable scale factor of order unity.
The $f_s$ parameter is enough to rescale the velocity dispersion for the BHs,
however, the factor involving the mass ratio contributes substantially
and $f_s$ remains of order unity and does not vary wildly across the
GCs we consider.

\begin{figure}
  \includegraphics[width=1.0\columnwidth]{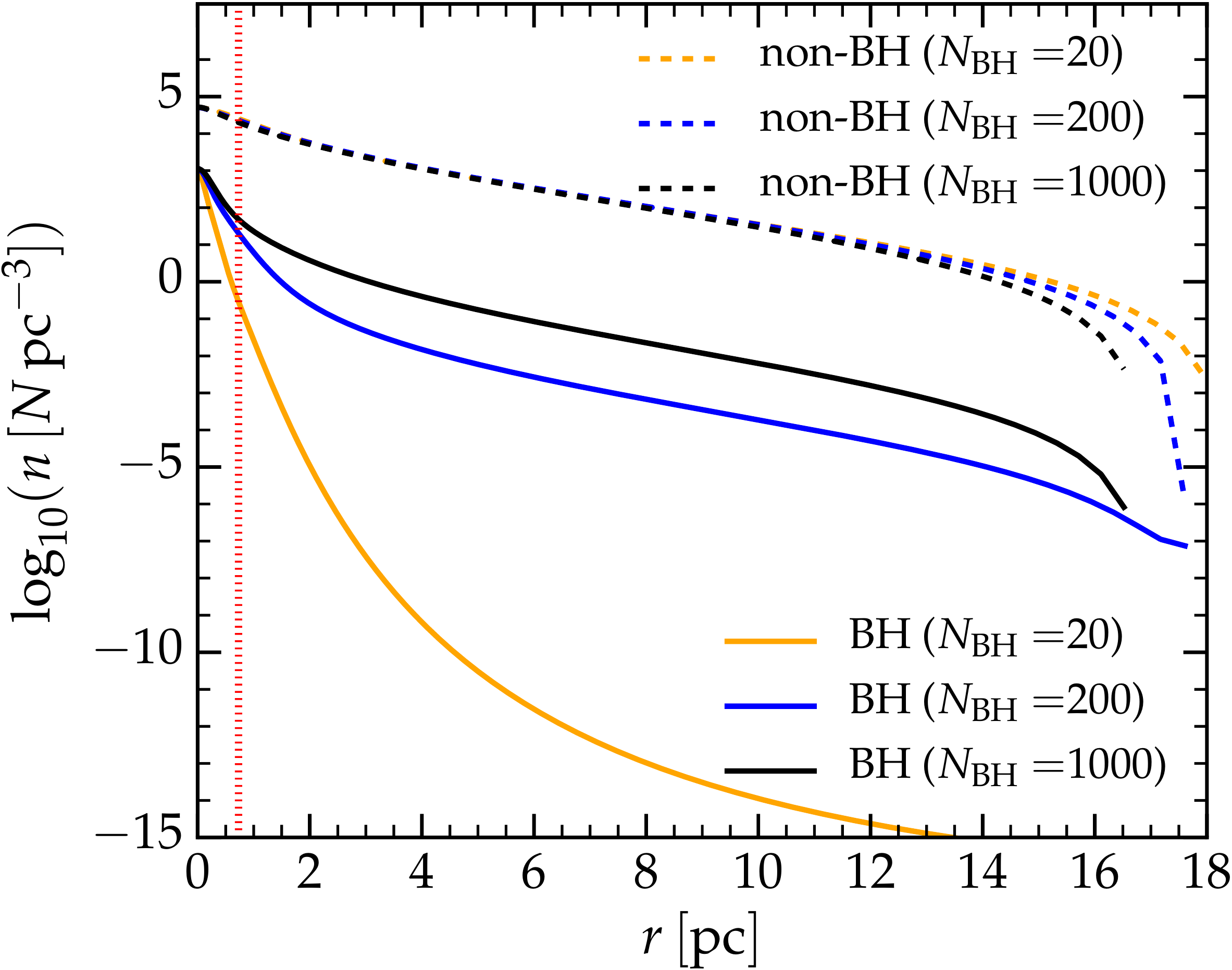}
  \caption{Radial number density profiles for the BH subgroup (solid lines)
    and the non-BH objects (dashed lines) in NGC 6656 for the three
    considered values of $N_\mathrm{BH}$.
    The vertical line (red, dashed), at
    $r_\mathrm{c} = 0.73 \,\mathrm{pc}$, marks
    the core radius for this cluster.
    The non-BH objects are largely
    unaffected by the different numbers of BHs added to the cluster
    and the necessary modification to the velocity dispersion.
    For $N_\mathrm{BH} = 20$,
    the BHs are concentrated in the core region, whereas
    to accommodate $N_\mathrm{BH} \ge 200$, the modified velocity
    dispersion spreads the BHs throughout the cluster with a profile
    similar to that of the non-BH objects.
    \label{fig:BH_dens}
  }
\end{figure}

With this modified BH velocity dispersion in place, we find that we can match
the observed structural parameters of a specific cluster for zero BHs up
to $\sim 20$ BHs, in the case of more massive clusters up to $\sim 100$ BHs,
and in the most massive clusters up to $\sim 1000$ BHs. We vary the number of
BHs residing in the cluster by adjusting the scale factor $f_s$ in
Equation~\ref{eq:bhvel}
and the fraction retained, $f_\mathrm{r_{BH}}$, introduced in section~\ref{sec:emf}.
To illustrate the spreading of the BHs, we present in Figure~\ref{fig:BH_dens}
the radial density profiles for the BHs and the non-BH objects for different populations
of retained BHs in the cluster model representing NGC 6656.
In the case of minimal BH retention, the BH number density falls off quickly outside of the core,
which for our model of NGC 6656 is located at $r_\mathrm{c} = 0.73 \,\mathrm{pc}$
and is marked by a vertical line in Figure~\ref{fig:BH_dens} for reference.
However, in the case of many BHs, the modified velocity dispersion extends the number density
profile radially, spreading the BHs throughout the cluster, without affecting the
central density. The distribution of non-BH objects is largely unaffected by the
change in BH numbers.

\subsubsection{Binary initialization}\label{sec:bininit}
We choose the initial masses for our `test binary'
by randomly sampling from the evolved mass
distribution and reject those that do not contain at least one BH.
If one of the component masses falls within a
mass bin with a non-zero luminous population,
we then sample from the luminous mass fraction 
to determine whether the low-mass object is an MS star or WD.
Additionally, if the selected mass is
in the turnoff group, $0.63\,M_\odot \le m \le 0.8\,M_\odot$, then the object is chosen
to be a giant with probability $P = 0.095f_\mathrm{L}$, where $f_\mathrm{L}$ is the luminous
fraction for the turnoff-mass group. The probability for giants is adopted from~\cite{Sigurdsson:1995} and represents the approximate fraction of the cluster age that giants in this
mass range survive.
Once the masses and object types are established,
the BH radii are set to the Schwarzschild radius $R_\mathrm{BH} = 2GM/c^2 $,
while the stellar radii are determined as described in~\cite{Sigurdsson:1995}.
The eccentricity of the binary $e$ is specified by sampling from the 
probability density function $f(e) = 2e$~\citep{Jeans:1919}, commonly referred to
as a `thermal' eccentricity distribution.
The semi-major axis $a$ is obtained
from a distribution uniform
in $\log_{10} a$ in the range $-3 \le\log_{10} (a ~\mathrm{au}^{-1})\le1 $.
To avoid an immediate merger of the objects in
our initial binary, we enforce $a > f_\mathrm{tid}(R_1 + R_2)/(1-e)$,
where $R_i$ are the radii of each component of the binary and $f_\mathrm{tid}=3.1$,
by letting
$a \rightarrow 2a$ until this condition is satisfied. The factor $f_\mathrm{tid}$
is chosen based on the separation at which tidal effects would induce a merger
~\citep{Lee:1986}.
Once the binary parameters
are set, we sample the primary-mass number density profile $n_{\alpha}(r)$ to
determine the binary placement within the cluster and obtain a velocity from the
primary-mass velocity distribution function at $r$.

\subsection{Evolution of the `test binary'}
\label{sec:clustev}
Once we have an appropriate model, which satisfies the structural
parameters for a specific cluster and
an initial binary, we then evolve this single binary within the
cluster background. In addition to the static potential, we
include the interaction terms discussed in section~\ref{sec:model}.
To account for dynamical friction,
the diffusion coefficient $D(\Delta v_{\parallel})$ is added to the potential
gradient to create a smooth effective acceleration $\bm{a}_\mathrm{eff} = \nabla \Psi(r) + D(\Delta v_{\parallel})$.
This smooth force is integrated using a $4\mathrm{th}$ order Runge-Kutta integrator,
which is discussed in detail in section~\ref{sec:timestep}.
The quadratic scattering terms, or random `kicks', are implemented
by discretely updating the corresponding velocity components at each time step $\Delta t$.
As discussed in section~\ref{sec:model}, the diffusion coefficient for $\Delta v^2$, of the
  form $D(\Delta v^2)$, represents the change in this quantity per unit time, i.e. $\Delta v^2/ \Delta t$.
  We update the velocity at each time step by sampling from the normal distribution of kicks
  through
\begin{equation}
  \begin{split}
  \Delta v_{\parallel} &= X \sqrt{D(\Delta v_{\parallel}^2) \Delta t} \,\,, \\
  \Delta v_{\perp_1} &= Y \sqrt{\frac{1}{2} D(\Delta v_{\perp}^2) \Delta t} \,\,, \\
  \Delta v_{\perp_2} &= Y \sqrt{\frac{1}{2} D(\Delta v_{\perp}^2) \Delta t} \,\,,
  \end{split}
\end{equation}
where $X$ and $Y$ are random numbers with mean values of zero and standard deviations of one.

At each time step, we also consider the evolution of the binary's semi-major axis $a$
and its eccentricity $e$ due to gravitational wave (GW) emission. If the BH is in a binary with
another compact object --- which includes BHs, NSs, and WDs --- then we implement the evolution of $a$
and $e$ according to the gravitational radiation formalism of~\cite{Peters:1964}.
In these cases, we also calculate the time until coalescence $t_\mathrm{d}$ due to the
decay of $a$, and if this will occur within the current time step, $t_\mathrm{d}<\Delta t$,
we consider this a GW merger. If the merger is of a BH-BH or BH-NS binary, we
add a recoil velocity, or `kick', based on the fits to numerical relativity simulations
given by~\cite{Campanelli:2007} with initial spin magnitudes and orientations assigned 
as in \citet{Clausen:2013}.

\subsubsection{Short-range encounters}
\label{sec:encounters}
As the binary moves throughout the cluster, at each time step,
we check for the possibility of a short-range encounter with a single star.
Since the effects of long-range interactions are accounted for by the diffusion
coefficients (section~\ref{sec:model}), here we focus on capturing the effects 
due to strong three-body interactions with much smaller impact parameters.
We limit the range of encounters to include only those three-body interactions
that result in a resonance, exchange, ionization, or the occasional flyby.
We accomplish this by
choosing the maximum impact parameter to be
\begin{equation}\label{eq:maxp}
  p=a[B + C(1+e)] \,\,,
\end{equation}
where we have set $B=4$ and $C=0.6$ following~\cite{Hut:1983a}. The choice
of these coefficients is intended to limit the number of weak encounters
that have minimal impact on the binary, as these still require full resolution
of the encounter, which is one of the more computationally intensive tasks
during evolution.
However, the coefficients only provide an approximate
contour in the space of initial conditions, hence the occasionally flyby.
The cross section for an
encounter to take place between the binary and a star of mass $m_\mathrm{\alpha}$
with velocity $\bm{v}_{\alpha}$ is
\begin{equation}\label{eq:cs}
  \sigma(\bm{v},\bm{v}_{\alpha}) = \pi p^2
  + \frac{2\pi G (m_\mathrm{b} + m_{\alpha})p}{|\bm{v}-\bm{v}_{\alpha}|^2} \,\,,
\end{equation}
(see, e.g.,~\citealt{Spitzer:1987}).
We then calculate the expected encounter rate between the binary
and each mass group
\begin{equation}\label{eq:rate}
  \Gamma(r,v,\alpha) = \int \sigma(\bm{v},\bm{v}_{\alpha}) |\bm{v}-\bm{v}_{\alpha}|
  f_{\alpha}(v_{\alpha}) dv_{\alpha} \,\,,
\end{equation}
and from this assign the probability of interacting with mass group $\alpha$ to be
\begin{equation}
  P_{\alpha} = \Gamma(r,v,\alpha) \Delta t \,\,.
  \label{eq:prob}
\end{equation}
An encounter is deemed to have occurred, based on a random generated number $Z$ from
a uniform distribution between 0 and 1, if $Z$ is less than the total
probability $P = \sum_{\alpha} P_{\alpha}$. The total probability is implicitly
constrained to be less than unity by controlling the time-step size $\Delta t$,
which is discussed in more detail in the subsequent section.
In the case that $Z < P$, we select
the third star $m_3$ based on the relative probabilities $P_{\alpha}$ and initiate
our three-body integration scheme explained in section~\ref{sec:threeb}.

\subsubsection{Time stepping}\label{sec:timestep}
We use a $4\mathrm{th}$ order Runge-Kutta integrator
to evolve the effective acceleration introduced in section~\ref{sec:clustev}
as well as the three-body
interactions described in section~\ref{sec:threeb}.
During integrations,
we utilize a time step reduction scheme requiring that the accuracy of the solution
does not vary by more than a tolerance of $\epsilon_\mathrm{rk}=10^{-5}$ when
the time step is halved. 
The initial integration time step $\Delta t = \lambda (1+r)/(1+v)$
is dynamically determined to account for the position
and velocity of the binary in the cluster, with $\lambda=0.1$
chosen to produce a time step that is a fraction of the
core dynamical time $r_\mathrm{c}/\bar{\sigma}$
for a binary at rest in the core. This time stepper accounts
for the higher density in the core and the
enlarged cross section at small velocities. Although this
choice of time step is usually sufficient,
some extra care needs to be taken
when using $\Delta t$ in Equation~\ref{eq:prob}
to determine the encounter probability, so that
the total probability does not exceed unity. To ensure
that we correctly sample the encounter probabilities,
by satisfying the constraint $P \ll 1$, we set $P_\mathrm{max} = 0.1$
and enforce $P<P_\mathrm{max}$ by reducing the time step $\Delta t$
when necessary.
For the case $P>P_\mathrm{max}$, we decrease the succeeding time step
by letting $\lambda \rightarrow 0.9(P_\mathrm{max}/P)\lambda$.
During the subsequent step, if $\lambda < \lambda_\mathrm{o}$, where
$\lambda_\mathrm{o} = 0.1$ is the fiducial value, and $P<P_\mathrm{max}$,
we allow the time step to increase slowly by setting
$\lambda \rightarrow 1.1\lambda$. Once $\lambda > \lambda_\mathrm{o}$
and the probability is satisfactorily small, which often occurs
once the binary migrates out of the problematic dense region,
we reset the time step factor to $\lambda = \lambda_\mathrm{o}$.

\subsubsection{Three-body interactions}\label{sec:threeb}
In the case of an encounter, the relative probabilities described
in section~\ref{sec:encounters} determine the mass and velocity of the
third object. We take this sampled velocity $v_\mathrm{3}$ to be the velocity of
the third body at infinity and calculate the relative velocity at
infinity for the encounter from 
$v_{\infty} = |\bm{v}-\bm{v}_{3}| = \sqrt{v^2 + v_{3}^{2} - 2v v_{3} \, \mathrm{cos} \chi}$.
Given $v$ and the sampled $m_3$ and $v_3$,
the relative velocity at infinity is
determined up to the $\mathrm{cos} \chi$ term,
which for an isotropic King model distribution function can be sampled
from an analytic expression for $\chi \in [0,\pi]$ as in~\cite{Sigurdsson:1995}.
With the mass of the third body and the relative velocity known,
the maximum impact parameter is obtained from the cross section for the encounter
\begin{equation}
  \pi p_\mathrm{max}^2 = \sigma(\bm{v},\bm{v}_{3}) = \pi p^2 \bigg(1 + \frac{2G(m_\mathrm{b} + m_3)}{p v_{\infty}^2} \bigg) \,\,,
\end{equation}
with $p$ defined in Equation~\ref{eq:maxp}.
The actual impact parameter for the encounter is sampled from a uniform distribution
in the area spanned by the maximum impact parameter $\pi p_\mathrm{max}^2$.
The angles that comprise the remaining free variables necessary to specify the
initial conditions are the projected true anomaly $f$ of the binary at
the time that the incoming third body reaches pericenter, two angles $\theta$
and $\phi$ specifying the initial location of the third body with respect
to the binary center-of-mass, and the impact orientation $\psi$, which
specifies the angle of the impact parameter in a plane transverse to the incoming
velocity of the third body. Theses four angles
are sampled in a manner consistent with~\cite{Hut:1983a}.
With the initial conditions specified, the explicit integration is
performed with a modified scheme based on~\cite{Sigurdsson:1993B}.

We modify the original method of a fixed initial distance of the third star,
at $R_\mathrm{in} = 20a$, to one of variable distance to improve
efficiency and to prevent the case of long three-body interactions that
can exceed the cluster time step. The addition of massive BHs introduces
the possibility for wide binaries with orbital separations much greater
than those for which the previous method was suited to handle. With a fixed
choice for the distance of the third star from the binary, interactions such
as distant flybys, which are the quickest to resolve computationally and
have little impact on the binary, often take a time that exceeds the cluster
evolution time step and leads to the possibility of missing other
probable encounters.

To represent the three-body system as an isolated
one, and to reduce excessive time spent integrating long approaches,
we require that $R_\mathrm{in} \le R_\mathrm{max}(n)$,
where $R_\mathrm{max}(n) = (4 \pi n/3)^{-1/3}$ is the `interparticle' distance
and is a function of the local density $n(r)$. Once $R_\mathrm{in}$ is
specified, we determine the relative velocity $v_\mathrm{in}$ at $R_\mathrm{in}$
based on the relative velocity at infinity. With these two quantities specified,
we approximate the time for a flyby as $\delta t = 2 R_\mathrm{in}/v_\mathrm{in}$.
For the case in which $\delta t > \Delta t$,
we let $R_\mathrm{in} \rightarrow (\Delta t / \delta t) R_\mathrm{in}$, calculate
$v_\mathrm{in}$ at the new initial distance and recompute the new estimated time.
We repeat this procedure until the estimated time is roughly the
same as the cluster time step, $0.9 < \delta t / \Delta t < 1.1$.
One important caveat is that this could lead to placing the third object
too close to the binary, spoiling the assumption of an object at infinity
approaching a well defined binary. To address this issue,
we maintain one extra condition on the initial distance specification,
a consideration for which we are willing to forgo our time step restrictions:
that $(a/R_\mathrm{in})^3 \le 0.01$.

To increase the speed of the three-body integration, we move from a constant
integration time step to one that is dynamical. We choose a maximum time step
$\delta T_\mathrm{max}$
to be an arbitrarily small fraction $\epsilon=6.25\times10^{-3}$ of the binary period $T_\mathrm{b}$,
i.e. $\delta T_\mathrm{max} = \epsilon T_\mathrm{b}$.
At the end of each integration step,
we update the time step to
$\delta T = \epsilon (r_\mathrm{min}/v_\mathrm{max})$,
where $r_\mathrm{min}$ is the minimum separation between any pair of
the three objects and $v_\mathrm{max}$ is the largest velocity of the three bodies.
This sets the time step to the maximum allowable value in 
consideration of the need to resolve the dynamics of the three objects
or any potentially bound pair.
In some instances, a resonance can form a temporarily bound triple system,
causing the integrator to reach the maximum number of steps
$N_\mathrm{max} = 2\times10^{6}$ or to exceed the arbitrarily specified
maximum allowable time of $5\Delta t$.
Under these rare circumstances, we
reinitialize the system with newly sampled initial angles and restart
the integration.
In addition to the occasional long-lasting semi-stable triples that form,
there are also instances when a binary makes its way to the core
where the average timescale necessary to resolve the three body encounters
begins to approach the timescale for the evolution of the binary in the
cluster.
Since we calculate three-body encounters decoupled from the binary's
evolution in the cluster, we are forced to terminate the
run in such cases.
As the cluster timescale is inversely proportional to the cluster density,
this situation is most likely to occur in the densest clusters.
As a result of this timescale termination criterion, although
a similar number of realizations are performed for each cluster,
the highest density clusters have noticeably fewer runs than the
lower density clusters, as is observable in the rightmost column
of Table~\ref{table:runs}.
For standard encounters, which are often much shorter than
the cluster time step,
we periodically check whether the
interaction has resolved
--- according to the criteria discussed in the following section --- 
and in the case that a new binary has formed, even temporarily, we
update $\delta T_\mathrm{max}$ with the period of this new binary.
\begin{table*}
\begin{tabular}{l|llllrr}
Name & M \, $[M_{\odot}]$ & $\sigma^2$ \, $[\mathrm{cm}^2 \mathrm{s}^{-2}]$ & $\rho_\mathrm{L}$ \, $[L_{\odot} \mathrm{pc}^{-3}]$ & $c$ & $N_\mathrm{BH}$ & $N_\mathrm{runs}$ \\
\hline \hline
Pal 13 & 5.12$\times 10^{3}$ & 8.10$\times 10^{9}$ & 1.45 & 0.66 & 20       & 15232 \\

NGC 6838 & 3.67$\times 10^{4}$ & 5.29$\times 10^{10}$ & 6.76$\times 10^{2}$ & 1.15 & 20    & 18364 \\
                                   & & & & & 200    & 20430 \\

NGC 6535 & 5.93$\times 10^{4}$ & 5.76$\times 10^{10}$ & 5.19$\times 10^{2}$ & 1.33 & 20    & 35865 \\
                                   & & & & & 200    & 33561 \\

NGC 6362 & 1.17$\times 10^{5}$ & 7.84$\times 10^{10}$ & 1.95$\times 10^{2}$ & 1.09 & 20    & 32544 \\
                                   & & & & & 200    & 33798 \\

NGC 5053 & 1.66$\times 10^{5}$ & 1.96$\times 10^{10}$ & 3.47 & 0.74 & 20    & 69058 \\
                                   & & & & & 200    & 74681 \\

NGC 6121 & 2.25$\times 10^{5}$ & 1.60$\times 10^{11}$ & 4.37$\times 10^{3}$ & 1.65 & 20    & 14429 \\
                                   & & & & & 200    & 17884 \\
                                   & & & & & 1000   & 24667 \\

NGC 5694 & 2.92$\times 10^{5}$ & 3.36$\times 10^{11}$ & 8.91$\times 10^{3}$ & 1.89 & 20    & 14029 \\
                                   & & & & & 200    & 13382 \\
                                   & & & & & 1000   & 17445 \\

NGC 6093 & 3.67$\times 10^{5}$ & 1.54$\times 10^{12}$ & 6.17$\times 10^{4}$ & 1.68 & 20    & 7435 \\
                                   & & & & & 200    & 7019 \\
                                   & & & & & 1000   & 4645  \\

NGC 5286 & 4.80$\times 10^{5}$ & 6.56$\times 10^{11}$ & 1.26$\times 10^{4}$ & 1.41 & 20    & 6761 \\
                                   & & & & & 200    & 10032 \\
                                   & & & & & 1000   & 8196 \\

NGC 6656 & 5.36$\times 10^{5}$ & 6.08$\times 10^{11}$ & 4.27$\times 10^{3}$ & 1.38 & 20    & 12539 \\
                                   & & & & & 200    & 20993 \\
                                   & & & & & 1000   & 14832 \\

NGC 1851 & 5.61$\times 10^{5}$ & 1.08$\times 10^{12}$ & 1.23$\times 10^{5}$ & 1.86 & 20    & 7189 \\
                                   & & & & & 200    & 6950 \\
                                   & & & & & 1000   & 4563  \\

NGC 6205 & 6.27$\times 10^{5}$ & 5.04$\times 10^{11}$ & 3.55$\times 10^{3}$ & 1.53 & 20    & 13444 \\
                                   & & & & & 200    & 24899 \\
                                   & & & & & 1000   & 23583 \\

NGC 6441 & 1.30$\times 10^{6}$ & 3.24$\times 10^{12}$ & 1.82$\times 10^{5}$ & 1.74 & 20    & 2388 \\
                                   & & & & & 200    & 2439  \\
                                   & & & & & 1000   & 2463  \\

NGC 104 & 1.45$\times 10^{6}$ & 1.21$\times 10^{12}$ & 7.59$\times 10^{4}$ & 2.07 & 20     & 9545 \\
                                   & & & & & 200    & 10467 \\
                                   & & & & & 1000   & 8559 \\

NGC 5139 & 2.64$\times 10^{6}$ & 2.82$\times 10^{12}$ & 1.41$\times 10^{3}$ & 1.31 & 20    & 13197 \\
                                   & & & & & 200    & 17466 \\
                                   & & & & & 1000   & 23513 \\

\end{tabular}
\caption{Summary of simulations. Listed are the 15 GCs modeled
  for evolution along with the total cluster mass $M_\mathrm{c}$, squared velocity dispersion $\sigma^2$,
  the luminous core density $\rho_\mathrm{L}$ and concentration $c$. The clusters are
  ordered by total mass. There are 39 independent
  models after taking into account the number of BHs retained by the cluster.
  Medium to high mass clusters can accommodate large BH populations without
  disrupting the listed structural
  parameters. The size of the BH population in lower-mass clusters is either (1) limited in number
  by the IMF or (2) by the ability
  of the cluster to maintain the model structural parameters in their presence; in these cases, the
  cluster is not used for evolutions and is omitted from the table.
  In the final column we list the total number of evolutions performed for each case.}
\label{table:runs}
\end{table*}

\subsubsection{Encounter resolution}\label{sec:encres}
We first identify a potential binary among the triple system composed
of the original binary, $m_1$ and $m_2$, and the third mass $m_3$,
by selecting the pair with the largest gravitational binding energy.
We refer to the masses in the potential binary
as $\bar{m}_1$ and $\bar{m}_2$,
which may no longer correspond to the original binary composed of $m_1$ and $m_2$.
The remaining object, which is not part of the potential binary, is labeled $\bar{m}_3$
which is distinct from $m_3$.
All unbarred variables
represent the initial configuration where the third object is incoming,
while barred variables refer to the system where a binary has been
identified and the encounter is nearly resolved.
The encounter can be resolved in three ways: (I) there is a well defined bound binary
system with the third object unbound and moving off to infinity,
(II) a merger has occurred or (III) the system is completely ionized.

For case (I),
we terminate
the integration once the following criteria are all satisfied:
(i) the third body has achieved the minimum required separation from
the binary, $|\bar{r}_3 - (\bar{m}_1\bar{r}_1 + \bar{m}_2\bar{r}_2)/(\bar{m}_1+\bar{m}_2)|
> \mathrm{max}\{R_\mathrm{max}(n),1.1 \, R_\mathrm{in} \}$,
(ii) the eccentricity $\bar{e}$ of $\bar{m}_1$ and $\bar{m}_2$ is less than unity,
(iii) $\bar{m}_1$ and $\bar{m}_2$ are bound, specifically $\bar{E}_\mathrm{b} < 0$,
and (iv) $\bar{m}_3$ is unbound, i.e. $\bar{E}_3 > 0$.
Here, $\bar{E}_\mathrm{b}$ is the total energy of the final binary and
$\bar{E}_3$ is the total energy of the third body.
In addition to the above requirements, to determine the final state
of the `isolated' binary, we continue the integration until the total
potential energy between $\bar{m}_3$ and each mass in the binary is a fraction
of the total energy of the system $E$, specifically
\begin{equation}
  \frac{G\bar{m}_1\bar{m}_3}{|\bar{r}_1-\bar{r}_3|}
  + \frac{G\bar{m}_2\bar{m}_3}{|\bar{r}_2-\bar{r}_3|} > 0.05E \,\,.
\end{equation}

In case (II),
two of the bodies merge and the third body
is either unbound or forms a new binary with the merger product.
The criteria for mergers is based on the distance of nearest approach $d$
between two bodies during the three body encounter. In the case of a
potential merger between two BHs, the merger criterion is
$ d \le R_1 + R_2$.
For the remaining
merger situations, the criterion remains $d = f_\mathrm{tid} (R_1+R_2)$, 
as adopted from~\cite{Sigurdsson:1995}, using the same
value for $f_\mathrm{tid}$ as introduced in section~\ref{sec:bininit}.
When a merger occurs between
the BH companion and the third body, if the merger product remains bound to
the BH, this dynamically formed binary becomes our new `test binary',
which we continue to follow and evolve within the cluster.
Similarly, if the BH merges with a third
body and we still have a bound binary system, we again continue to follow this binary.
However, if the BH becomes unbound by merging with another body
or becomes unbound from a merger product, we handle the newly single BH
as described in the subsequent section.
In each of these cases, the position of the new binary, or single BH,
is updated by continuing along the original binary trajectory and the velocity
is updated by converting from the three-body center-of-mass
frame, where the three-body integration is performed, back to the
cluster frame.

The result of the encounter can also end in complete ionization, case (III).
Ionization occurs in the case of ill-defined binaries that will inevitably be unbound 
if, given that all previous criteria are satisfied, either
(a) the eccentricity of $\bar{m}_1$ and $\bar{m}_2$ satisfies $1-\bar{e} < 1\times 10^{-7}$ or
(b) $|\bar{r}_1 - \bar{r}_2| > R_\mathrm{max}(n)$ is satisfied. Additionally,
ionization occurs
if $\bar{m}_i\bar{v}_i^{2} > 2 \big( \bar{m}_i \bar{m}_j/|\bar{r}_i-\bar{r}_j|
+ \bar{m}_i \bar{m}_k/|\bar{r}_i-\bar{r}_k| \big)$ is true for all masses at any time,
with $i\ne j\ne k$ taking on values $\{1,2,3\}$. This last criterion is a
straightforward definition for a totally unbound triple.
In addition to these choices for ionization during three-body encounters,
there is one other instance in which the binary can be dissociated.
For very wide binaries, the encounters are dominated by
repeated grazing encounters with low mass stars, which
tend to further widen the orbital separation.
As a result, strong interactions become less likely and
the binary will inevitably be dissociated by the increasing
occurrence of these slowly ionizing encounters.
For this reason, we use the encounter rate
to define a maximum semi-major axis of
dynamically formed binaries as 
\begin{equation}
  a_\mathrm{max}(\Gamma) = \bigg(\frac{G m_\mathrm{b}}{3(2\pi \Gamma)^2} \bigg)^{1/3} \,\,,
\end{equation}
which is equivalent to requiring a minimum of three orbits between encounters.
Here, the total encounter rate $\Gamma = \sum_{\alpha} \Gamma(r,v,a)$ is
a sum over the rate associated with each mass group defined by Equation~\ref{eq:rate}).
The final criterion for ionization is then 
$a> \mathrm{min}\{a_\mathrm{max}(\Gamma),R_\mathrm{max}(n)\}$.

\subsubsection{Single black holes}\label{sec:ion}
As described in the previous section, a BH can become
single due
to three-body dynamics such as exchange, merger,
or through the dismantling of a binary that exceeds our
large $a$ or large $e$ criteria.
In the case of a single BH, we allow for the solitary
BH to form a new binary by interacting with existing
binaries within the cluster.

In order to accomplish this, we need to know the probability for
the following encounter,
\begin{equation}\label{eq:exch}
  (m_\mathrm{1},m_\mathrm{2}) + m_\mathrm{BH} \rightarrow (m_\mathrm{BH},m_\mathrm{2}) + m_\mathrm{1} \,\,,
\end{equation}
in which the BH exchanges with $m_1$ into a binary originally composed of masses $m_1$ and $m_2$.
We also consider the possibility that $m_\mathrm{BH}$ and $m_2$ undergo an exchange, which
contributes to the total probability that the BH will exchange into the binary. However,
for conciseness in deriving the probability of exchange,
we will focus specifically on the encounter described by Equation~\ref{eq:exch},
later adding the contribution from the reaction where the subscripts are
interchanged.
Unfortunately, we can no longer compute the probability for 
encounter as in section~\ref{sec:encounters}, since we do not possess a
distribution function for binaries.
However, by considering the reverse reaction of Equation~\ref{eq:exch}, given by
\begin{equation}\label{eq:rev}
  (m_\mathrm{BH},m_\mathrm{2}) + m_\mathrm{1} \rightarrow (m_\mathrm{1},m_\mathrm{2}) + m_\mathrm{BH} \,\,,
\end{equation}
and relating this to the one of interest, we can obtain the encounter probability
for the BH to exchange into an existing binary in the same way that we compute
encounters for a binary composed of a BH and a companion.

We use the seminumerical fit of~\cite{Heggie:1996}, 
\begin{equation}
  \bar{\sigma}_{1,2} = \bigg(\frac{M_{23}}{M_{123}} \bigg)^{1/6} \bigg(\frac{m_{3}}{M_{13}} \bigg)^{7/2}
  \bigg(\frac{M_{123}}{M_{12}} \bigg)^{1/3} \bigg(\frac{M_{13}}{M_{123}} \bigg) \ g(2,3,1) \,\,,
  \label{eq:dimcs}
\end{equation}
as the dimensionless cross section for a generically labeled single mass $m_3$ to exchange into a binary
of masses $m_1$ and $m_2$ to form a new binary composed of $m_3$ and $m_2$, with
$m_1$ being ejected. In this notation, uppercase masses represent the sum
of the mass subscripts, i.e. $M_{ij} = m_{i} + m_{j}$.
The coefficient $g(2,3,1)$ is a numerical fitting factor designed
to improve the analytically derived fit. This dimensionless cross section
$\bar{\sigma}_{1,2}$ is related to the dimensionful
cross section for exchange $\Sigma_{1,2}$
through
\begin{equation}\label{eq:exchcs}
  \bar{\sigma}_{1,2} = \frac{2 |\bm{v}_\mathrm{1,2} - \bm{v}_\mathrm{3}|^2 }{\pi G M_{123} a_{1,2}}
  \Sigma_{1,2} \,\,.
\end{equation}
The existing binaries that the BH is likely to encounter,
which have remained intact in the cluster over long timescales,
can be considered `hard'. These `hard' binaries are characterized
by having a binding energy $U_\mathrm{bin}$ that exceeds the average
energy of the other stars in the cluster $|U_\mathrm{bin}| > \frac{1}{2}\bar{m} \bar{\sigma}^2$
and this is what allows them to stay intact over such long timescales.
In this case, we approximate the total encounter cross section by the dominant
gravitational focusing term in Equation~\ref{eq:cs}, explicitly:
\begin{equation}\label{eq:hard}
  \sigma_\mathrm{1,2} \simeq \frac{2 \pi G M_\mathrm{123} a_\mathrm{1,2}}{|\bm{v}_\mathrm{1,2}
    - \bm{v}_\mathrm{3}|^2} \,\,.
\end{equation}
Finally, relating Equation~\ref{eq:exchcs} and Equation~\ref{eq:hard}
allows us to express the cross section for exchange in terms of
the total encounter cross section $\sigma_\mathrm{1,2}$ through
\begin{equation}\label{eq:dimcross}
  \Sigma_\mathrm{1,2} = \big( \bar{\sigma}_\mathrm{1,2}/4 \big) \sigma_\mathrm{1,2} \,\,.
\end{equation}
Evidently, the dimensionless cross section for exchange is related to
the fractional probability that the total encounter ends in the specific exchange
we previously described.
Considering Equation~\ref{eq:hard}
and assuming the relative velocities are similar for the forward and reverse reactions, we can 
relate the forward and backward total cross sections through
$\sigma_\mathrm{1,2} = (\frac{a_\mathrm{1,2}}{a_\mathrm{2,3}})\sigma_\mathrm{2,3}$.
Since the energy given to the binary is comparable to the energy
required to destroy it,
$m_\mathrm{1} m_\mathrm{2}/a_\mathrm{1,2} \sim  m_\mathrm{2} m_\mathrm{3}/a_\mathrm{2,3}$,
we can recast the relation in terms of the masses alone:
\begin{equation}\label{eq:relate}
  \sigma_\mathrm{1,2} = \bigg( \frac{m_\mathrm{1}}{m_\mathrm{3}} \bigg) \sigma_\mathrm{2,3} \,\,.
\end{equation}
The cross section for the specific exchange of $m_\mathrm{3}$ for $m_\mathrm{1}$ in terms of the
total encounter cross section of the original binary
is found by substituting Equation~\ref{eq:relate} into Equation~\ref{eq:dimcross}, yielding 
\begin{equation}
  \Sigma_\mathrm{1,2} = \bigg(\frac{\bar{\sigma}_\mathrm{1,2} m_\mathrm{1}}{4 m_\mathrm{3}} \bigg)
  \sigma_\mathrm{2,3} \,\,.
\end{equation}
By writing the exchange probability in terms of the post-exchange binary, we can now
utilize the same procedure described in section~\ref{sec:encounters}. In this formalism,
$m_\mathrm{3}$ represents the BH and we return to referring to this body as $m_\mathrm{BH}$,
while $m_\mathrm{1}$ goes to $m_{\alpha}$, a variable companion used for computing
the relative probabilities for each mass group $\alpha$.
First we select a companion object $m_\mathrm{2}$ for the BH on the left-hand
side of Equation~\ref{eq:rev}.
We obtain $m_\mathrm{2}$ by sampling from the local number density
and determine $a$ and $e$ for the binary as in section~\ref{sec:bininit}.
The probability of the encounter described by Equation~\ref{eq:exch},
where the BH exchanges places with $m_{\alpha}$
in a binary composed of $m_\mathrm{2}$ and $m_{\alpha}$ is then,
\begin{equation}
  P_{\alpha,2} = \Delta t \int \bigg(\frac{\bar{\sigma}_{\alpha,\mathrm{2}} m_{\alpha}}{4 m_\mathrm{BH}}
  \bigg)
  \sigma_\mathrm{2,BH}(\bm{v},\bm{v}_{\alpha}) |\bm{v}-\bm{v}_{\alpha}|
  f_{\alpha}(v_{\alpha}) dv_{\alpha} \,\,.
\end{equation}
The usefulness of the manipulations in this section is most clearly seen by writing this in
terms of Equation~\ref{eq:prob}:
\begin{equation}
P_{\alpha,\mathrm{2}} = \bigg(\frac{\bar{\sigma}_{\alpha,\mathrm{2}} m_{\alpha}}{4 m_\mathrm{BH}} \bigg)
P_{\alpha} \,\,,
\end{equation}
which in practice makes computing the exchange probabilities as easy as rescaling
our standard encounter computations by the parenthetical factor.
Since we also allow for the BH to exchange
with $m_\mathrm{2}$, we also consider the probability 
$P_{\mathrm{2},\alpha} = \bigg(\frac{\bar{\sigma}_{\mathrm{2},\alpha} m_\mathrm{2}}{4 m_\mathrm{BH}} \bigg)
P_\mathrm{2}$.

We apply one final rescaling to account for the density of binaries
that are of type $m_\mathrm{2}$ and $m_{\alpha}$. We assume that the fraction of objects
that are binaries $f_\mathrm{b}$ is constant throughout the cluster with the
value specified by Equation~\ref{eq:binf}.
The density of binaries is then $n_\mathrm{b}(r) = (\frac{f_\mathrm{b}}{1+f_\mathrm{b}}) n(r)$, which
is derived from Equation~\ref{eq:binfrac}.
Additionally, we also assume
that the fraction of binaries of a given type is
constant at all cluster radii, 
$n_{i j}(r) = f_{i/j} n_\mathrm{b}(r)$.
Here, $f_{i /j}$ represents the fraction of binaries
that have a star of type $i$ and a star of type $j$,
e.g. $f_{\mathrm{NS} / \mathrm{MS}}$ is the fraction of all binaries that are
composed of an NS and an MS star.
For binaries composed of only MS or WD we use values of 
$f_\mathrm{MS/MS} = 0.23$, $f_\mathrm{MS/WD} = 0.44$,
and $f_\mathrm{WD/WD} = 0.32$~\citep{Fregeau:2009}.
The remaining one percent of binaries contain at least one BH or NS, for which we compute the
binary fraction through $f_{i/j} = 0.01(\frac{N_i}{N})(\frac{N_\mathrm{j}}{N_\mathrm{BH+NS}})$,
where $i$ can be any object type, $j$ is limited to BH or NS, $N$ is the total number
of objects in the cluster, and $N_\mathrm{BH+NS}$ is the total number of BHs and NSs.

The final total probability for the BH to exchange into a binary,
given the sampled mass $m_\mathrm{2}$, is then
\begin{equation}
  P_\mathrm{exch}(r) = \sum_{\alpha} n_{\alpha \mathrm{2}}(r)
  \bigg( \frac{P_{\alpha,\mathrm{2}}}{n_{\alpha}(r)} + \frac{P_{\mathrm{2},\alpha}}{n_{2}(r)} \bigg) \,\,.
\end{equation}
Here we divide out the respective local density picked up in the integration
of the distribution function in order to enforce our assumption of a uniform binary fraction.
If an exchange is determined to occur based on this total probability, we select a specific
binary for the encounter based on the relative probabilities of exchange for each mass group $m_\alpha$.
With a binary in hand, we initiate our three-body system,
which is run until we get the proper outcome dictated
by the encounter cross section --- i.e. that $m_\mathrm{BH}$ exchanges
with the appropriate mass in the binary.

\section{Simulations}\label{sec:simulations}

We present 698,486 realizations from 15 GC models
with total masses in the range of $5.12\times 10^{3}$ -- $2.64\times10^{6}$
$M_\odot$, velocity dispersions
covering $9\times 10^{4}$ -- $1.8\times 10^{6}$ $\mathrm{cm}\,\mathrm{s}^{-1}$,
core densities of
$1.45$ -- $1.23\times 10^{5}$ $\mathrm{pc}^{-3}$,
and concentrations spanning $0.66$ -- $2.07$.
The simulations are summarized in Table~\ref{table:runs}, which
includes the catalog name for the modeled cluster, total mass,
velocity dispersion squared, central luminosity density, concentration,
the number of retained BHs in the model, and the total number of completed runs.
The simulations are run for $t=10^{10}$ years or until the single/binary is
ejected from the cluster, when $r>r_\mathrm{t}$.

\subsection{Structural parameters}\label{sec:structp}
\begin{figure}
  \includegraphics[width=1.0\columnwidth]{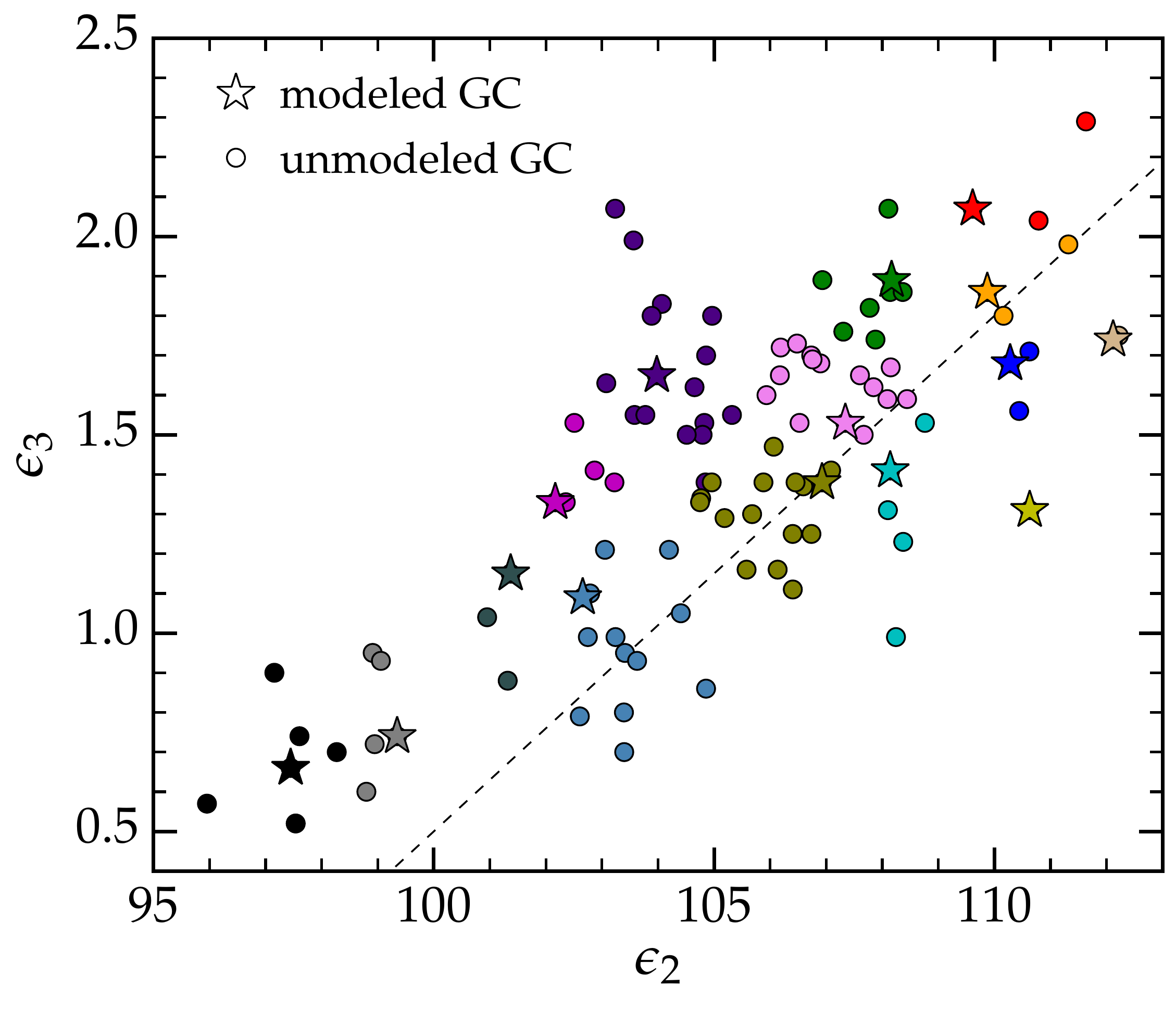}
  \caption{The distribution of non core-collapsed
    Milky Way GCs in a face-on view of the fundamental plane.
    The color of each unmodeled GC (marked by circles) indicates the
    corresponding modeled GC (marked by stars) that serves as its proxy 
    for determining the properties of the ejected binaries.
    The plane is defined by $\epsilon_2 = 2.05 \, \log_{10} E_\mathrm{b}^{*} + \log_{10} L$ and
    $\epsilon_3 = c$, with the dashed line corresponding
    to the fit $\epsilon_3 = -12.5 + 0.13\epsilon_2$.
    Here $c$ is the concentration, $L$ is the total luminosity,
    and $E_\mathrm{b}^{*}$ is an additional parameter related to
    $L$ (see section~\ref{sec:galev} for additional details).
    \label{fig:fp}
  }
\end{figure}

In our framework, a GC's structure is determined
by four parameters: the total cluster mass $M_\mathrm{c}$, the core
velocity dispersion $\sigma$, the core luminosity density $\rho_\mathrm{L}$,
and the concentration $c$.
~\cite{McLaughlin:2000} finds that GCs described
by single-mass isotropic King models are fully defined by four
independent physical parameters:
the mass-to-light ratio $\Upsilon_{\mathrm{v},0}$,
total binding energy $E_\mathrm{b}$, central concentration $c$,
and total luminosity $L$.
Furthermore,~\cite{McLaughlin:2000} shows
that Milky Way GCs
lie in a `fundamental plane' and thus can be fully described by just
two independent parameters, $c$ and $L$.
A face-on view of the fundamental plane is defined by the axes
$\epsilon_2 = 2.05 \, \log_{10} E_\mathrm{b}^{*} + \log_{10} L$ and
$\epsilon_3 = c$.
The apparent dependence on the third quantity $\log_{10} E_\mathrm{b}^{*}$
is due to a rotation in the larger three dimensional space in order to remove
projection effects. However, this is reconciled by showing that this third
parameter, $E_\mathrm{b}^{*}$, is fully described by the luminosity,
such that $E_\mathrm{b}^{*}(L)$~\citep{McLaughlin:2000}.
With the space of physical
clusters reduced to the fundamental plane, we determine a representative
group of 15 Milky Way clusters by sampling from the two-dimensional
distribution. A face-on view of the fundamental plane is
given in Figure~\ref{fig:fp}, which includes all GCs from the
Harris catalog~\citep[][2010 edition]{Harris:1996} for which observed concentrations
are available. We omit clusters identified in the catalog
as core-collapsed, since these are not generally well described by King
models. This includes those with $c = 2.5$, an arbitrary value
assigned to clusters in the catalog
with central density cusps indicative of
core collapse.
There are 125 Milky Way GCs remaining after core-collapse pruning;
of these, 15 GCs are chosen as representative models, in an attempt
to properly cover the fundamental parameter space.
The 15 Milky Way GC models representative of the 125 Milky Was GCs
are described in Table~\ref{table:runs}
and represented by stars in Figure~\ref{fig:fp} to visualize our
coverage of the fundamental parameter space. 

As stated in section~\ref{sec:model}, our input parameters for
specifying the structure of a cluster are the core velocity dispersion $\bar{\sigma}$,
the central density $n_\mathrm{o}$, and the King parameter $W_\mathrm{o}$.
The mean core velocity
dispersion $\bar{\sigma}$ is chosen to be the observed value listed in the
Harris catalog. The core number density $n_\mathrm{o}$ is adjusted until
the central luminosity density $\rho_\mathrm{L}$
is consistent with observation. Finally, the King parameter
$W_\mathrm{o}$, which sets the depth of the potential, is varied
until the cluster has the desired total mass $M_\mathrm{c}$ and
concentration $c$. Once we have a model for a given GC,
we add BHs by increasing the fraction of retained BHs
$f_\mathrm{r_{BH}}$, where a value of unity corresponds to
retention of all BHs produced according to the IMF.
For a given number of BHs in the cluster,
we use the parameter $f_s$ in Equation~\ref{eq:bhvel}
to adjust the BH velocity dispersion such that
the overall structure of the cluster is unaffected by the presence
of a significant number of BHs. However, we find that
there is a limit to the number of BHs each cluster can
harbor.
For the lowest-mass clusters, such as Pal 13, setting
the retention factor to unity, $f_\mathrm{r_{BH}}=1$, in order to maximize
the number of BHs retained by the cluster produces a peak number
of $\sim 20$ BHs.
In this case, the number of BHs retained by
the cluster is inherently limited by its structure.
More generally, for lower-mass clusters that allow for
more BHs, the large number of BHs can become problematic
as they become a more significant part of the total mass of the cluster.
As the fraction of the total mass in BHs increases, 
the BHs begin to affect the structural parameters
such that no set of initial parameters exists
that satisfy the observed structure of the GC.
We find that for many of the lower-mass clusters we are only able
to simulate populations of 20 or 200 BHs (cf. Table~\ref{table:runs}).

\subsection{Galactic evolution}\label{sec:galev}
The GC evolution models, described in detail in section~\ref{sec:methods},
compute the properties of the BH binaries at the moment they are ejected from a GC.
Determining the present day properties of potentially observable, ejected BH binaries
requires further modeling that tracks both the evolution of ejected binaries in
the Milky Way potential and the internal evolution of each binary. In this section, we
describe Monte Carlo models for the subsequent evolution of the ejected binaries that
are seeded with results from our GC models.

\subsubsection{Globular cluster orbits}
We first build a sample of GCs to include in our galactic evolution simulations.
The orbit of a cluster is specified by its location on the sky (right ascension
and declination), distance from the Sun $D_\odot$, radial velocity $v_r$, and proper
motion $\mu_\alpha$ and $\mu_\delta$. Of the 125 non core-collapsed GCs in the
Harris catalog~\citep[][2010 edition]{Harris:1996}, we are
able to find literature values for the orbital parameters of 106 of these clusters in
the catalogs of~\cite{Moreno:2014} and~\cite{Kharchenko:2013}.
For clusters appearing in both catalogs we use the values given in~\citet{Moreno:2014}. 

To begin each realization in our Monte Carlo ensemble, we initialize the GC orbits by
sampling the uncertainty in their current positions and velocities. We assume normally
distributed errors and use the quoted uncertainties in $v_r$, $\mu_\alpha$,
and $\mu_\delta$. Following~\citet{Krauss:2003}, we assume a 6 per cent error in $D_\odot$.
After the orbit is specified, we integrate it 10 Gyr backward in time, corresponding
to the duration of our GC dynamical simulations.      

The orbits of the GCs, and the ejected binaries, are integrated using the~\verb#python#
galactic dynamics library~\verb#galpy#\footnote{\url{http://jobovy.github.io/galpy/}}~\citep{Bovy:2015}. We model the Milky Way gravitational
potential using the built in~\verb#MWPotential2014#. The potential includes contributions
from the galactic bulge, disk, and halo, which have been fit to observational data to
provide a realistic model of the Milky Way potential. The physical scale of the potential is
set using the distance from the center of the Galaxy to the Sun and the circular velocity
of the Sun, which we set to 8 kpc and 220 km s$^{-1}$, respectively. For all calculations,
we use the~\verb#dopr54_c# integrator, a fast implementation of a high order
Dormand-Prince method included with~\verb#galpy#.    

Now that we have calculated the positions and velocities of the Milky Way GCs during the
past 10 Gyr, the next step is to determine the properties of any potential BH-LMXBs
ejected by these clusters. Since our dynamical simulations only include a subset of
the galactic GCs, we use the results from the 15 GCs simulated in Table~\ref{table:runs}
as proxies for the ejected binary populations produced by the remaining 110 clusters
in our galactic evolution models.
For each of the unmodeled clusters,
a proxy cluster is selected by finding the nearest simulated cluster in the fundamental
plane (see section~\ref{sec:structp}). Specifically, we find
$\min \Big[ (\epsilon'_\mathrm{2,i}-\epsilon'_\mathrm{2,j})^2
+(\epsilon'_\mathrm{3,i}-\epsilon'_\mathrm{3,j})^2 \Big]$,
where the $i$ index runs over all 106 clusters in the galactic evolution models,
the $j$ index runs over the 15 clusters included in our GC dynamics models,
and the primes denote the normalized versions of $\epsilon_2$ and $\epsilon_3$
restricted to the range $[0,1]$.
Figure~\ref{fig:fp} shows the proxy cluster chosen for each GC, by assigning the
same color marker to each GC as the color of the proxy cluster used, which are
marked by colored stars.
To ensure the robustness of this method for choosing a proxy cluster,
we assign a proxy by two additional methods.
One secondary method is to assign the proxy cluster based on the 
minimum distance in the fundamental plane using the unnormalized axes
$\epsilon_2$ and $\epsilon_3$.
The second alternative is by identifying 
the most similar cluster using the structural
parameters $M_\mathrm{c}$, $\sigma$, and $\rho_\mathrm{L}$ weighted according
to the strengths of the correlations between these parameters and the ejected
binary populations, which are explored in~\ref{sec:ejects}.
Selecting the proxy cluster by any of these three methods gives
similar results in our galactic evolution models.
In fact, all three methods will select the same proxy cluster
for all but $\sim$15 of the 110 unmodeled GCs in our study.
In what follows, we discuss models that use the scaled distance in the fundamental plane to
assign the proxy cluster.

\subsubsection{The ejected binaries}\label{sec:ejbin}
The output of our GC dynamical simulations describes the properties of the BH-binaries
ejected from GCs. To model the present day population of BH-LMXBs that are ejected
from GCs, we use as inputs for our galactic evolution models:
the ejection time $t_\mathrm{ej}$, ejection velocity $v_\mathrm{ej}$,
and the properties of the binary, the semi-major axis $a$, eccentricity $e$,
the mass of the BH primary $m_\mathrm{1}$, and the mass of the companion $m_\mathrm{2}$.
This is accomplished by constructing
empirical cumulative distribution functions (CDFs) of these quantities for each
of the 37 sets of parameters listed in Table~\ref{table:runs}, and then sampling these
distributions in our Monte Carlo models. We assume that
the ejection time, ejection velocity, and binary properties are independent and sample the marginal distributions of each.

In the GC dynamical models, $a$, $e$, $t_\mathrm{ej}$, and $v_\mathrm{ej}$ are
treated as continuous variables. As such, we are able to sample the CDFs for these
quantities directly. We fit cubic splines to the empirical CDFs and invert the
distributions by interpolation. The GC dynamical models treat $m_\mathrm{1}$
and $m_2$ as discrete quantities, which fall into the mass bins shown in
Table~\ref{table:remnts}. In our galactic evolution models, however, we want to
consider continuous masses. To accomplish this, we first determine an object's mass bin by sampling the discrete CDF output by the dynamical simulations. Next, we sample
the mass distribution within that bin using the evolved mass function described
in section~\ref{sec:emf}. Using these CDFs, we are able to generate sample populations
of the BH-binaries ejected by the 106 GCs in our galactic evolution simulations.

During each realization, for each cluster, we first determine the number of
binaries that the cluster will eject during the 10 Gyr simulation by sampling a Poisson
distribution with rate parameter  $\langle N_\mathrm{ej} \rangle$ (third column of Table~\ref{table:ejections}).
Once we have determined the number $N_\mathrm{bin}$ of ejected binaries,
we draw $N_\mathrm{bin}$ samples
from the $a$, $e$, $m_\mathrm{1}$, $m_2$, $t_\mathrm{ej}$, and $v_\mathrm{ej}$
distributions.

Since the internal evolution of a binary is independent of its orbit in
the Galaxy, we separately compute the full internal evolution of the binary
using the rapid binary population synthesis code~\verb#BSE# described
in~\cite{Hurley:2002} with the updates described
in~\cite{Clausen:2012b} and~\cite{Lamberts:2016}.~\verb#BSE#
combines interpolated stellar evolution models with recipes for mass-transfer and
other binary evolution processes to enable rapid modeling of a binary system's
lifetime. Binary population synthesis calculations employ parameterized models
to describe poorly understood processes in binary evolution. In our ~\verb#BSE# runs, we assume that stable mass transfer is
conservative. Additionally, we use a common-envelope efficiency parameter of $1.0$
and include the effects of tidal circularization.

We use each set of $a$, $e$, $m_\mathrm{1}$, $m_2$ as the initial conditions
for a~\verb#BSE# run. Furthermore, we set the companion star's metallicity to
that of its parent GC and its age to $t_\mathrm{ej}$. The latter has little
effect because most of the ejected stars have lifetimes that exceed 10 Gyr.
The binary is evolved for $t_\mathrm{evol} = 10 \, \mathrm{Gyr} - t_\mathrm{ej}$, i.e.,
to the present day. Systems are discarded if the companion star is not
overflowing its Roche-lobe and transferring mass to the BH at the end of
the simulation. For each mass transferring binary, we determine the position
$\bm{r}_\mathrm{GC}$ and velocity $\bm{v}_\mathrm{GC}$ of its parent GC
at $t_\mathrm{ej}$. We initialize an orbit for the ejected binary
at $\bm{r}_\mathrm{GC}$ and $\bm{v}_\mathrm{GC} + \bm{v}_\mathrm{ej}$,
assuming that the binaries are ejected isotropically. With the initial conditions
determined, we then evolve these binaries using~\verb#galpy# to determine
their positions at the present day.

Our galactic evolution models consider three BH-retention scenarios. In the
first, we assume that most BHs are ejected and use the results from our GC
dynamics models with $N_\mathrm{BH}=20 M$. We refer to this set of models
as MIN. In the second case, referred to as 200, we assume moderate BH retention,
using the results from our GC dynamics models with $N_\mathrm{BH}=200$.
Finally, in a case denoted MAX, we consider significant BH retention by utilizing
the GC dynamics models with $N_\mathrm{BH} = 1000$. In cases where we are
unable to generate a background cluster model with the appropriate $N_\mathrm{BH}$,
we use the results from the model with nearest $N_\mathrm{BH}$ simulated for that same
cluster.
We compute $10^{4}$ realizations for the MIN and 200 cases and
$5\times 10^{3}$ realizations for the MAX case.

\section{Results}\label{sec:results}
Our simulations of binary-single star interactions in GCs provide
us with statistical properties of the ejected BH binaries
they produce including ejection time $t_\mathrm{ej}$, ejection
velocity $v_\mathrm{ej}$, the orbital properties $a$ and $e$,
and the component masses $m_1$ and $m_2$.
Combining these results with the methods described in
section~\ref{sec:galev},
we obtain predictions for the distribution
and properties of the galactic population of BH-LMXBs produced by GCs.
Additionally, the simulations
allow us to explore merger events involving BHs
such as gravitational radiation driven mergers, both
in the cluster and post-ejection,
as well as those mergers that occur during three-body encounters.
We describe these results in detail below.

\subsection{Ejected black-hole binaries}\label{sec:ejects}

\begin{table*}
  \begin{tabular}{l|r|lll}
    Name & $N_\mathrm{BH}$ & BH-NC & BH-NS & BH-BH \\ 
\hline \hline
Pal 13 & 19.64 & 3.14 & $7.74 \times 10^{-3}$ & $1.40 \times 10^{-1}$ \\
NGC 6838 & 20.61 & $6.33 \times 10^{-1}$ & $3.59 \times 10^{-2}$ & $5.08 \times 10^{-1}$ \\
 & 174.55 & $2.56 \times 10^{1}$ & $2.39 \times 10^{-1}$ & 2.67 \\
NGC 6535 & 19.89 & $2.35 \times 10^{-1}$ & $1.72 \times 10^{-2}$ & $3.64 \times 10^{-1}$ \\
 & 198.95 & 5.12 & $1.24 \times 10^{-1}$ & 2.08 \\
NGC 6362 & 20.22 & $1.61 \times 10^{-1}$ & $6.83 \times 10^{-3}$ & $2.31 \times 10^{-1}$ \\
 & 199.33 & 1.07 & $2.36 \times 10^{-2}$ & 1.55 \\
NGC 5053 & 21.71 & $2.04 \times 10^{-2}$ & $3.14 \times 10^{-4}$ & $7.31 \times 10^{-2}$ \\
 & 199.65 & $1.79 \times 10^{-1}$ & $2.67 \times 10^{-3}$ & $4.96 \times 10^{-1}$ \\
NGC 6121 & 20.70 & $3.11 \times 10^{-1}$ & $6.31 \times 10^{-2}$ & $4.96 \times 10^{-1}$ \\
 & 200.53 & 1.74 & $3.03 \times 10^{-1}$ & 2.66 \\
 & 1039.16 & $1.02 \times 10^{2}$ & 1.43 & 8.17 \\
NGC 5694 & 20.49 & $2.29 \times 10^{-1}$ & $1.18 \times 10^{-1}$ & $7.49 \times 10^{-1}$ \\
 & 200.39 & 1.54 & 1.02 & 4.19 \\
 & 1001.94 & $3.21 \times 10^{1}$ & 2.87 & $1.54 \times 10^{1}$ \\
NGC 6093 & 19.85 & $1.01 \times 10^{-1}$ & $4.81 \times 10^{-2}$ & $3.42 \times 10^{-1}$ \\
 & 198.31 & 1.13 & $3.67 \times 10^{-1}$ & 2.66 \\
 & 1004.51 & $1.23 \times 10^{1}$ & 2.38 & $1.31 \times 10^{1}$ \\
NGC 5286 & 12.29 & $6.00 \times 10^{-2}$ & $2.36 \times 10^{-2}$ & $1.91 \times 10^{-1}$ \\
 & 198.28 & $9.29 \times 10^{-1}$ & $5.93 \times 10^{-2}$ & 2.08 \\
 & 787.45 & 4.42 & $3.84 \times 10^{-1}$ & 5.48 \\
NGC 6656 & 19.80 & $6.79 \times 10^{-2}$ & $1.42 \times 10^{-2}$ & $2.57 \times 10^{-1}$ \\
 & 205.86 & $4.22 \times 10^{-1}$ & $8.83 \times 10^{-2}$ & 1.74 \\
 & 1000.35 & 3.10 & $2.02 \times 10^{-1}$ & 5.09 \\
NGC 1851 & 20.76 & $8.37 \times 10^{-2}$ & $4.91 \times 10^{-2}$ & $4.74 \times 10^{-1}$ \\
 & 203.71 & $8.79 \times 10^{-1}$ & $4.98 \times 10^{-1}$ & 3.09 \\
 & 1039.94 & $1.98 \times 10^{1}$ & 1.82 & $1.03 \times 10^{1}$ \\
NGC 6205 & 20.10 & $6.13 \times 10^{-2}$ & $1.79 \times 10^{-2}$ & $2.62 \times 10^{-1}$ \\
 & 199.58 & $4.25 \times 10^{-1}$ & $5.61 \times 10^{-2}$ & 1.70 \\
 & 998.62 & 1.61 & $1.27 \times 10^{-1}$ & 5.36 \\
NGC 6441 & 20.98 & $3.51 \times 10^{-2}$ & $1.76 \times 10^{-2}$ & $3.16 \times 10^{-1}$ \\
 & 212.57 & $9.59 \times 10^{-1}$ & $8.72 \times 10^{-2}$ & 1.57 \\
 & 1010.37 & 3.69 & $8.20 \times 10^{-1}$ & 4.72 \\
NGC 104 & 22.49 & $6.60 \times 10^{-2}$ & $3.06 \times 10^{-2}$ & $4.49 \times 10^{-1}$ \\
 & 222.95 & 1.09 & $4.47 \times 10^{-1}$ & 2.89 \\
 & 979.55 & 3.09 & 2.52 & 8.41 \\
NGC 5139 & 20.84 & 0.00 & 0.00 & $2.53 \times 10^{-2}$ \\
 & 207.50 & $1.19 \times 10^{-2}$ & 0.00 & $1.19 \times 10^{-1}$ \\
 & 1009.04 & 0.00 & 0.00 & $2.57 \times 10^{-1}$ \\

  \end{tabular}
  \caption{Expected number of binary ejections. For each cluster and number of
    retained BHs, we list the exact number of BHs in the cluster along with the
    expected number of ejections over the cluster lifetime for three binary types:
    BH-NC, BH-NS, and BH-BH.
    The clusters follow the same order as Table~\ref{table:runs}, sorted
    according to increasing total cluster mass. The values of $N_\mathrm{BH}$
    are non-integer values as a consequence of modeling the population with
    a smooth distribution function.
  }
  \label{table:ejections}
\end{table*}

\begin{figure}
  \includegraphics[width=1.0\columnwidth]{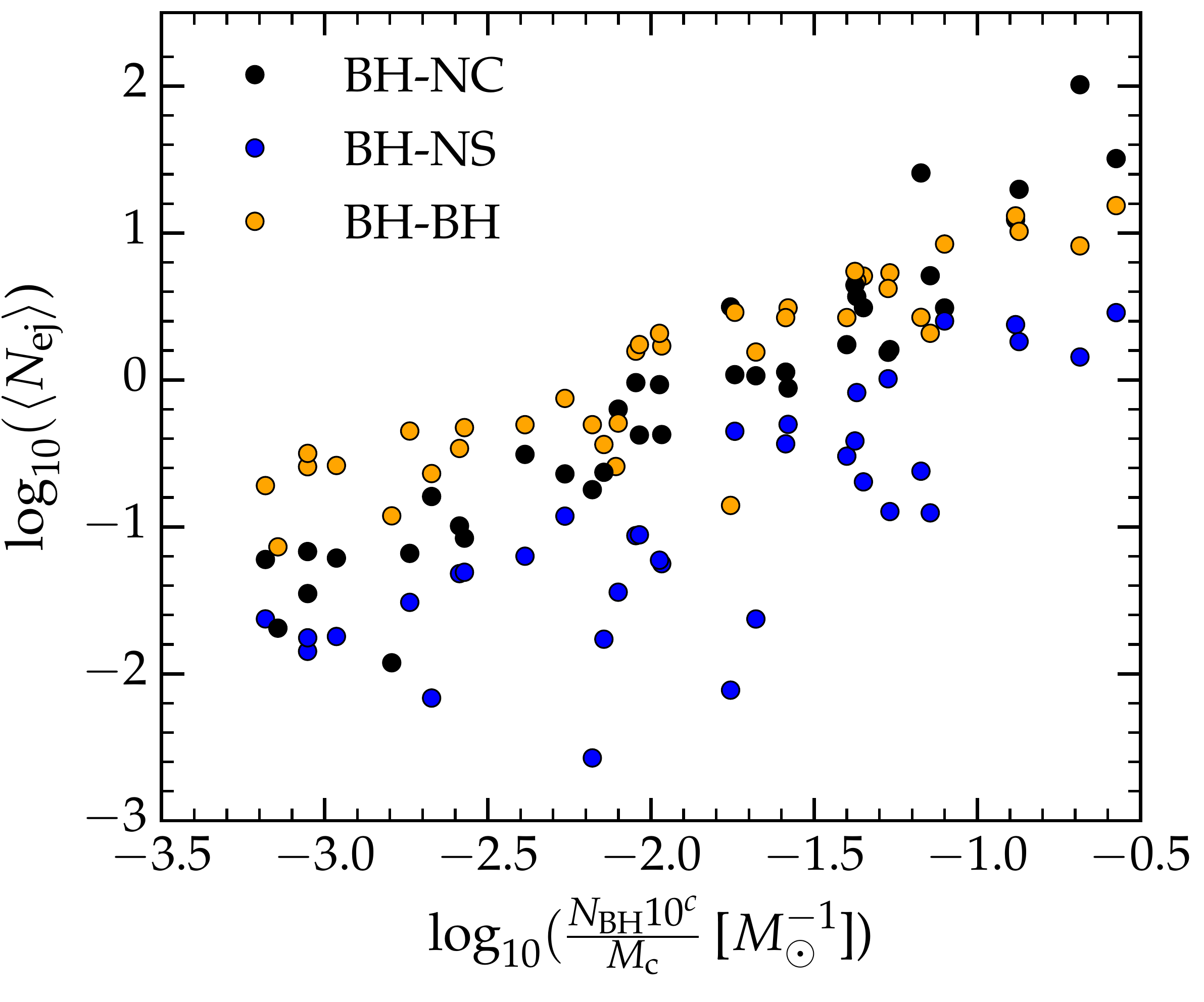}
  \caption{Expected number of binary ejections $\langle N_\mathrm{ej} \rangle$
    as a function of the number of retained BHs $N_\mathrm{BH}$,
    concentration $c$, and total cluster mass $M_\mathrm{c}$. The
    number of binaries ejected over the life of the cluster is well
    described by the two characteristic variables of the
    fundamental plane, $c$ and $M_\mathrm{c}$,
    along with the number of BHs retained by the cluster.
    \label{fig:GC_ALLejects}
  }
\end{figure}

We find that the number of ejected
binaries and the properties of these binaries are strongly affected
by the GC structure and the number of retained BHs.
In Table~\ref{table:ejections},
we list the expected number of ejected BH binaries
over the life of each cluster,
listed in order of increasing mass,
including the exact number of BHs in each cluster.
The ejected BH-binary expectation value is well described by
the number of retained BHs $N_\mathrm{BH}$ and the two characteristic variables that define the
fundamental plane of GCs (see Figure~\ref{fig:fp}),
namely the total cluster mass $M_\mathrm{c}$ and the concentration $c$.
In Figure~\ref{fig:GC_ALLejects}, we plot the expected number of ejected BH binaries
as a function of the three characteristic variables: $N_\mathrm{BH}$, $M_\mathrm{c}$, and
$c$.

\begin{figure}
  \includegraphics[width=1.05\columnwidth]{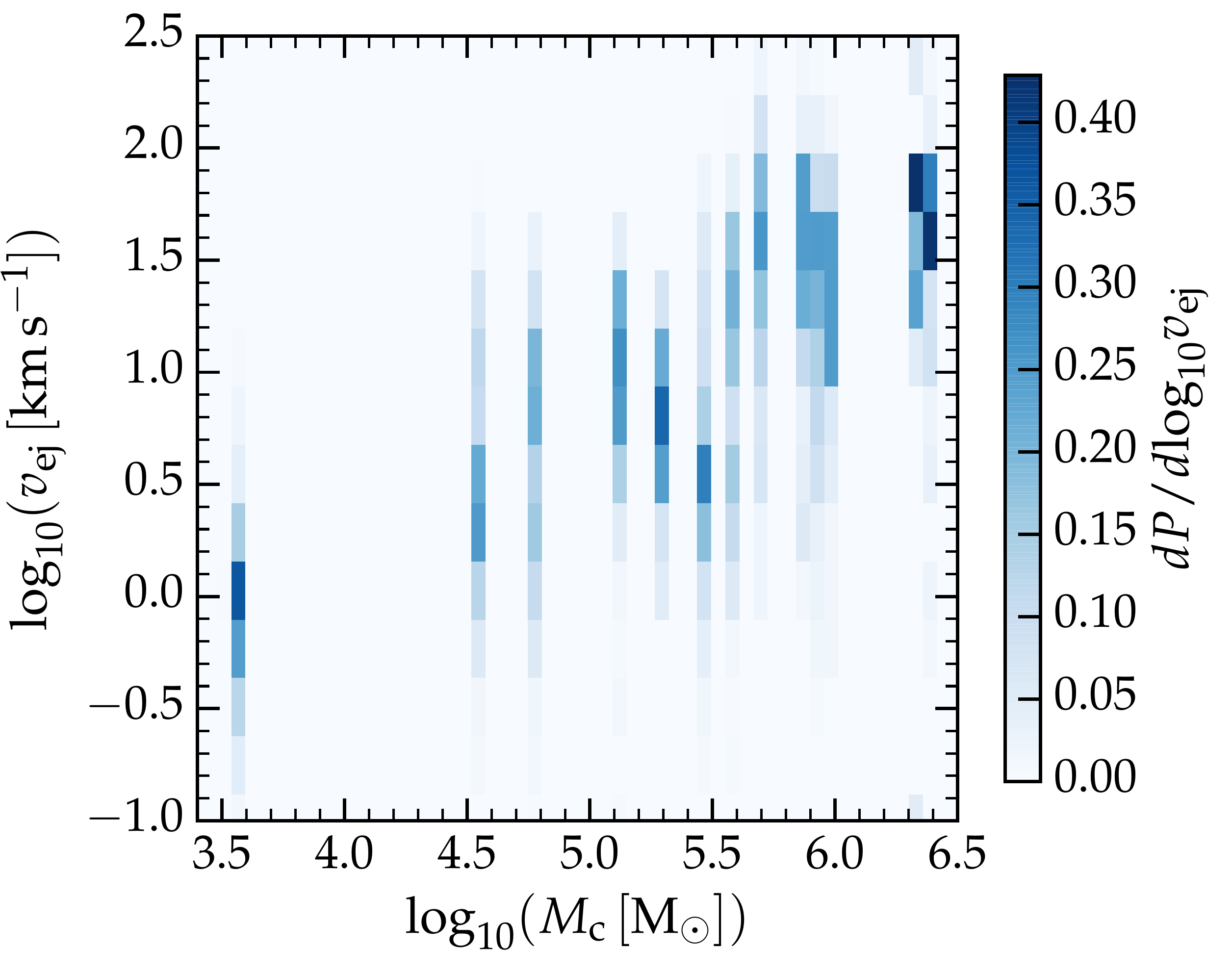}
  \caption{
    The distributions of ejection velocities $v_\mathrm{ej}$
    as a function of the total cluster mass $M_\mathrm{c}$
    for the ejected binaries.
    Each vertical bar represents the distribution of $v_\mathrm{ej}$
    for the corresponding mass $M_\mathrm{c}$ and is normalized
    such that the integral over $\mathrm{log}_{10} v_\mathrm{ej}$ in each
    mass bin yields unity.
    The binary velocity fluctuates due to random encounters with other
    stars in the cluster until the binary acquires a high enough
    recoil velocity to exceed the minimum ejection velocity,
    which is determined by the cluster mass.
    The increase in the necessary velocity for escape is apparent
    in the increasing mean value of each $v_\mathrm{ej}$ distribution.
  }
    \label{fig:vm}
\end{figure}

The most important structural variable that impacts
the ejected binary properties is the cluster mass.
The total cluster mass enforces a minimum energy needed to escape, which
the binary must gain through repeated encounters.
In order for a binary to escape from the cluster, it must acquire a
recoil velocity from a final three-body encounter
high enough to climb out of
the cluster gravitational potential.
In Figure~\ref{fig:vm}, we show the distribution 
of the ejected binary velocities as a function of cluster mass,
where the influence of the mass of the cluster on the
ejection velocity is apparent.
The expected number of ejections is then higher for lower-mass clusters due to the
lower escape velocities associated with these clusters, as is visible
in Figure~\ref{fig:GC_ALLejects}.
To decouple this statement from the
additional variables in Figure~\ref{fig:GC_ALLejects},
it can also be observed in Table~\ref{table:ejections} (which is ordered by increasing mass)
that for a fixed number of retained BHs,
the expected number of ejections scales with the cluster mass.

The mechanism through which the
binary converts binding energy to kinetic energy is easiest to understand
in the three-body center of mass frame, where we perform our integration
for encounters. After an encounter, the final relative velocity at infinity is given by
\begin{equation}\label{eq:relvel}
  \bar{v}_{\infty}^2 = \frac{m_3(m_1 + m_2)}{\bar{m}_3(\bar{m}_1 + \bar{m}_2)}v_{\infty}^2
  + \frac{2M_{123}}{\bar{m}_3(\bar{m}_1 + \bar{m}_2)}(U_\mathrm{bin} - \bar{U}_\mathrm{bin}) \,\,,
\end{equation}
where $U_\mathrm{bin} = -\frac{G m_1 m_2}{a}$ is the binding energy of the
binary and all unbarred quantities represent the initial binary
before encountering $m_3$, while barred quantities represent the final
binary and $\bar{m}_3$ is the ejected mass.
In the case of no exchange, and utilizing $\Delta a \equiv \bar{a} - a$,
Equation~\ref{eq:relvel} reduces to
\begin{equation}
  \bar{v}_{\infty}^2 = v_{\infty}^2 - \frac{2M_{123}}{m_3 m_\mathrm{b}} \bigg(\frac{G m_1 m_2 \Delta a}{a^2} \bigg) \,\,.
\end{equation}
In this frame, the binary velocity is related, through conservation of momentum,
to the relative velocity simply by $v_b = \frac{m_3}{M_{123}} v_{\infty}$.
The change in the kinetic energy, $\Delta T \equiv \bar{T} - T$, of the binary is then
\begin{equation}\label{eq:ke}
  \Delta T = - \frac{G m_1 m_2 m_3}{M_{123}} \bigg(\frac{\Delta a}{a^2} \bigg) \,\,.
\end{equation}
The amount by which the semi-major axis changes in an average encounter,
where the semi-major axis is reduced without exchange, is proportional
to the semi-major axis, $\Delta a \approx -\epsilon a$,
with $\epsilon$ in
the range $\sim [0,0.6]$~\citep{Sigurdsson:1993B}.
Using this relation, and assuming a binary with constant $m_1$ and $m_2$,
Equation~\ref{eq:ke} reduces to 
\begin{equation}\label{eq:delke}
  \Delta T \propto \frac{m_3}{M_{123}} \frac{\epsilon}{a} \,\,,
\end{equation}
yielding a simple relation that describes the gain in kinetic energy
in terms of the constant fractional change in the semi-major axis $\epsilon$
and the ratio of the third body to the total mass of the three-body system.
Additionally, Equation~\ref{eq:delke}
shows that this change in kinetic energy becomes more efficient as
the semi-major axis decreases, converting
more energy from binding to kinetic after each encounter that shrinks
the binary's orbit.
After repeated interactions,
the increase in velocity due to the decrease in $a$ becomes more substantial
and the binary can eventually reach the necessary velocity to escape.

\begin{figure}
  \includegraphics[width=1.0\columnwidth]{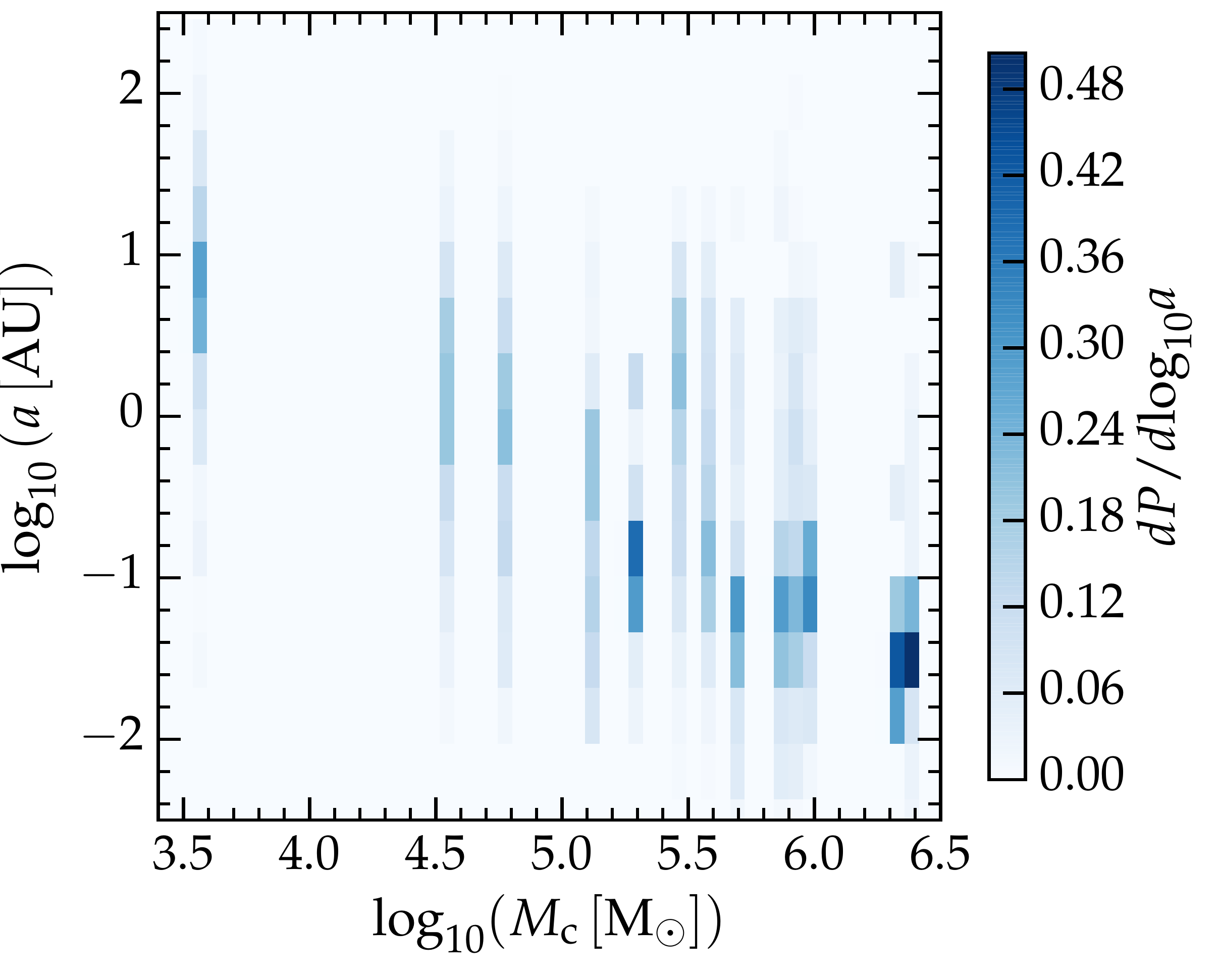}
  \caption{
    The distributions of the semi-major axes at ejection $a$
    as a function of the total cluster mass $M_\mathrm{c}$
    for the ejected binaries.
    Each vertical bar represents the distribution of $a$ for the corresponding
    mass $M_\mathrm{c}$ and is normalized such that the integral over
    $\mathrm{log}_{10} a$ in each mass bin yields unity.
    High mass clusters require a high velocity for escape,
    which a binary must acquire through three-body interactions
    in order to be ejected. 
    The energy needed to escape is more easily gained once
    the orbital separation has decreased sufficiently (see Equation~\ref{eq:delke}).
    As a consequence, the mean value of $a$ at ejection shifts to smaller
    separation with increasing cluster mass $M_\mathrm{c}$.
    \label{fig:semiaM}
  }
\end{figure}

We can directly relate the necessary gain in kinetic energy to
the change in binding energy
$\Delta U = \bar{U}_\mathrm{bin} - U_\mathrm{bin}$,
by simply rearranging Equation~\ref{eq:relvel} and assuming no exchange of masses, 
which yields
\begin{equation}
  \Delta T = - \frac{m_3}{M_{123}} \Delta U \,\,.
\end{equation}
In the process of the binary increasing its kinetic energy,
the binding energy becomes more negative.
Since the higher-mass clusters tend to hold on to the binaries longer,
this strict minimum kinetic energy for ejection is manifest in the 
more negative-valued binding energy of the binaries it ejects.
It follows from this, that on average, the semi-major axes of the binaries
ejected from more massive clusters tend to be smaller.
This is confirmed by Figure~\ref{fig:semiaM}, which depicts the distribution of
orbital separations as a function of cluster mass.

In addition to the increase in the expected number of ejected binaries
in lower-mass clusters, the total number of expected ejections also
increases with an increase in the number of BHs.
While the number of ejections is expected to increase with the number of BHs,
interestingly, the fraction of ejected binaries that
are BH-NC also grows with the number of BHs (see Figure~\ref{fig:GC_ALLejects}
and Table~\ref{table:ejections}).
This behavior can be attributed to the
fact that the BHs are not in energy equipartition with the rest of the cluster. Adding
more BHs without affecting the distribution of the luminous cluster members requires that
the BHs are spread out farther from the core, where they
have traditionally been expected to reside.
Accordingly, the mean density of BHs goes down,
and they are less likely to interact with each other.
However, because they are well mixed with the stars at larger radii, the number
of BH-NC binaries that form in three-body
exchanges grows.
Additionally, since these binaries form farther from the core,
they also have the benefit of a shallower potential to climb out of.

\begin{figure}
  \includegraphics[width=1.0\columnwidth]{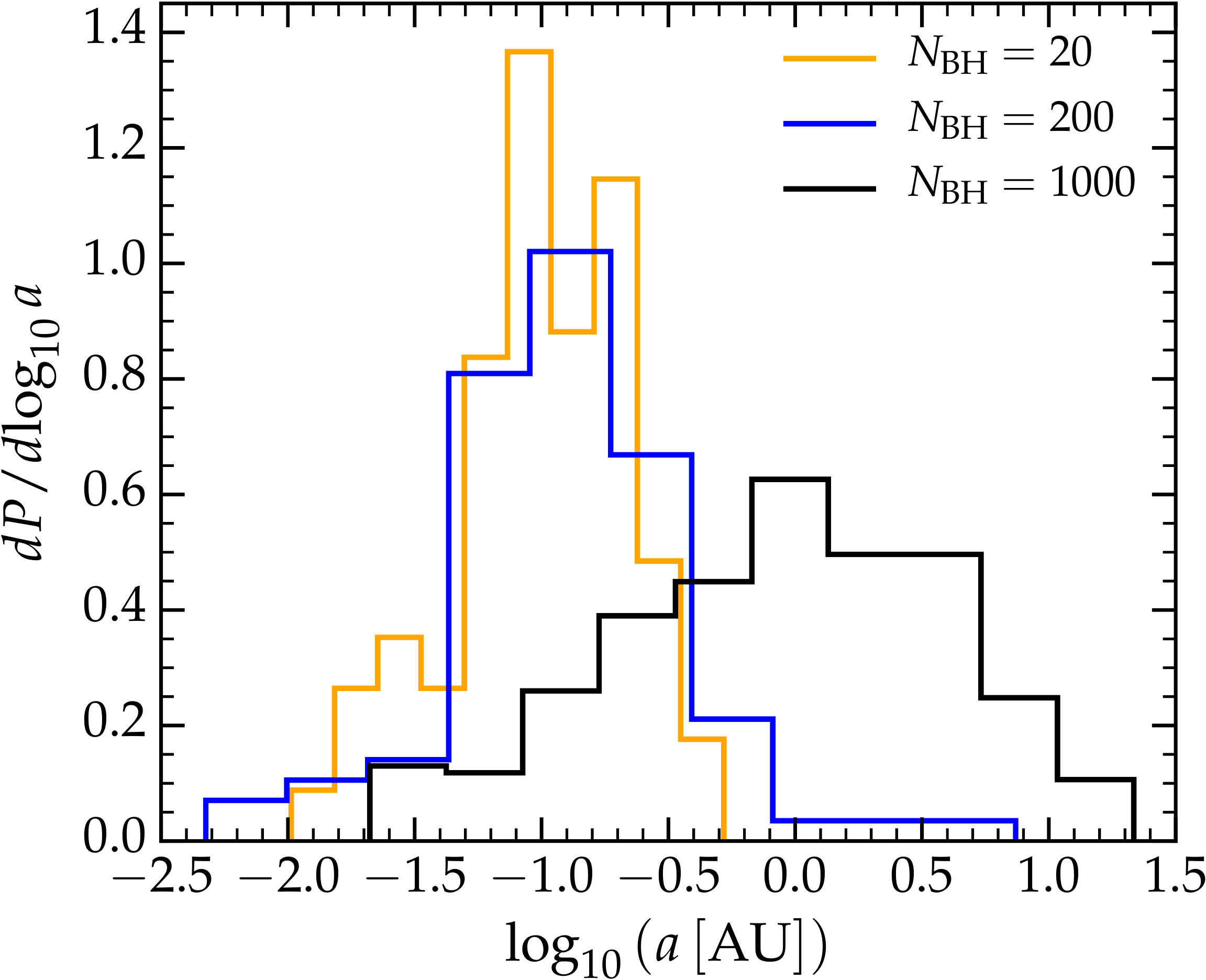}
  \caption{The probability distribution for the ejected BH-NC binary semi-major axes
    from NGC 5694, a representative case,
    with a population of 20, 200, and 1000 BHs.
    An increase in the number of BHs requires spreading the BHs
    outside of the core, where they are more likely to form binaries
    with NC objects. In the outskirts,
    the energy necessary to escape is much smaller,
    allowing the binary to escape before it has had
    sufficient time to harden. These binaries escape
    with comparatively low magnitude binding energy and wide
    orbital separations.
    \label{fig:semimajora}
  }
\end{figure}

Besides influencing the number of ejected binaries, the
number of retained BHs also affects the distribution of the semi-major axes of the ejected binaries.
In Figure~\ref{fig:semimajora}, we show the distribution of semi-major axes
for the ejected BH-NC binaries in our cluster model for NGC 5694 for
the three different choices of BHs retained. We choose this cluster
since it is representative of the effect that the number of retained BHs
has on the population of ejected BH-NC binaries. Figure~\ref{fig:semimajora}
displays an increase in the width of the distribution of
semi-major axes for larger populations of BHs.
This is again related to the necessary spreading of the BHs as we increase the
number of BHs harbored by the cluster.
Therefore, the BH-NC binaries that form outside of the core, where the escape velocity
drops rapidly as a
function of radius, can be ejected while their binding energies are of
comparably lower magnitudes.
Although the more widely separated binaries are
less likely to become mass-transferring systems,
the simulations with large BH numbers tend to have much higher
ejection rates. The higher ejection rates still produce enough
tight binaries in the tail of distribution to outnumber
those produced with fewer BHs present.

\begin{figure}
  \includegraphics[width=1.0\columnwidth]{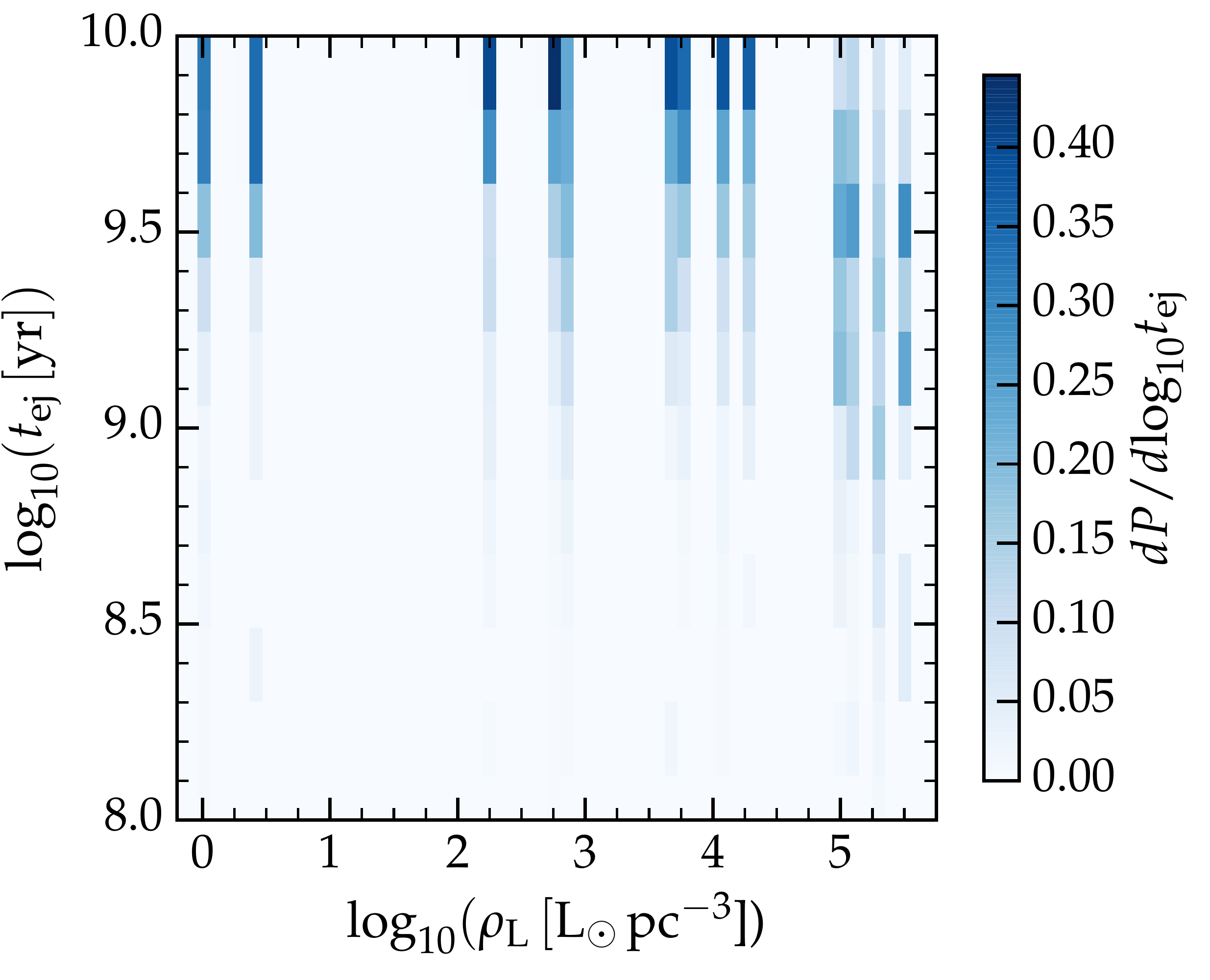}
  \caption{
    The distributions of time of ejection $t_\mathrm{ej}$
    as a function of the luminous central density $\rho_\mathrm{L}$
    for the ejected binaries.
    Each vertical bar represents the distribution of $t_\mathrm{ej}$ for the corresponding
    core luminosity density $\rho_\mathrm{L}$ and is normalized such that the integral over
    $\mathrm{log}_{10} t_\mathrm{ej}$ in each density bin yields unity.
    In higher density clusters, where encounters occur more frequently,
    many binaries are ejected after only a few Gyr, while in the lower
    density clusters most ejections occur near the end of the 10 Gyr
    evolution.
    \label{fig:tdens}
  }
\end{figure}

The remaining structural property of GCs that has a clear effect on
the population of ejected binaries is the cluster density.
In Figure~\ref{fig:tdens}, we plot the distribution of ejection times
as a function of the luminous central density, which is related to
the core density as discussed in section~\ref{sec:coredens}.
The distribution establishes that
the cluster density has some impact on the time at which binaries
are ejected from their host GC.
\begin{figure*}
  \includegraphics[width=2.0\columnwidth]{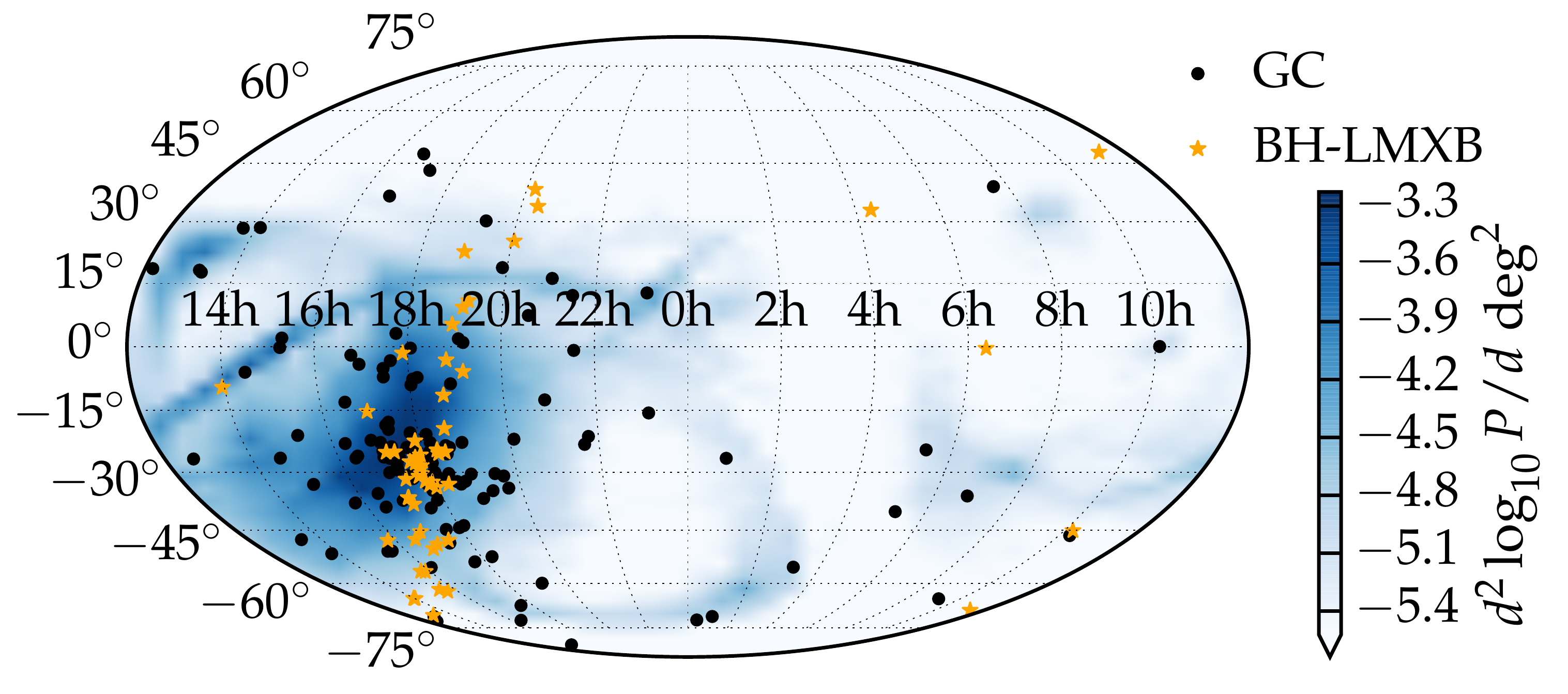}
  \caption{
    The spatial probability distribution of the simulated population of
    BH-LMXBs from GCs with $N_\mathrm{BH} = 1000$.
    The populations of Milky Way GCs (marked by black circles)
    and known BH-LMXBs (marked by orange stars) are included
    for reference.
    The map is a Mollweide projection of a heliocentric equatorial
    galactic coordinate system. The galactic center is located at
    $17^{\mathrm{h}} 45.6^{\mathrm{m}}, -28.94^{\circ}$,
    where the high density of objects
    explains the clustering of BH-LMXBs and GCs.
  }
  \label{fig:map_1000}
\end{figure*}
The time between binary-single encounters can be approximated by
\begin{equation}\label{eq:tenc}
  t_\mathrm{enc} = \Gamma^{-1} = \frac{v_\mathrm{m}}{2 \pi G (m_\mathrm{b} + \bar{m}) n_\mathrm{o} a} \,\,,
\end{equation}
where $v_\mathrm{m}$ is the mean velocity of stars in the cluster,
$n_\mathrm{o}$ is its core density, and $\bar{m}$ is the mean mass.
Combining this result with Equation~\ref{eq:ke}, we can obtain an
approximation for the rate at which a binary increases its kinetic
energy $\Delta T/\Delta t$. As encounters approximately occur
in increments of the encounter timescale, letting $\Delta t = t_\mathrm{enc}$,
we find that the rate at which the binary increases its kinetic energy, 
\begin{equation}
  \frac{\Delta T}{\Delta t} = \bigg( \frac{2 \pi G^2 m_1 m_2 m_3 \epsilon }{v_\mathrm{m}} \bigg) n_\mathrm{o} \,\,,
\end{equation}
scales with the cluster core density.
Therefore, the time it takes for a binary to acquire a high enough velocity
to escape is reduced for higher density clusters. As can be
seen in Figure~\ref{fig:tdens}, in clusters of higher density,
where encounters occur more frequently,
most BH-NC systems are ejected after only 3 Gyr of evolution whereas in
lower density clusters most
ejections take place near the end of the 10 Gyr simulation (i.e. the present day), 

\subsection{Black-hole low-mass X-ray binaries}\label{sec:bhlmxbs}
Here we focus strictly on the population of the present-day
mass-transferring systems that have successfully become BH-LMXBs.
These results reflect the contribution to the BH-LMXB population
from the entire population of non-core collapsed Milky Way GCs.
The production of BH-LMXBs is based on a subset of 15 simulated GCs
and the methods detailed in section~\ref{sec:galev}.
In the following section, we discuss the distribution and the
properties of this population of BH-LMXBs from GCs.

As discussed at the end of section~\ref{sec:structp}, some clusters
require choosing a BH retention fraction of unity, $f_\mathrm{r_\mathrm{BH}} = 1$,
in order to obtain the desired quantity of BHs.
This occurs in the lowest-mass cluster for each set of $N_\mathrm{BH}$,
i.e. Pal 13 for $N_\mathrm{BH} = 20$, NGC 6838 for $N_\mathrm{BH} = 200$,
and NGC 6121 for $N_\mathrm{BH} = 1000$.
These specific parameter sets are not used in determining the population of BH-LMXBs.
Although the results from these three sets are included in the previous discussions,
they are excluded here due to the unphysical nature of complete BH retention.
During BH formation, natal kicks ensure that at least some
fraction of the BHs formed from the IMF are ejected from the cluster. This
makes complete BH retention essentially unattainable. In consideration
of this, we include only those models with $f_\mathrm{r_\mathrm{BH}} < 1$.

\subsubsection{Population}\label{sec:popsec}
The number of mass transferring systems that develop
from the BH-NC binaries that are ejected from our model clusters
strongly depends on the assumed BH retention in GCs.
We employ the same notation as in section~\ref{sec:ejbin} for BH retention:
MIN refers to $N_\mathrm{BH} = 20$, 200 refers to $N_\mathrm{BH} = 200$,
and MAX refers to $N_\mathrm{BH} = 1000$.
The MIN case produces zero observable BH-LMXB systems. The 200 case
produces $25^{+10}_{-6}$ mass-transferring BH low-mass systems
and the MAX case yields an expectation value of $156^{+26}_{-24}$ ejected BH-LMXBs,
with the stated uncertainties bounding the $95$ per cent confidence interval.

The clusters that contribute the largest number of BH-LMXBs are
those with the highest BH-NC ejection rates (see Table~\ref{table:ejections}).
As is visible in Figure~\ref{fig:GC_ALLejects}, the expected number of ejections
can be approximated as a function of the number of retained BHs $N_\mathrm{BH}$ and the
two fundamental parameters describing the cluster: the concentration $c$ and
the total cluster mass $M_\mathrm{c}$.
While the initial semi-major axis at ejection $a$,
which is sensitive to the cluster mass (Figure~\ref{fig:semiaM}),
is an important factor in determining whether a BH-NC will lead to mass transfer,
surprisingly, the fraction of BH-NCs that become BH-LMXBs appears nearly
constant across clusters. Equivalently stated, $\langle N_\mathrm{BH-LMXB} \rangle \sim f_\mathrm{LMXB} \langle N_\mathrm{ej} \rangle$
appears to hold true for the set of clusters modeled, where $f_\mathrm{LMXB} \sim 0.25$ represents
the fraction of ejected BH-NC binaries that evolve into BH-LMXBs.
Although the distributions of most orbital parameters, which determine whether a
system will evolve into a BH-LMXB, vary from cluster to cluster,
the thermal eccentricity distribution shared by all clusters ensures
that a roughly equal proportion of the ejected binaries will become BH-LMXBs.
For clusters that tend to eject wider binaries, it is only the highly eccentric
systems that become BH-LMXBs, and vice versa.

For a given BH retention,
the number of successfully formed BH-LMXBs from GCs is potentially
a function of the ejection time, initial separation, initial eccentricity,
primary and companion masses, and the complex internal evolution of the
binary. Yet, since we find that the ejection properties are largely
determined by the cluster properties, namely the quantities defining the
fundamental plane, the size of the BH-LMXB population from GCs is well
approximated by the cluster properties alone.

\subsubsection{Distribution}
As GCs generally have low escape velocities, 
the ejected BH-LMXBs typically escape with relatively low velocities.
Due to this, the distribution of BH-LMXBs closely mimics the distribution
of GCs in the Milky Way galaxy. 
In Figure~\ref{fig:map_1000}, we present the 
spatial probability distribution of BH-LMXBs from GCs, for the MAX case, 
on a Mollweide projection of the galactic map in an equatorial coordinate system.
Additionally, we include the distribution of galactic GCs and known BH-LMXBs
from BlackCAT~\citep{Corral-Santana:2016}, a catalog of candidate BH-LMXBs, which
we use in all figures including an observed population, unless stated otherwise.
Although the 200 case produces fewer BH-LMXBs, the distribution is qualitatively
similar to the MAX case. The highest probability density region is near the
galactic center, where the majority of GCs reside. However, as Figure~\ref{fig:vm}
illustrates, the distributions of the ejection velocities
have widths that span an order of magnitude or more. As a consequence, some
fraction of the binaries have ejection velocities that allow them to separate
from their parent cluster. Additionally, the binaries that are ejected at an earlier
time in the GC's orbit have sufficient time to diverge from the host GC orbit.
The higher density streaks in Figure~\ref{fig:map_1000} can be attributed
to these binaries that have drifted from the parent GC.

\begin{figure}
  \includegraphics[width=1.035\columnwidth]{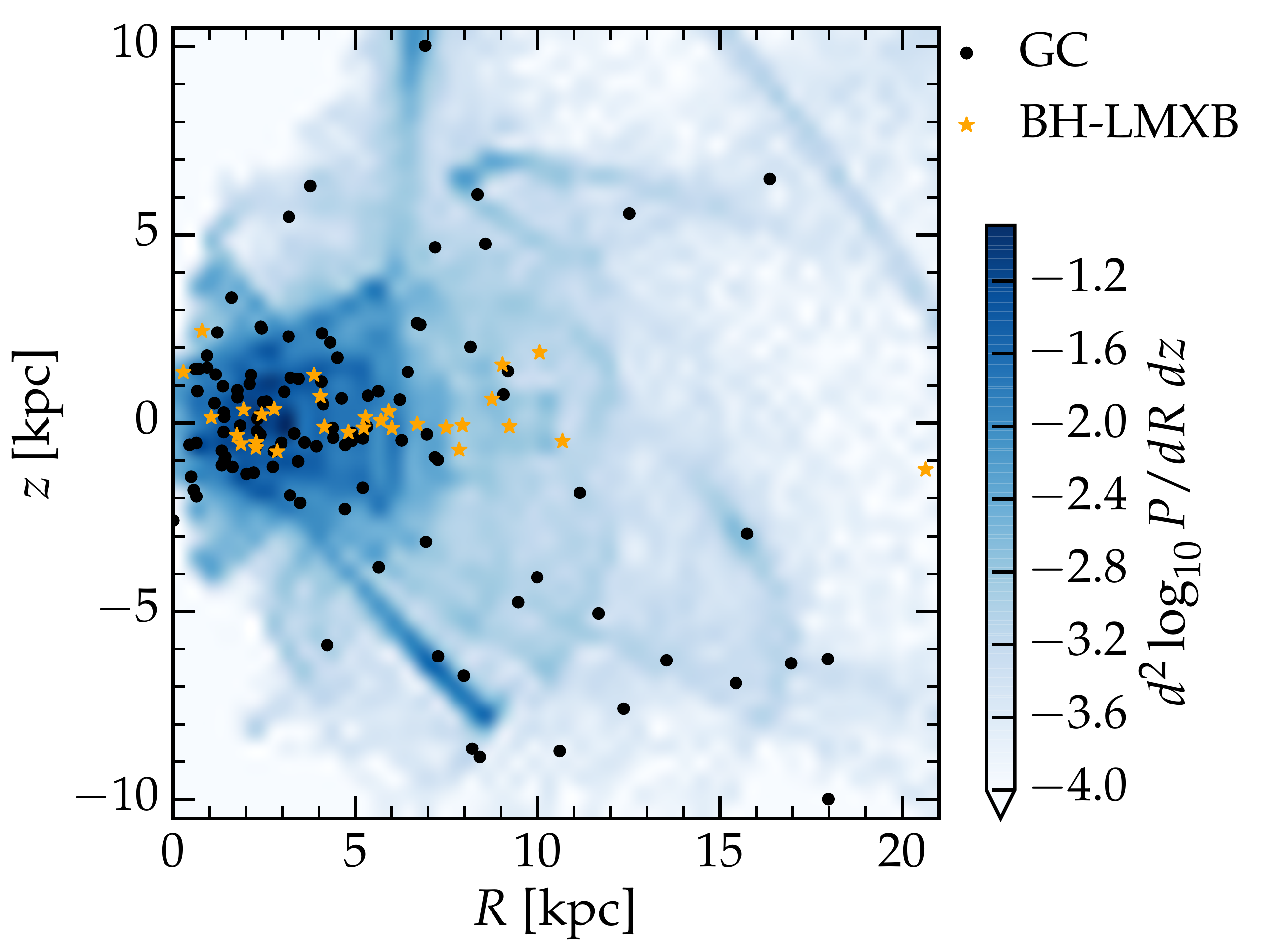}
  \caption{
      The spatial probability distribution of the simulated population of
      BH-LMXBs from GCs with $N_\mathrm{BH} = 1000$ in the $R-z$ plane.
      The coordinate $z$ specifies the distance perpendicular to
      the galactic plane and $R$ is the in-plane distance from
      the galactic center at the origin.
      The populations of Milky Way GCs
      (marked by black circles) and known BH-LMXBs (marked by orange stars)
      are included for reference.
      While many of the BH-LMXBs from GCs populate the galactic disk,
      the distribution extends well out of the galactic
      plane into the high-$|z|$ region.
  }
  \label{fig:z_plot}
\end{figure}

\begin{figure}
  \includegraphics[width=0.99\columnwidth]{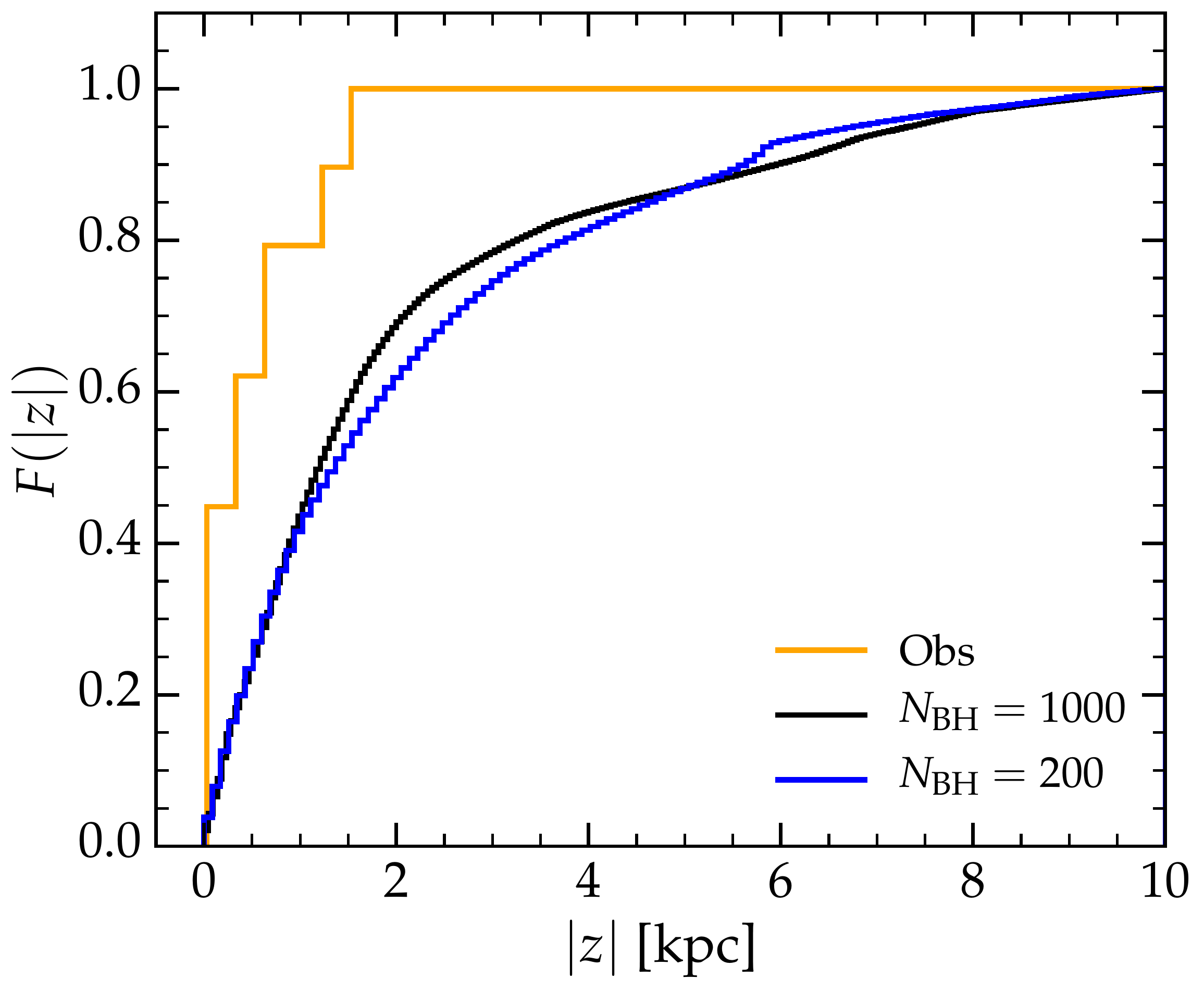}
  \caption{
    The normalized cumulative distribution function of the
    absolute distance perpendicular to the galactic plane
    $|z|$.
    The included distributions are the BH-LMXBs produced in our GC simulations for the cases of
    $N_\mathrm{BH} = 200$, $N_\mathrm{BH} = 1000$,
    and the observed population.
    Note that in the case that GCs have minimal BH retention ($N_\mathrm{BH} = 20$),
    no mass-transferring systems are produced.
  }
  \label{fig:z_hist}
\end{figure}

As GCs primarily follow halo orbits that extend well out of the
galactic plane, the GCs are easily able to populate this space with BH-LMXBs.
In Figure~\ref{fig:z_plot}, we provide the spatial
probability distribution for BH-LMXBs from the MAX case in the $R-z$ plane.
Again, we present only the MAX case, as the 200 case is similarly distributed
but with a lower overall probability density.
The median absolute distance from the galactic plane is $|z| = 1.63 \, \mathrm{kpc}$
and the median distance from the galactic center in the plane is 
$R = 4.51 \, \mathrm{kpc}$.
While it is clear from Figure~\ref{fig:z_plot} that many of the BH-LMXBs
from GCs are located in the galactic disk, the distribution extends well out
of the galactic plane into the lower density regions above and below the disk.
BH-LMXBs that form in the field will generally reside in the high
density galactic plane, unless they receive substantial kicks at birth, which
might eject them into the `high-z' regions.
However, the magnitude of BH-LMXB kicks is still uncertain 
and the magnitude necessary to reach the highest of BH-LMXBs
from GCs is considered unlikely~(see, e.g.,~\citealt{Repetto:2015};~\citealt{Mandel:2016}).
In Figure~\ref{fig:z_hist}, we show the cumulative distribution
function of the absolute distance $|z|$ perpendicular to the galactic plane
for the MAX case, the 200 case, and the observed population of BH-LMXBs.
The observed population terminates at a maximum $|z| \sim 2 \, \mathrm{kpc}$,
while the BH-LMXB population from GCs extends well beyond this point.
This produces a region of space that is unique to a population of BH-LMXBs from GCs,
a population distinct from those forming in the field.

\subsubsection{Properties}\label{sec:props}
A typical BH-LMXB with a GC origin has an initial semi-major axis of $5.71 \, R_\mathrm{\odot}$,
initial BH mass of $8.09 \, M_\mathrm{\odot}$, and an initial companion mass
of $0.4 \, M_\mathrm{\odot}$. The median present-day period is $4.48$ h and
the median present-day BH mass is $8.25 \, M_\mathrm{\odot}$, which
has increased above the initial median BH mass due to accretion from the companion.
As discussed in section~\ref{sec:ejbin}, the masses used in the Monte Carlo
models for the ejected binaries are sampled according to the EMF
from the mass bin corresponding to the mass in the ejected BH-NC. This is done
for both the primary BH mass $M_\mathrm{BH}$ and the companion mass $m_2$ to
obtain the mass distributions, which we discuss below.

\begin{figure}
  \includegraphics[width=1.0\columnwidth]{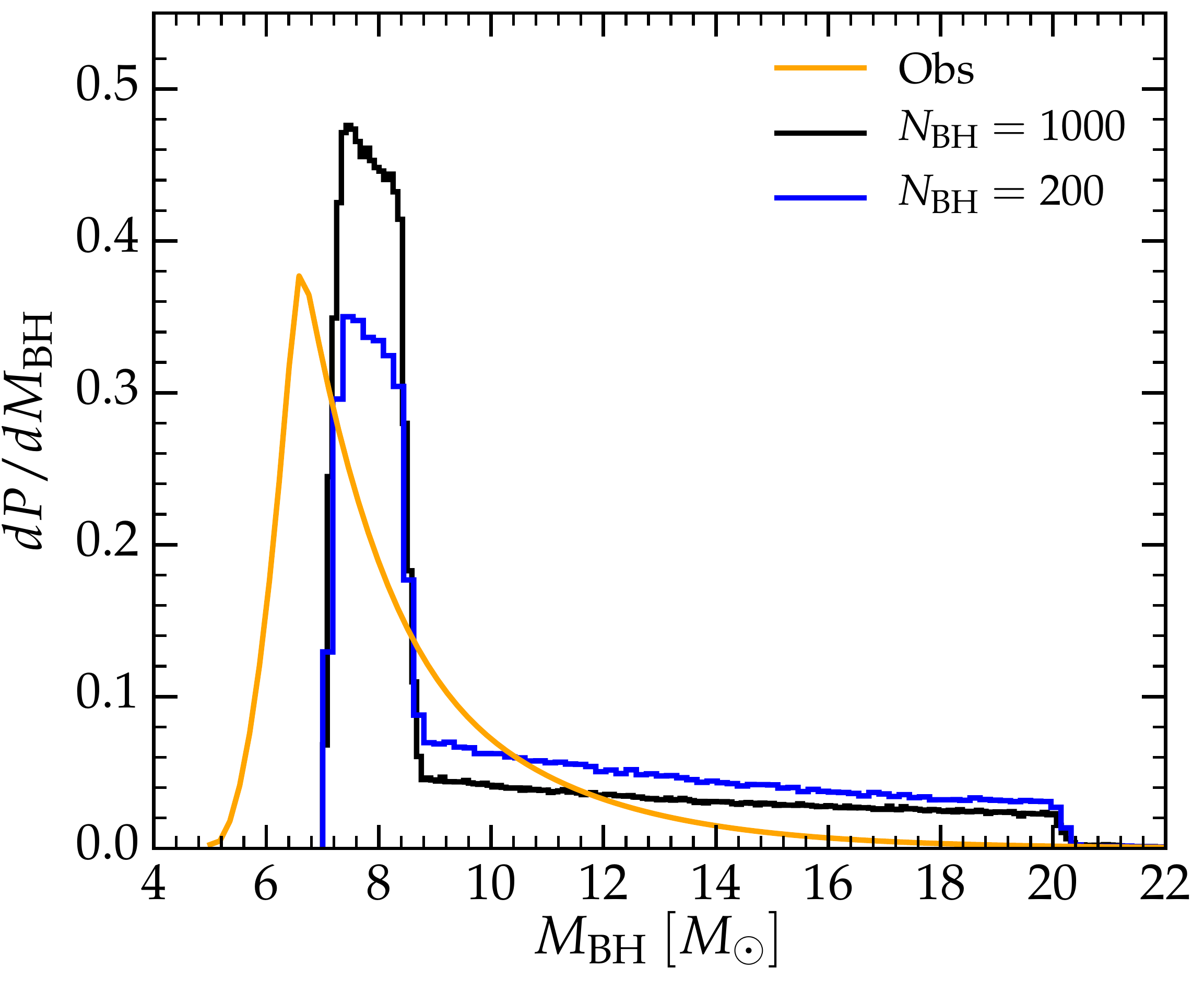}
  \caption{
    The probability distributions of BH masses in BH-LMXBs
    for the observed population~\citep{Ozel:2010}
    and for the BH-LMXBs produced
    in our GC simulations for the cases of $N_\mathrm{BH} = {200,1000}$.
    Note that in the case that GCs have minimal BH retention ($N_\mathrm{BH} = 20$),
    no mass-transferring systems are produced.
    The discontinuous
    jumps in the distribution correspond to the mass bin minimum and maximum,
    with a power law distribution in-between determined by the evolved mass function.
    The lowest BH mass bin was truncated at $7 M_\mathrm{\odot}$.
    }
  \label{fig:lmxb_mbh}
\end{figure}

In Figure~\ref{fig:lmxb_mbh}, we show the distribution of the
BH mass in the population of BH-LMXBs from GCs for both cases that
produce mass transferring systems.
Along with the BH mass distributions
for the 200 and MAX cases, we include the inferred BH mass distribution
from observations~\citep{Ozel:2010}.
Although the observed mass distribution reaches down to $\sim 5 \, M_\mathrm{\odot}$,
our EMF does not produce BH masses in the range $M_\mathrm{BH} < 7 \, M_\mathrm{\odot}$.
The BH primary mass is peaked at $7.4 \, M_\mathrm{\odot}$ and displays a preference
for the lower-mass BHs. The lack of systems at high BH mass can be attributed to
two contributing factors. The leading contribution is the distribution of BH masses
in the ejected BH-NCs, which is dominated by the two lowest BH mass bins (i.e. $8.87 \, M_\mathrm{\odot}$ and
$20.48 \, M_\mathrm{\odot}$). Although these are produced in nearly equal numbers, the preference
for the lowest mass bin that arises in the BH-LMXBs is
due to a secondary effect introduced during the binary stellar evolution.
High mass ratio systems are prone to disrupting the companion star, ending the possibility
of evolving into a stable BH-LMXB. Despite these barriers to forming BH-LMXBs with high
mass BHs, there remains a small population of high mass present-day BH-LMXBs,
with $M_\mathrm{BH} > 40 \, M_\mathrm{\odot}$,
which accounts for $\sim 1$ per cent of the population.

\begin{figure}
  \includegraphics[width=1.0\columnwidth]{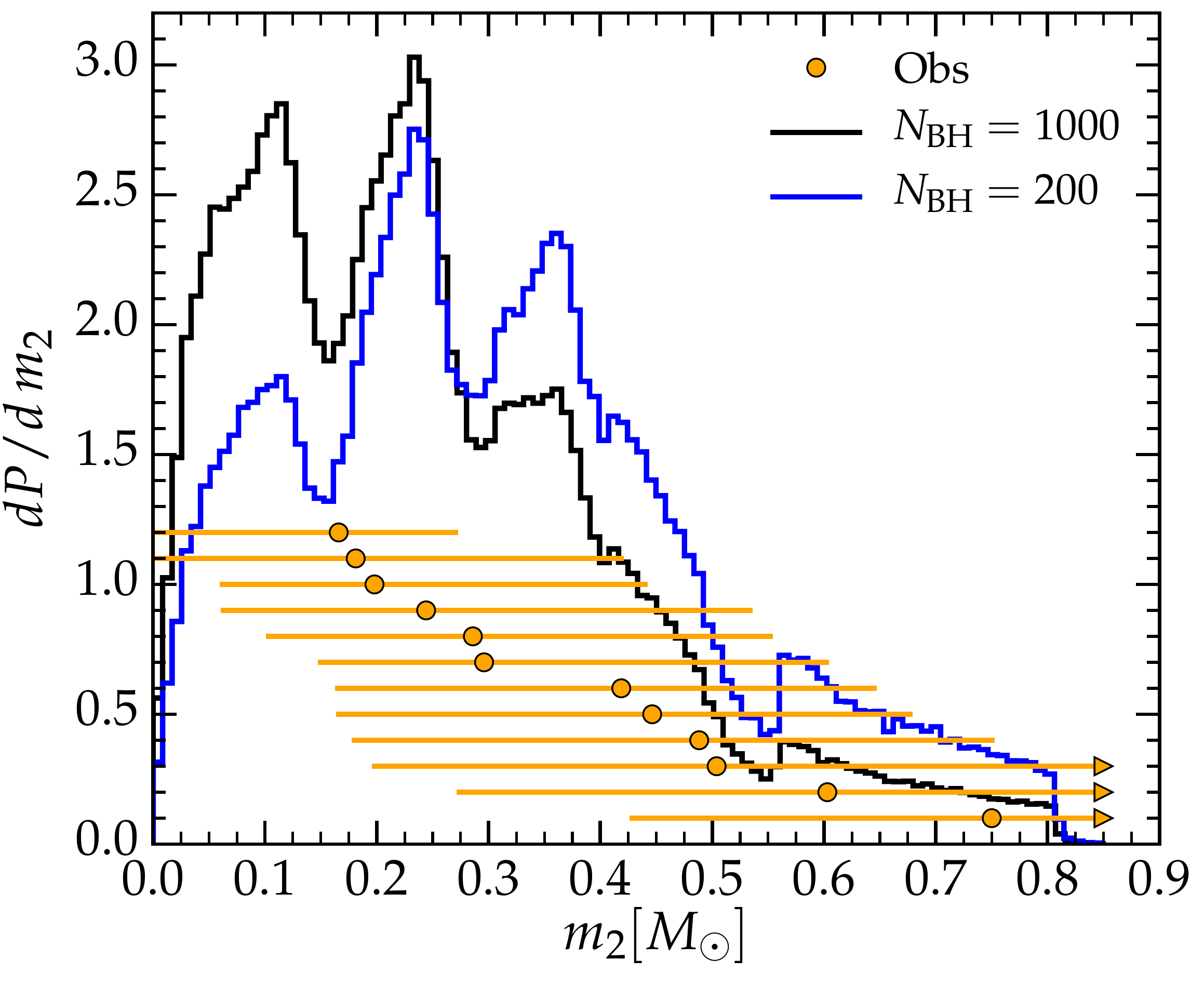}
  \caption{
    The probability distributions of the companion masses in BH-LMXBs
    for the cases $N_\mathrm{BH} = {1000}$ and $N_\mathrm{BH} = {200}$. 
    The observed population includes 12 of the 18 confirmed BH-LMXBs
    in BlackCAT~\citep{Corral-Santana:2016} that have the necessary observational
    quantities (see section~\ref{sec:props} for a description of the observed population)
    and are included for reference; the circles indicate the mean value, the
    line represents the uncertainty in the observations, and the inclusion of an
    arrow indicates that the uncertainty is only bounded on one side.
    The peaks in the simulated distributions are due to
    the sampling of companion masses from the evolved mass function (EMF)
    within each mass bin.}
  \label{fig:lmxb_m2}
\end{figure}

The low-mass companions are restricted to the
range $m_2 < 0.85 \, M_{\odot}$,
where the
maximum mass is constrained by the MS
turnoff-mass, $m_{\mathrm{to}} = 0.85\,M_\odot$. The present-day companion mass
is a function of the mass-transfer rate and the time since the onset of mass transfer.
The majority of the companion masses are MS stars, however there exists
a subpopulation of WD companion masses which account for $\sim 10$ per cent
of the companions in the MAX case and $\sim 20$ per cent in the 200 case.
In Figure~\ref{fig:lmxb_m2}, we display the companion mass distribution
for the MAX case, 200 case, and the observed population of BH-LMXBs.
The lack of lower-mass companions in the 200 case relative to the MAX case 
is due to the higher fraction of WDs, which have masses $m_\mathrm{WD} \gtrsim 0.4\,M_\odot$.
In the MAX case there is a larger number of BHs in the outskirts where the lowest masses
reside, whereas the 200 case is more centrally concentrated where there is an increase in the
probability of picking up a higher mass companion and which includes a larger population of WDs.
The observed population in Figure~\ref{fig:lmxb_m2} is generated from the observational
data in the candidate BH-LMXB catalog BlackCAT.
There are $18$ confirmed BH-LMXBs in the catalog that have a measurement of the BH
mass $M_\mathrm{BH}$ and the mass ratio $q$, which we use to estimate the
companion mass $m_2 = q\, M_{\mathrm{BH}}$. The companion masses in the observed
population have large error bars due to the uncertainty in the measurements of the
BH mass and the mass ratio.

\begin{figure}
  \includegraphics[width=0.97\columnwidth]{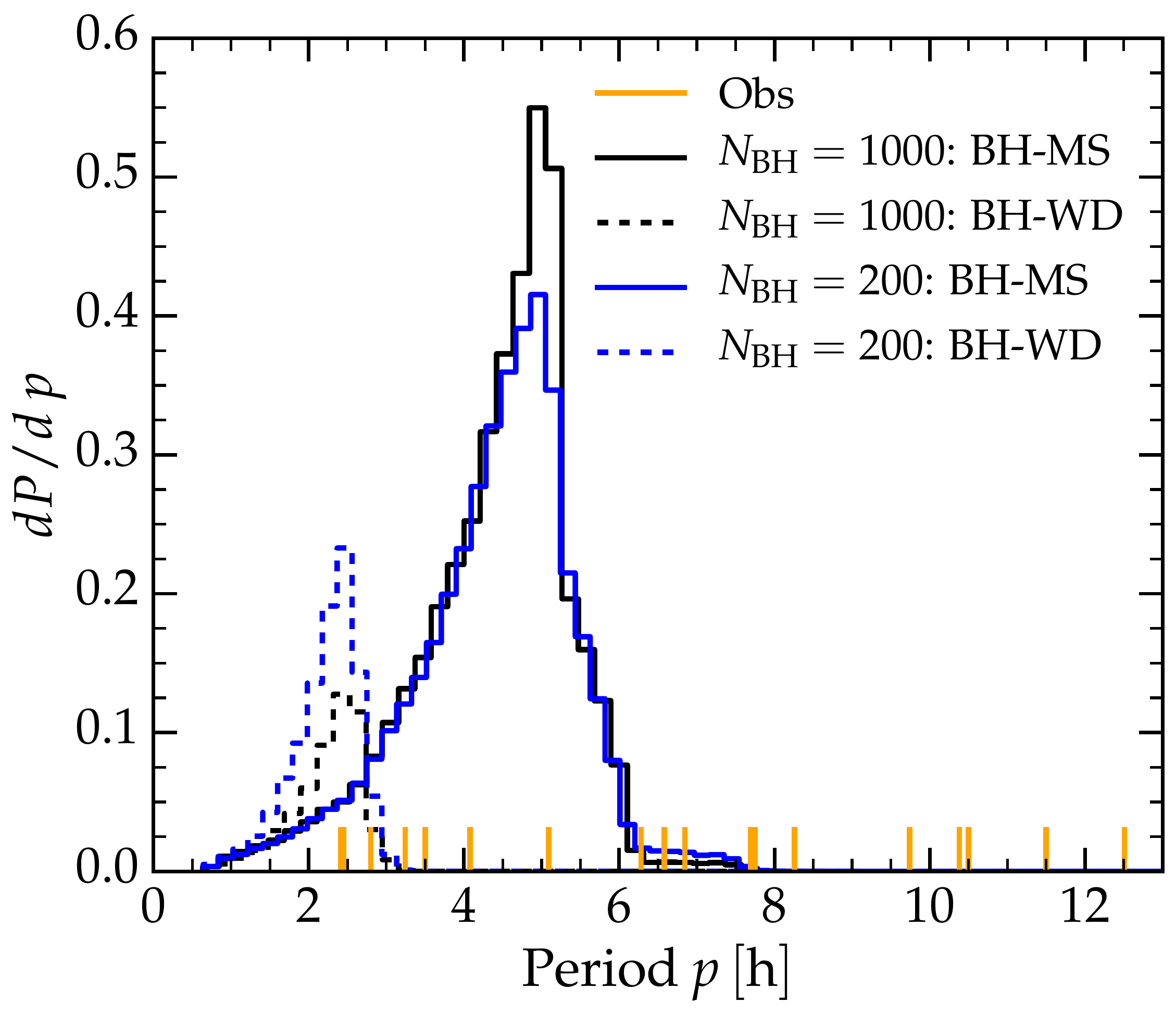}
  \caption{
    The probability distribution of orbital periods in the simulated
    BH-LMXBs from GCs for the two
    stellar companion sub-populations: WD and MS. The periods for the
    observed population of BH-LMXBs that are less
    than $13$ h are included for reference and are identified by
    orange tick marks (18 of the 28 candidate BH-LMXBs
    from BlackCAT).
    To preserve the relative size of the MS and WD companion
    populations, each distribution is independently
    normalized and then multiplied
    by the factors $N_\mathrm{BH-MS}/N$ and $N_\mathrm{BH-WD}/N$,
    respectively, with $N = N_\mathrm{BH-MS} + N_\mathrm{BH-WD}$.
    This normalization is applied to each $N_\mathrm{BH}$ case
    independently.
    }
  \label{fig:period}
\end{figure}

The initial eccentricity of the binaries follows a thermal distribution, while
the initial semi-major axis, as discussed in~\ref{sec:popsec}, is typically
$(a/\mathrm{AU}) \ll 1$, due to their GC origin. The small initial
separation of the BH-NCs leads to a distribution of periods $p$ where
$\sim 99$ per cent of the BH-LMXBs have $p\lesssim 6.2 \, \mathrm{h}$ for the MAX case and
$p\lesssim 6.8 \, \mathrm{h}$ in the 200 case. The subpopulation of BH-LMXBs with a
WD companion have a qualitatively similar distribution but with a reduced period such
that $\sim 99$ per cent of the population have $p\lesssim 3 \, \mathrm{h}$ for
both cases, MAX and 200. The reduced period for the WD companions is due to the
smaller separations necessary to induce mass transfer for these compact objects.
In Figure~\ref{fig:period}, we display the bi-modal distribution of the orbital period
for our population of BH-LMXBs
along with a subset of the observed population with periods less than $\sim 1/2$ day.

The mass transfer in these systems is primarily driven by angular momentum
loss due to tidal circularization. As the companion star passes the BH
at periastron, the tidal forces from the BH deform the star and dissipate
energy. This tidal torque efficiently removes eccentricity from the system
and eventually leads to circularization of the orbit with a reduced period. 
Once the period reaches some critical separation, the companion star
overfills its Roche lobe and transitions to a state of mass transfer.
This is the same mechanism operating on the BH-LMXBs with a WD companion,
however due to the compact nature of WDs, the critical separation which
leads to Roche lobe overflow occurs at smaller separations, hence the
shorter orbital periods. 
The binary evolution for the BH-LMXBs from GCs is significantly different
from the evolution of field binaries. In the standard binary evolution picture,
the companion evolves to overfill its Roche lobe, which can lead to mass transfer
at relatively large separations. The MS stars in BH-LMXBs from GCs have not evolved
significantly within the cluster, but evolve on much longer timescales, preventing
them from achieving mass transfer at wide separations.

\begin{figure}
  \vspace{5pt}
  \includegraphics[width=1.03\columnwidth]{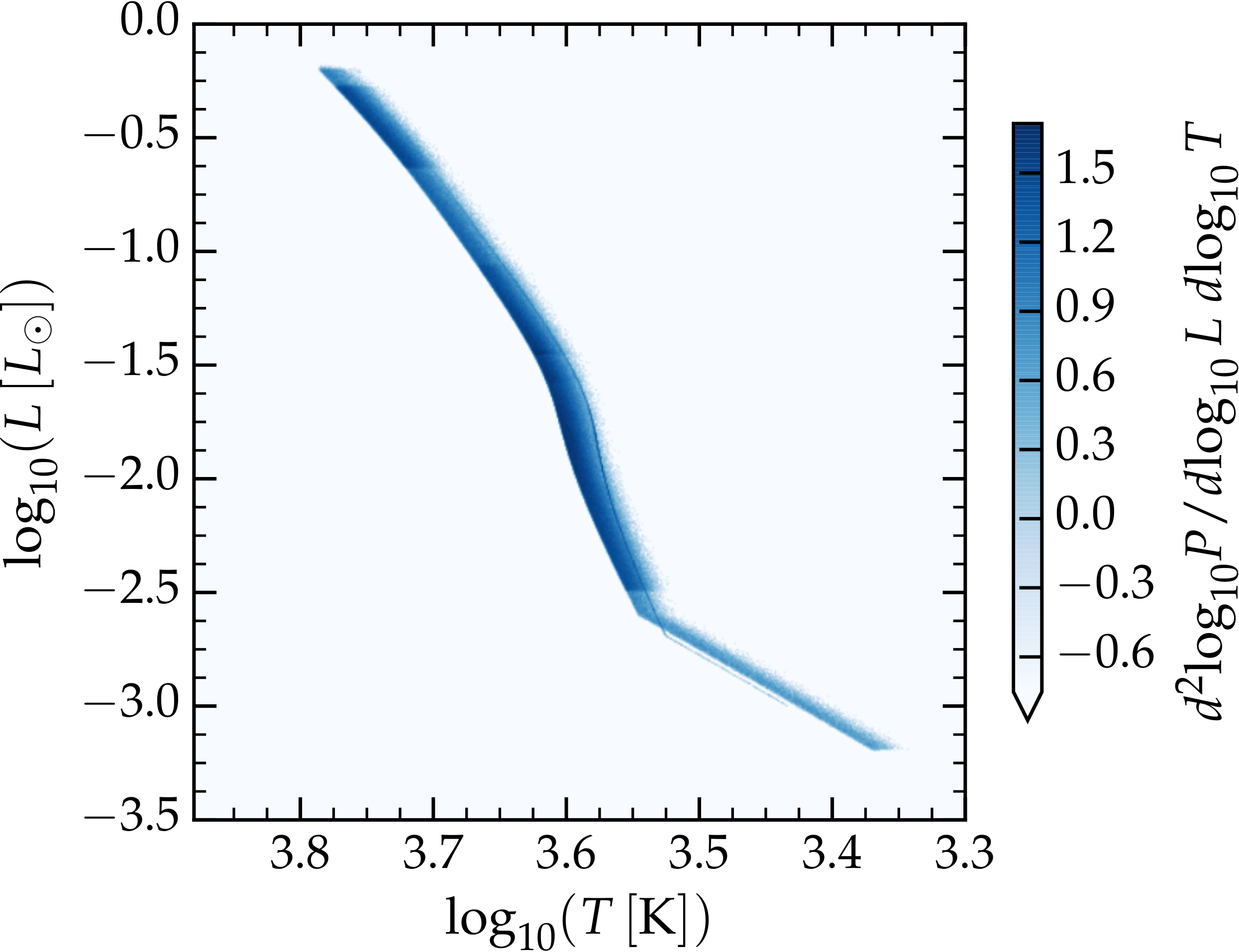}
  \vspace{-5pt}
  \caption{
    Temperature-luminosity diagram for the BH-LMXB companion mass
    in the simulated population of BH-LMXBs from GCs with $N_\mathrm{BH} = 1000$.
    The low-luminosity WD companions are excluded from the figure, 
    leaving only the mass-transferring MS companions. Since the MS companions
    from GCs are unevolved stars, the companion temperature-luminosity diagram is essentially
    the portion of the Hertzsprung-Russell MS branch with $m_2 < m_\mathrm{to}$.
  }
  \label{fig:lum_temp}
\end{figure}

In Figure~\ref{fig:lum_temp}, we provide a temperature-luminosity
diagram for the mass-transferring MS companions. We exclude the WD systems
from the diagram, since they are likely too faint for observation. 
The MS companions have temperatures $\sim 1500$ -- $6300 \, \mathrm{K}$
and luminosities $\sim 6\times 10^{-4}$ -- $0.5 \, \mathrm{L}_{\odot}$,
making these identifiable as K/M late-type MS stars below the MS turnoff.

\begin{figure}
  \vspace{22pt}
  \includegraphics[width=1.0\columnwidth]{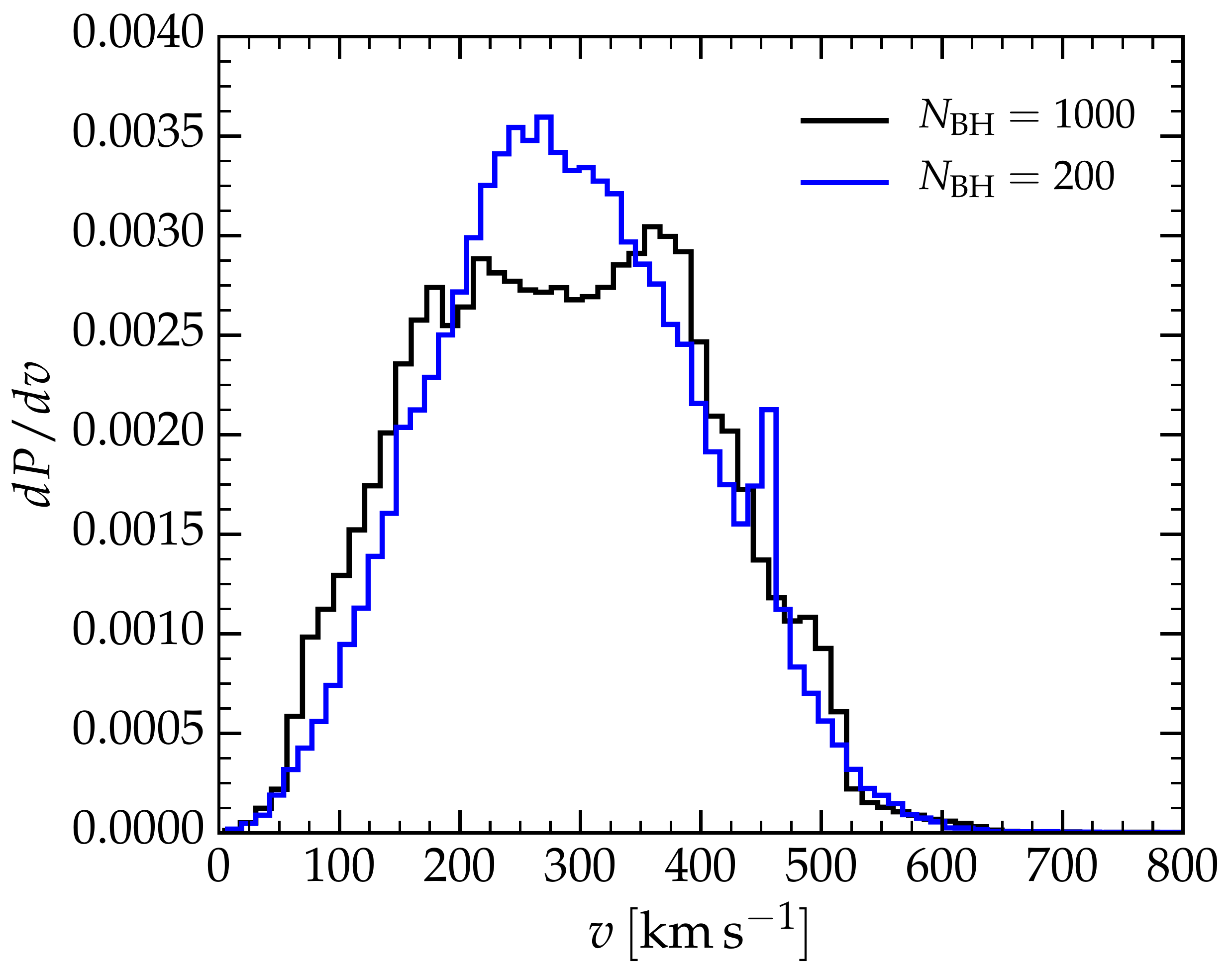}
  \vspace{-10pt}
  \caption{
    The probability distributions for the space velocity $v$ of
    the simulated BH-LMXB population 
    for the two BHs retention values $N_\mathrm{BH} = {1000}$
    and $N_\mathrm{BH} = {200}$. The BH-LMXB space velocity
    is $\bm{v} = \bm{v}_\mathrm{ej} + \bm{v}_\mathrm{GC}$,
    where $\bm{v}_\mathrm{ej}$ is the ejection velocity
    and $\bm{v}_\mathrm{GC}$ is the velocity of the host GC.
    Since $\bm{v}_\mathrm{ej}$ is approximately the GC escape velocity,
    the magnitude $v$ is dominated by the relatively large
    contribution from $\bm{v}_\mathrm{GC}$. As such, the velocity
    distribution of BH-LMXBs is consistent with the velocity
    distribution of GCs, which is reflected in the high mean
    velocities.
  }
  \label{fig:vdist}
\end{figure}

A distinct characteristic of these systems are their kinematic properties.
In Figure~\ref{fig:vdist}, we show the distribution of the magnitude of
the velocity $v$ of the BH-LMXBs from GCs.
\begin{table*}
  \begin{tabular}{@{\extracolsep{4pt}}l|r|llll|lll@{}}
    & & \multicolumn{4}{c}{3-body mergers} & \multicolumn{3}{c}{GW mergers}  \\ \cline{3-6} \cline{7-9}\vspace{-5pt} & & & & & & & & \\Name & NBH & BH-NC & BH-WD & BH-NS & BH-BH & BH-WD & BH-NS & BH-BH \\ 
\hline \hline
Pal 13 & 19.64 & $2.53 \times 10^{-1}$ & $1.81 \times 10^{-2}$ & 0.00 & 0.00 & $1.42 \times 10^{-2}$ & $1.42 \times 10^{-2}$ & $7.74 \times 10^{-3}$ \\
NGC 6838 & 20.61 & 8.27 & 1.02 & $8.98 \times 10^{-3}$ & $1.68 \times 10^{-3}$ & $9.27 \times 10^{-1}$ & $6.85 \times 10^{-2}$ & $2.99 \times 10^{-1}$ \\
 & 174.55 & $4.40 \times 10^{1}$ & 4.76 & $1.71 \times 10^{-2}$ & $4.27 \times 10^{-3}$ & 4.15 & $2.22 \times 10^{-1}$ & 1.54 \\
NGC 6535 & 19.89 & 5.32 & $5.96 \times 10^{-1}$ & $1.28 \times 10^{-2}$ & $2.77 \times 10^{-4}$ & $5.15 \times 10^{-1}$ & $6.38 \times 10^{-2}$ & $1.97 \times 10^{-1}$ \\
 & 198.95 & $3.29 \times 10^{1}$ & 3.30 & $1.19 \times 10^{-2}$ & 0.00 & 2.88 & $1.90 \times 10^{-1}$ & 1.19 \\
NGC 6362 & 20.22 & 4.77 & $4.81 \times 10^{-1}$ & $4.97 \times 10^{-3}$ & $1.24 \times 10^{-3}$ & $5.33 \times 10^{-1}$ & $4.97 \times 10^{-2}$ & $1.83 \times 10^{-1}$ \\
 & 199.33 & $3.40 \times 10^{1}$ & 3.82 & $2.95 \times 10^{-2}$ & $2.95 \times 10^{-3}$ & 3.49 & $1.59 \times 10^{-1}$ & 1.11 \\
NGC 5053 & 21.71 & $5.63 \times 10^{-1}$ & $2.51 \times 10^{-2}$ & $3.14 \times 10^{-4}$ & $3.14 \times 10^{-4}$ & $3.21 \times 10^{-2}$ & $1.57 \times 10^{-3}$ & $1.79 \times 10^{-2}$ \\
 & 199.65 & 3.89 & $1.71 \times 10^{-1}$ & 0.00 & 0.00 & $2.33 \times 10^{-1}$ & $2.33 \times 10^{-1}$ & $1.22 \times 10^{-1}$ \\
NGC 6121 & 20.70 & $1.51 \times 10^{1}$ & 2.31 & $4.30 \times 10^{-2}$ & $5.02 \times 10^{-3}$ & 2.19 & $6.99 \times 10^{-1}$ & $9.91 \times 10^{-1}$ \\
 & 200.53 & $1.22 \times 10^{2}$ & $1.71 \times 10^{1}$ & $3.25 \times 10^{-1}$ & $2.80 \times 10^{-2}$ & $1.74 \times 10^{1}$ & 3.45 & 6.74 \\
 & 1039.16 & $3.85 \times 10^{2}$ & $4.32 \times 10^{1}$ & $5.90 \times 10^{-1}$ & $8.43 \times 10^{-2}$ & $5.50 \times 10^{1}$ & 4.80 & $1.70 \times 10^{1}$ \\
NGC 5694 & 20.49 & $2.21 \times 10^{1}$ & 4.62 & $9.93 \times 10^{-2}$ & $2.19 \times 10^{-3}$ & 4.36 & 2.34 & 2.34 \\
 & 200.39 & $1.98 \times 10^{2}$ & $3.83 \times 10^{1}$ & $8.39 \times 10^{-1}$ & $4.49 \times 10^{-2}$ & $3.53 \times 10^{1}$ & $1.57 \times 10^{1}$ & $1.69 \times 10^{1}$ \\
 & 1001.94 & $6.90 \times 10^{2}$ & $1.10 \times 10^{2}$ & 2.87 & $2.87 \times 10^{-2}$ & $1.14 \times 10^{2}$ & $2.75 \times 10^{1}$ & $5.06 \times 10^{1}$ \\
NGC 6093 & 19.85 & $3.70 \times 10^{1}$ & 9.09 & $1.23 \times 10^{-1}$ & $1.33 \times 10^{-3}$ & 6.19 & 4.46 & 5.21 \\
 & 198.31 & $3.96 \times 10^{2}$ & $9.52 \times 10^{1}$ & 1.33 & $2.83 \times 10^{-2}$ & $6.85 \times 10^{1}$ & $3.67 \times 10^{1}$ & $4.64 \times 10^{1}$ \\
 & 1004.51 & $2.01 \times 10^{3}$ & $4.60 \times 10^{2}$ & 3.68 & $1.08 \times 10^{-1}$ & $3.81 \times 10^{2}$ & $1.42 \times 10^{2}$ & $2.03 \times 10^{2}$ \\
NGC 5286 & 12.29 & $1.10 \times 10^{1}$ & 1.50 & $3.09 \times 10^{-2}$ & $1.82 \times 10^{-3}$ & 1.45 & $6.40 \times 10^{-1}$ & 1.06 \\
 & 198.28 & $2.07 \times 10^{2}$ & $3.19 \times 10^{1}$ & $3.95 \times 10^{-1}$ & $3.95 \times 10^{-2}$ & $3.38 \times 10^{1}$ & 5.97 & $1.45 \times 10^{1}$ \\
 & 787.45 & $7.43 \times 10^{2}$ & $1.14 \times 10^{2}$ & 1.06 & $9.61 \times 10^{-2}$ & $1.23 \times 10^{2}$ & $1.20 \times 10^{1}$ & $3.96 \times 10^{1}$ \\
NGC 6656 & 19.80 & $1.53 \times 10^{1}$ & 2.18 & $4.90 \times 10^{-2}$ & $7.90 \times 10^{-4}$ & 2.37 & $4.74 \times 10^{-1}$ & 1.13 \\
 & 205.86 & $1.52 \times 10^{2}$ & $2.32 \times 10^{1}$ & $2.94 \times 10^{-1}$ & $1.96 \times 10^{-2}$ & $2.46 \times 10^{1}$ & 3.55 & 9.43 \\
 & 1000.35 & $5.92 \times 10^{2}$ & $7.79 \times 10^{1}$ & 1.01 & 0.00 & $9.36 \times 10^{1}$ & 8.36 & $2.90 \times 10^{1}$ \\
NGC 1851 & 20.76 & $2.40 \times 10^{1}$ & 4.28 & $9.53 \times 10^{-2}$ & 0.00 & 2.74 & 2.69 & 3.41 \\
 & 203.71 & $2.77 \times 10^{2}$ & $4.80 \times 10^{1}$ & $8.21 \times 10^{-1}$ & $4.40 \times 10^{-2}$ & $3.62 \times 10^{1}$ & $2.41 \times 10^{1}$ & $2.85 \times 10^{1}$ \\
 & 1039.94 & $1.42 \times 10^{3}$ & $2.45 \times 10^{2}$ & 5.24 & $2.28 \times 10^{-1}$ & $2.30 \times 10^{2}$ & $8.11 \times 10^{1}$ & $1.17 \times 10^{2}$ \\
NGC 6205 & 20.10 & $1.42 \times 10^{1}$ & 2.06 & $4.34 \times 10^{-2}$ & $1.50 \times 10^{-3}$ & 2.28 & $5.73 \times 10^{-1}$ & 1.17 \\
 & 199.58 & $1.35 \times 10^{2}$ & $1.94 \times 10^{1}$ & $3.69 \times 10^{-1}$ & $1.20 \times 10^{-2}$ & $2.23 \times 10^{1}$ & 3.60 & 8.74 \\
 & 998.62 & $5.12 \times 10^{2}$ & $6.68 \times 10^{1}$ & $7.62 \times 10^{-1}$ & 0.00 & $7.95 \times 10^{1}$ & 7.92 & $2.49 \times 10^{1}$ \\
NGC 6441 & 20.98 & $2.57 \times 10^{1}$ & 3.95 & $7.91 \times 10^{-2}$ & $1.32 \times 10^{-2}$ & 2.26 & 2.76 & 5.12 \\
 & 212.57 & $3.54 \times 10^{2}$ & $6.14 \times 10^{1}$ & 1.57 & $4.36 \times 10^{-2}$ & $4.98 \times 10^{1}$ & $2.65 \times 10^{1}$ & $5.08 \times 10^{1}$ \\
 & 1010.37 & $2.07 \times 10^{3}$ & $3.32 \times 10^{2}$ & 7.38 & 0.00 & $3.06 \times 10^{2}$ & $1.06 \times 10^{2}$ & $1.99 \times 10^{2}$ \\
NGC 104 & 22.49 & $2.51 \times 10^{1}$ & 4.83 & $1.88 \times 10^{-1}$ & $4.71 \times 10^{-3}$ & 3.36 & 4.29 & 4.34 \\
 & 222.95 & $2.92 \times 10^{2}$ & $5.64 \times 10^{1}$ & 1.90 & $1.07 \times 10^{-2}$ & $4.70 \times 10^{1}$ & $4.08 \times 10^{1}$ & $3.80 \times 10^{1}$ \\
 & 979.55 & $1.30 \times 10^{3}$ & $2.33 \times 10^{2}$ & 7.90 & $5.72 \times 10^{-2}$ & $2.21 \times 10^{2}$ & $1.29 \times 10^{2}$ & $1.33 \times 10^{2}$ \\
NGC 5139 & 20.84 & 7.15 & $8.37 \times 10^{-1}$ & $1.89 \times 10^{-2}$ & $7.90 \times 10^{-4}$ & 1.16 & $1.52 \times 10^{-1}$ & $4.63 \times 10^{-1}$ \\
 & 207.50 & $7.02 \times 10^{1}$ & 6.80 & $1.07 \times 10^{-1}$ & $1.78 \times 10^{-2}$ & $1.15 \times 10^{1}$ & $9.86 \times 10^{-1}$ & 3.45 \\
 & 1009.04 & $2.91 \times 10^{2}$ & $2.84 \times 10^{1}$ & $5.15 \times 10^{-1}$ & $2.15 \times 10^{-2}$ & $4.55 \times 10^{1}$ & 4.29 & $1.14 \times 10^{1}$ \\

  \end{tabular}
  \caption{Expected number of mergers. For each cluster and number of
    retained BHs, we list the exact number of BHs in the cluster along with the
    expected number of mergers over the cluster lifetime.}
  \label{table:mergers}
\end{table*}
The velocity $v$ is computed
from the components of the space velocity
in the heliocentric galactic 
coordinate system $(U,V,W)$, a right-handed coordinate system
with $U$ in the direction of the galactic center, $V$ along the direction of rotation,
and $W$ pointing toward the galactic north pole.
The median values of the velocity
components for the MAX case are $(U,V,W) = (-24.47,-211.31,-22.23) \, \mathrm{km} \, \mathrm{s}^{-1}$. The
large negative velocity in the $V$ component is indicative of this population
not participating in galactic rotation.
The peculiar velocity --- the velocity of a source relative to a local standard
of rest, obtained by removing the contribution of galactic rotation at the
source distance in the galactic plane $R$ --- is sometimes used to
infer a `natal kick' for BH-LMXBs. Although it is possible to convert the
Galactic space velocity to a peculiar velocity, this inferred `kick velocity'
is only justified in assuming the source was born in the galactic disk, where
it participates in galactic rotation.
For BH-LMXBs formed in the field, which is most likely to occur in the
disk, this is a reasonable assumption.
However, the $V$ component of the
BH-LMXBs from GCs indicate low rotational velocities,
which is consistent with the parent GC halo orbits, which are typically non-circular
and extend well out of the galactic plane. As the BH-LMXBs with GC origins
are ejected at relatively low velocities along the GC's orbit in the galaxy,
this population of BH-LMXBs has a velocity
distribution consistent with the high-velocity halo orbits of GCs.
As these systems have high apparent peculiar velocities, due to
their halo orbits and the lack of participation in galactic rotation,
attempting to infer a `natal kick' from the peculiar velocity
in such a case is ill-posed and leads to the conclusion of a large required
`natal kick.'

\subsection{Merger events}\label{sec:mergers}
\subsubsection{GW-driven mergers}
As briefly discussed in section~\ref{sec:clustev},
we allow for gravitational radiation driven mergers between compact objects.
Since all of our `test binaries' contain at least one BH, the allowable
set of GW merger pairs is limited to BH-NS, BH-WD, and BH-BH. 
In addition to those binaries that merge during their evolution within the cluster,
binaries of these types can also be ejected from the cluster.
In the case of the ejection of a compact pair, we calculate the expected
merger time $t_\mathrm{d}$ using the ejected binary parameters
and refer to these as post-ejection mergers if
$t_\mathrm{ej} + t_\mathrm{d} < t_\mathrm{H}$, where
$t_\mathrm{H} = 10^{10} \, \mathrm{yr}$ is approximately the Hubble time.
The total merger rate includes these post-ejection mergers
in addition to the in-cluster mergers. Here we present an estimate of
the merger rates averaged over the $10^{10} \, \mathrm{yr}$ simulations
for different BH retention values.

For notational convenience, we refer to a parameter set
as $x_i$, where the index $i$ runs over the 39 parameter sets
which make up each row of Table~\ref{table:runs} and corresponds
to a specific GC and value of $N_\mathrm{BH}$.
We compute the expected number of mergers for each parameter set
by considering the
probability of a BH being involved in a merger, defined simply
by $P_\mathrm{m}(x_i) = \frac{N_\mathrm{mergers}(x_i)}{N_\mathrm{runs}(x_i)}$,
multiplied by the BH population
\begin{equation}\label{eq:expmerge0}
  \langle N_\mathrm{m} \rangle_i = P_\mathrm{m}(x_i) \, N_\mathrm{BH}(x_i) \,\,.
\end{equation}
In the case of a merger involving two BHs, the expectation value
is calculated using $N_\mathrm{BH}(x_i)/2$ in order to avoid double
counting.
The rightmost three columns of Table~\ref{table:mergers} list the expected
number of GW-driven compact object mergers over the lifetime of
each cluster for a given BH population.
The number of BH-BH mergers is strongly correlated with the GC
core density $n_\mathrm{o}$.
Each population of BHs has a merger expectation value
that follows a power-law in the core density with exponent $\sim 0.58$.
Since we do not include primordial binaries, exchange encounters
are the only means to forming BH-BH binaries that can later merge.
The average rate of encounters is directly proportional to the
density, with the highest density clusters
providing the largest number of opportunities to successfully form
BH-BH binaries.
There are additional correlated variables,
such as the concentration $c$ and velocity dispersion $\sigma$, however
these are secondary to the density $n_\mathrm{o}$ and likely due to
their own correlation with $n_\mathrm{o}$.

Given the expected number of mergers for each cluster, we determine
a weighted average using the GC mass function,
since the total cluster mass of GCs is not uniformly distributed~\citep{McLaughlin:1996}.
We do this individually for each group of simulations
belonging to the sets $N_\mathrm{BH} = \{20,200,1000\}$,
utilizing the GC mass spectrum $dN(M_\mathrm{c})/dM_\mathrm{c}$
of~\cite{McLaughlin:1996}.
For each simulated cluster, 
we assign a weight $w_i = N(M_\mathrm{c}(x_i))$ and compute the expected number of mergers
per cluster in the Milky Way from
\begin{equation}\label{eq:expmerge}
  \langle N_\mathrm{m} (N_\mathrm{BH}) \rangle = \frac{\sum_i w_i \, \langle N_\mathrm{m} \rangle_i}{\sum_i w_i} \,\,.
\end{equation}
For clarity, to obtain the expected number of mergers for $N_\mathrm{BH} = 20$, we sum
over all parameter sets in Table~\ref{table:runs} with $N_\mathrm{BH} = 20$.
The resulting expected number of BH-BH mergers over the life of
a cluster for each choice of $N_\mathrm{BH}$ are 
$\langle N_\mathrm{m} (20) \rangle = 0.513$,
$\langle N_\mathrm{m} (200) \rangle = 5.08$,
and $\langle N_\mathrm{m} (1000) \rangle = 62.5$.

We convert the expected number of mergers to a merger rate density
by assuming that our simulations of Milky Way GCs are a fair representation
of GCs in other galaxies,
that the GCs are all approximately $t_\mathrm{GC} = 10^{10}$ yrs old,
and that the spatial density of GCs in the universe is 
$\rho_\mathrm{GC} = 0.77 \,\mathrm{Mpc}^{-3}$
(see supplemental materials of~\citealt{Rodriguez:2015}).
Using the weighted averages computed above as our `typical' cluster merger values
and assigning this value to each GC in the volume, we obtain
the merger rate density due to all GCs in the universe,
\begin{equation}\label{eq:rateeqn}
  \langle R (N_\mathrm{BH}) \rangle = \frac{\langle N_\mathrm{m} (N_\mathrm{BH}) \rangle}{t_\mathrm{GC}} \rho_\mathrm{GC} \,\,.
\end{equation}
In Table~\ref{table:gwmergers}, we provide the computed estimated 
merger rate densities for compact object mergers due to GCs
for the three populations of $N_\mathrm{BH}$ we consider.
\begin{table}
  \begin{centering}
  \begin{tabular}{r|l|l|l}
    $\langle R(N_\mathrm{BH}) \rangle $ & BH-BH & BH-NS & BH-WD \\
\hline \hline
$\langle R(20) \rangle$ & $3.95 \times 10^{-2}$ & $2.71 \times 10^{-2}$  & $7.15 \times 10^{-2}$ \\
$\langle R(200) \rangle$   & $3.91 \times 10^{-1}$ & $2.51 \times 10^{-1}$  & $7.73 \times 10^{-1}$ \\
$\langle R(1000) \rangle$  & 4.81 & 2.83  & 10.59 \\
\hline

  \end{tabular}
  \caption{The contribution to the compact merger rate density from all GCs in
    the universe, stated in $\mathrm{Gpc}^{-3} \, \mathrm{yr}^{-1}$.
    Each row corresponds to the merger rate contribution from GCs
    with the simulated BH population specified by $N_\mathrm{BH}$ in
    $\langle R(N_\mathrm{BH})\rangle$. The merger rate densities are averaged
    over the life of the cluster, weighted by the GC mass function to account
    for the non-uniform mass distribution of GCs,
    and assumes a GC spatial
    density of $\rho_\mathrm{GC} = 0.77 \,\mathrm{Mpc}^{-3}$.
  }
  \label{table:gwmergers}
  \end{centering}
\end{table}
Although there is an increased
interest in the BH-mass spectrum for BH-BH mergers in GCs,
stimulated by the larger than expected BH masses
recently detected by aLIGO~\citep{Abbott:2016},
the use of just three discrete BH masses precludes the possibility
of such an analysis.

Since BH-BH mergers from GCs only partially contribute to the total merger
rate, with the remaining mergers coming from the field, the rates due to
GCs should not exceed the upper bound of the total estimated merger rate.
The most recent observational evidence constrains the BH-BH merger rate density
to lie in the range $12-213 \; \mathrm{Gpc}^{-3} \, \mathrm{yr}^{-1}$~\citep{Abbott:2017}.
The BH-BH merger rate densities given in Table~\ref{table:gwmergers}
for the three different BH retention scenarios are well below
the upper bound, presenting no conflict with the observed rate.
However, since the fraction of the total mergers attributable
to GCs is still largely uncertain, none of the of GC rates presented
here can be ruled out based on the current observed rate of BH-BH mergers.

The bounds of our merger rates, which span a wide range of uncertainty
in BH retention, are consistent with previous studies that provide
estimates of the BH-BH merger rate from GCs~(\citealt{Oleary:2006};
~\citealt{Sadowski:2008};~\citealt{Downing:2011};~\citealt{Morscher:2015},
~\citealt{Rodriguez:2016}). However, we find that only $\sim 10$ per cent
of the BH-BH mergers occur outside of the cluster boundaries, which
differs from a subset of these previous studies.
In~\cite{Downing:2011} no mergers occur in-cluster, while
in~\cite{Morscher:2015}, $\sim 85$ per cent of BH-BH mergers occur post-ejection,
and~\cite{Rodriguez:2016} find that $\sim 90$ per cent merge outside the cluster.
In contrast to the small number of BH-BH binaries these studies find
merging in cluster,~\cite{Oleary:2006} finds that
only $\sim 24-72$ per cent of the BH-BH mergers are post-ejection.
Finally,~\cite{Sadowski:2008} is most closely aligned with our results,
with $\sim 10$ per cent of mergers occurring out of the cluster.

This discrepancy in merger location can be attributed to the
distribution of the BHs in the cluster and their interactions
with the lower-mass components.
In models with centrally clustered BHs, the BHs are segregated
from the remainder of the cluster, forming an isolated
and decoupled system. These self-interacting BHs efficiently
form BH binaries. Strong binary-binary interactions
can eject these binary BHs from the cluster,
where they might later merge in isolation.
In addition to the efficient removal of BH binaries from
the core, binary-single interactions are equally efficient
at ejecting single BHs from the cluster. 
Furthermore, these strong encounters are likely
to interrupt potential mergers of eccentric BH binaries which
would merge in-cluster if uninterrupted.
This channel leads to a majority of BH-BH mergers outside of the cluster and
eventually depletes the GC of BHs~(e.g.,~\citealt{Oleary:2006};~\citealt{Banerjee:2010};~\citealt{Downing:2011}).
We assume that in order for GCs to retain significant BH populations,
the BHs must avoid segregating in the core, which we accomplish
through a modified velocity dispersion for the BHs, as discussed in section~\ref{sec:equip}.
This modified velocity dispersion spreads the BHs throughout the cluster,
where they can interact with the lower-mass stars. This supposition is
similar to the assumptions made in~\cite{Sadowski:2008} and produces
qualitatively similar results.

\begin{figure}
  \includegraphics[width=1.0\columnwidth]{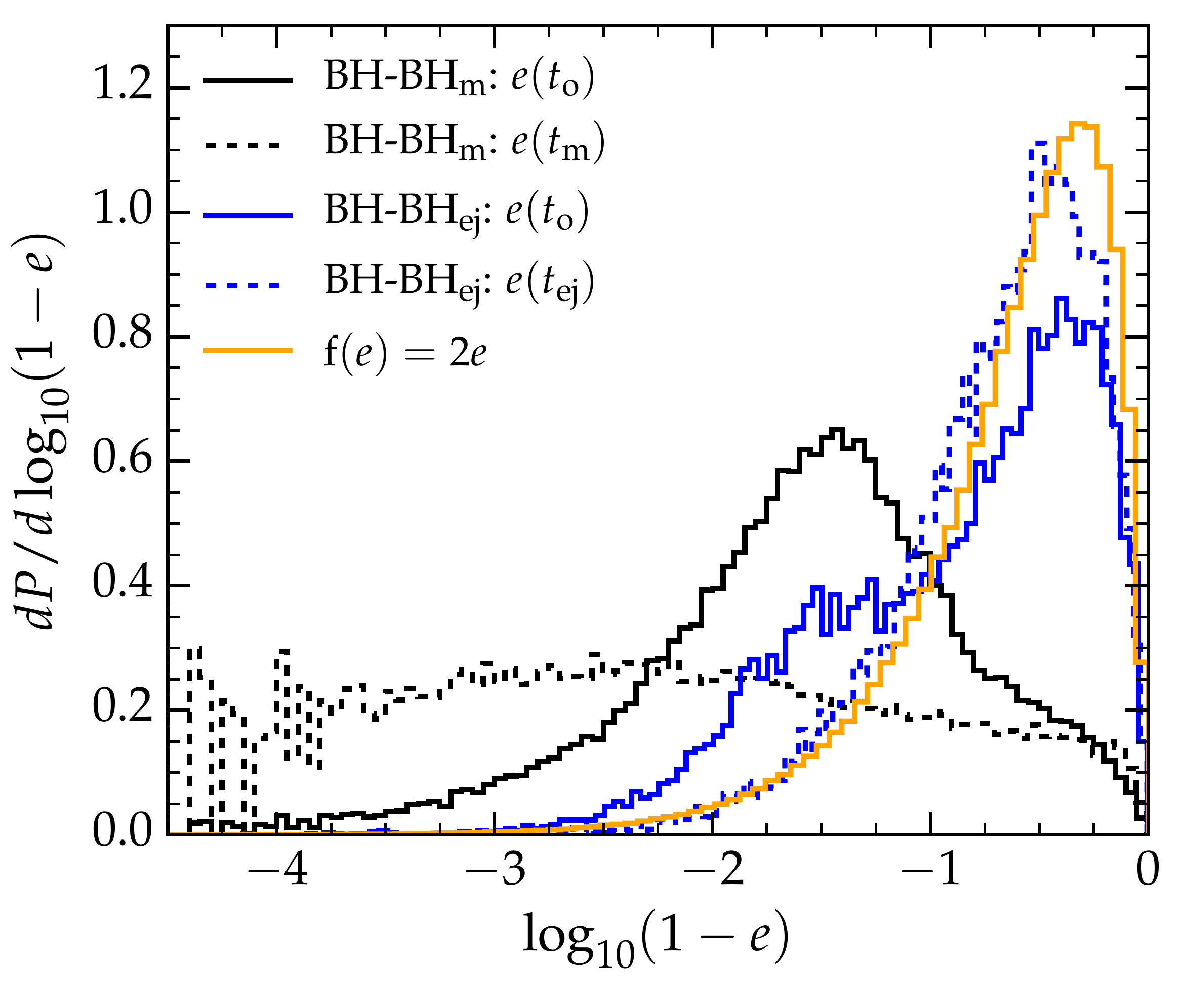}
  \vspace{-2pt}
  \caption{
    The probability distributions of eccentricity for two populations of
    BH-BH binaries in GCs:
    BH-BH binaries which form and merge in cluster (BH-BH$_\mathrm{m}$, black lines)
    and the BH-BH binaries which form and are ejected from the
    cluster (BH-BH$_\mathrm{ej}$, blue lines).
    For each population, we show the eccentricity distribution
    at the time the binary forms, $e(t_\mathrm{o})$ (solid lines), and the
    distribution of eccentricities at the binary's final state (dashed lines).
    The final state of the in-cluster mergers is at a time
    $t_\mathrm{m}$, the time at which the computed merger time is
    less than the cluster timestep. The final state for the
    ejected binaries is the time of ejection $t_\mathrm{ej}$.
    A thermal eccentricity distribution, with probability density $f(e)=2e$, is included
    for reference.
  }
  \label{fig:ecc_dist}
\end{figure}

In our simulations, a key channel for
producing BH-BH binaries is through the formation of a binary composed
of a BH and a non-BH outside of the core,
which eventually drift to the center where there is
a high density of BHs.
The non-BH will be preferentially exchanged with
one of the more massive BHs in the core, producing a BH-BH binary that
will realize one of three outcomes: (1) the BH-BH binary will be dismantled in
the high density region,
(2) given a sufficiently large eccentricity (hence a shorter orbital decay time),
will eventually merge in the core,
or (3) will harden and be ejected from the cluster.
This formation channel is similar to that described in~\cite{Sadowski:2008}.
As discussed in section~\ref{sec:ion}, we allow for single BHs to exchange into
existing binaries. The majority of binaries that a single BH encounters are
binaries composed of two low-mass stars. Successful exchanges of a more massive
BH for one of the lower-mass stars tend to produce high-eccentricity BH--non-BH binaries
following the relation
\begin{equation}\label{eq:ecc}
  \langle e \rangle \approx 1 - 1.3 \big(\frac{m_\mathrm{non-BH}}{m_\mathrm{BH}} \big) \,\,,
\end{equation}
which is independent of the initial
eccentricity and applicable when $m_\mathrm{non-BH} \ll m_\mathrm{BH}$~\citep{Sigurdsson:1993B}.
For the three BH masses considered,
$M_\mathrm{BH} = \{8.87,20.48,57.18\} \, \mathrm{M_{\odot}}$,
and a cluster non-BH star with an average 
mass of $\langle m_\mathrm{non-BH} \rangle \approx 0.3 \, \mathrm{M_{\odot}}$,
this leads to mean initial eccentricities of $\langle e \rangle \approx \{0.956,0.981,0.993\}$.
Once the binary makes it to the core, the non-BH is easily exchanged
for one of the many massive BHs, yielding a highly eccentric BH-BH
binary according to Equation~\ref{eq:ecc}.
In Figure~\ref{fig:ecc_dist}, we display the eccentricity distributions
for the BH-BH binaries at formation and at merger or ejection for those
binaries that have end states (2) and (3), as described above, respectively.
Some fraction of the eccentric binaries that form through this channel
are driven to high enough eccentricities that they can merge in-cluster
in-between encounters. The remainder are subject to further
encounters that drive their eccentricities toward a thermalized distribution,
are hardened in the process, and are eventually ejected.

The eccentricity distribution of merging BH-BH binaries is important
for the detection of the resulting gravitational waves. The eccentricity
tends to zero as the orbit shrinks,
however modern detectors are sensitive to the GW signal
at frequencies when the binary is still in the inspiral phase
and the eccentricity is finite.
The aLIGO~\citep{LIGO:2015} detectors are sensitive to $\sim 10 \, \mathrm{Hz}$, at
design sensitivity, while the future space-based detector LISA~\citep{Amaro-Seoane:2013}
will be sensitive to much lower frequencies $\sim 1 \, \mathrm{mHz}$.
We determine the eccentricity at a specific frequency by evolving
$a_\mathrm{o}$ and $e_\mathrm{o}$,
according to $\langle de/da \rangle$~\citep{Peters:1964},
up until some target value $a$ associated with the frequency in
consideration.
\begin{figure}
  \includegraphics[width=1.0\columnwidth]{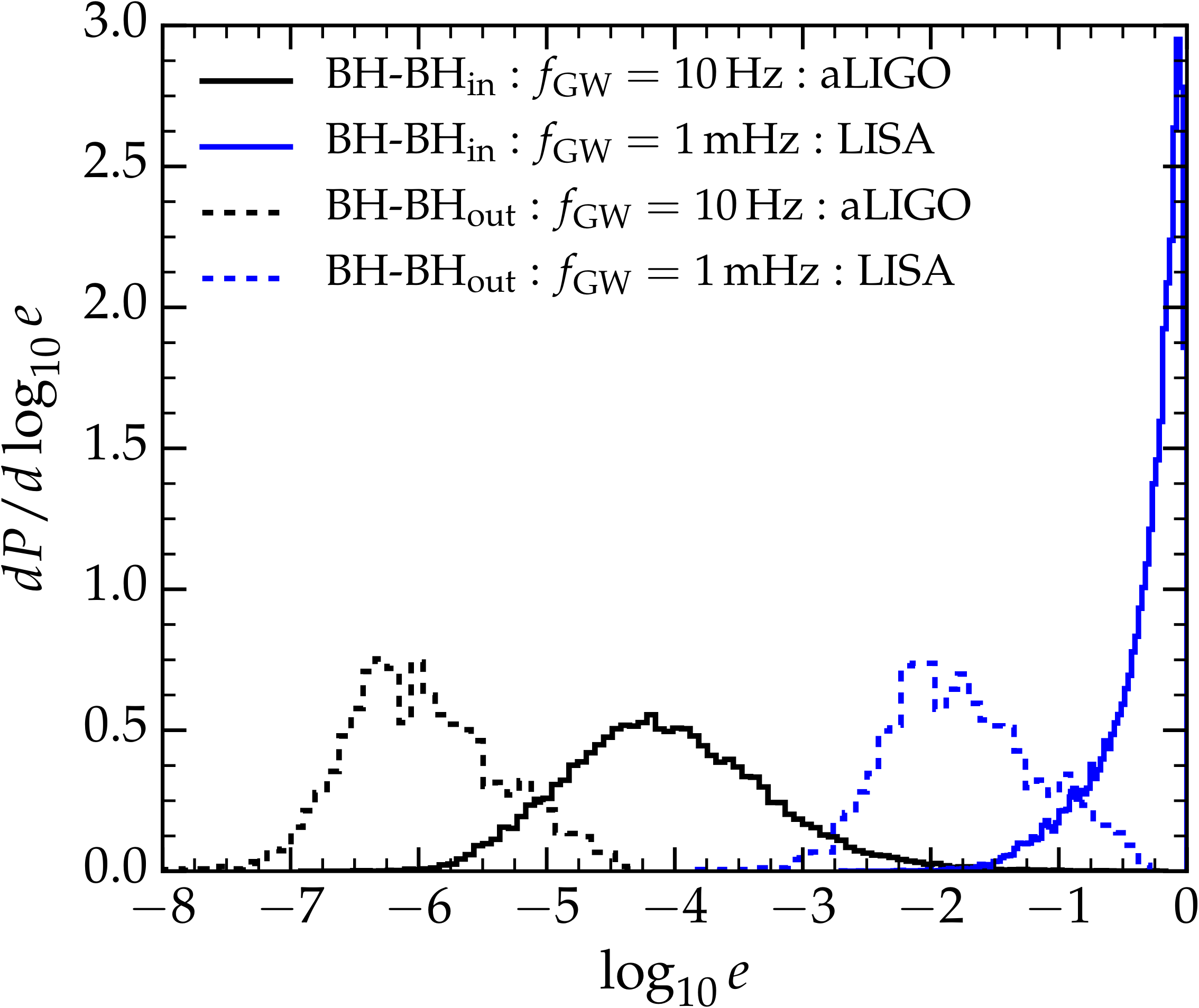}
  \caption{
    The eccentricity probability distributions for two populations of
    BH-BH mergers from GCs for the two detectors aLIGO and LISA.
    The two populations correspond to the BH-BH mergers occurring in-cluster (solid lines)
    and those that merge outside of the cluster, post-ejection (dashed lines).
    The black lines correspond to the eccentricity of each population
    when it reaches a corresponding gravitational wave frequency of
    $f_\mathrm{GW} = 10 \, \mathrm{Hz}$,
    the lower bound frequency of the aLIGO band at design sensitivity.
    The blue lines represent the eccentricity distribution at
    $f_\mathrm{GW} = 1 \, \mathrm{mHz}$, the proposed lower frequency bound for LISA.
  }
  \label{fig:ecc_detect}
\end{figure}
In Figure~\ref{fig:ecc_detect}, we display the residual eccentricity of the inspiraling
BH-BH binaries as they first enter the design-sensitivity frequency bands
for aLIGO and LISA.
It is apparent that for aLIGO, both the ejected mergers
and the initially high-eccentricity in-cluster mergers have residual eccentricity distributions
below $10^{-1}$, which has a negligible effect on detections using circularized
templates. However, in the case of LISA, while the ejected mergers result in
a small eccentricity at $1 \, \mathrm{mHz}$, the initially
highly eccentric in-cluster merger population remains significantly eccentric
at this frequency.

Utilizing $\langle de/da \rangle$ to determine the evolved eccentricity assumes that
the binary evolves in isolation. For the in-cluster mergers, we classify a BH-BH binary
as merged once the orbital decay time has fallen below the cluster timestep.
However, this could leave significant time for further dynamics
to modify the eccentricity such that the binary will not in fact merge in cluster
\citep{Banerjee:2010}. To account for this possibility, the in-cluster mergers
in Figure~\ref{fig:ecc_detect} only include those mergers which
satisfy the additional constraint $t_\mathrm{dec} < \langle t_\mathrm{enc} \rangle$,
which is satisfied for $\sim 70$ per cent of in-cluster mergers. 
Here, the average encounter time is
approximated by $\langle t_\mathrm{enc} \rangle = t_\mathrm{bin}/N_\mathrm{enc}$
with $t_\mathrm{bin}$ corresponding to the time since the binary's formation and $N_\mathrm{enc}$ is
the number of three-body encounters the binary has been subject to during the time $t_\mathrm{bin}$.
The remaining $\sim 30$ per cent of mergers are uncertain and are not further evolved;
they may be broken up, ejected, or merge after subsequent interactions.

\subsubsection{Three-body mergers}\label{sec:tbmergers}
In addition to the GW-driven mergers, we also calculate the
rate of tidally driven mergers or `collisions' that occur during three-body
encounters. The merger criteria are based on a minimum separation
between bodies, as discussed in section~\ref{sec:encres}.
We compute the expected number of three-body merger events only
for those that involve a BH. 
Although we track the number of three-body mergers for all object types,
including NS-NS, MS-WD, etc., we are missing a significant fraction of these
mergers by only tracking single BHs or binaries with at least one BH.
We compute the expected number of mergers in a manner similar to the
computation of GW mergers above.

The left columns of Table~\ref{table:mergers} list the expected
number of mergers involving a BH that occur during three-body
encounters over the lifetime of each cluster for a given BH population.
These three-body mergers are computed using Equation~\ref{eq:expmerge0}
to obtain an expected value for each cluster in the set.
As the majority of these events will only be observationally
relevant locally, we provide these rates solely for the Milky Way galaxy.
Using the computed values from
Table~\ref{table:mergers} we construct a cluster weighted
average with Equation~\ref{eq:expmerge}. From this we use a modified
version of Equation~\ref{eq:rateeqn}, with
$N_\mathrm{GC} \simeq 150$, for the approximate number of GCs in our galaxy,
in place of $\rho_\mathrm{GC}$ to obtain the final approximate
rate for each event: $\langle R (N_\mathrm{BH}) \rangle = \frac{\langle N_\mathrm{m} (N_\mathrm{BH}) \rangle}{t_\mathrm{GC}} N_\mathrm{GC}$. These computed rates for BH-BH, BH-NS, BH-WD and BH-NC are shown in
Table~\ref{table:tbmergers},
stated in terms of the number of expected events per Milky Way equivalent galaxy (MWEG) per Myr.
The BH-NC merger rate includes the three-body mergers of both BH-RG and BH-MS.
\begin{table}
  \begin{centering}
  \begin{tabular}{r|l|l|l|l}
    $\langle R(N_\mathrm{BH}) \rangle $ & BH-NC & BH-WD & BH-NS & BH-BH \\
\hline \hline
$\langle R(20) \rangle$ & $1.02 \times 10^{-1}$ & $1.60 \times 10^{-2}$  & $2.65 \times 10^{-4}$ & $1.64 \times 10^{-5}$\\
$\langle R(200) \rangle$   & $1.08$ & $1.72 \times 10^{-1}$  & $2.53 \times 10^{-3}$ & $1.40 \times 10^{-4}$\\
$\langle R(1000) \rangle$  & $12.27$ & $2.14$ & $3.03 \times 10^{-2}$ & $1.11 \times 10^{-3}$\\
\hline

  \end{tabular}
  \caption{
    The rate of three-body mergers in GCs computed for the
    Milky Way galaxy and stated in $\mathrm{MWEG}^{-1} \mathrm{Myr}^{-1}$.
    Each row corresponds to the three-body merger rate in Milky Way GCs
    with the simulated BH population specified by $N_\mathrm{BH}$ in
    $\langle R(N_\mathrm{BH})\rangle$. The merger rates are averaged
    over the life of the cluster, weighted by the GC mass function to account
    for the non-uniform mass distribution of GCs,
    and assumes $N_\mathrm{GC} \simeq 150$ for the number of GCs in the galaxy.
    }
  \label{table:tbmergers}
  \end{centering}
\end{table}

These rates are included to ensure that a large population of retained
BHs in GCs does not lead to a conflict with observations.
Even in the case of maximal BH retention, the occurrence of these events
is relatively infrequent. The most commonly occurring three-body collision is
that between a BH and a NC star. The interaction of a NC object with a BH,
commonly referred to as a tidal disruption event (TDE), is often
studied in the context of supermassive BHs rather than stellar-mass BHs.
However, there is some interest in GC-relevant NC collisions
with stellar-mass BHs, which are referred to as micro-TDEs~\citep{Perets:2016}.
These events lead to full or partial tidal disruption of the NC star and
are accompanied by long-duration energetic flares. There is large uncertainty
in the signals associated with these events as the strength and duration of
the signal depends heavily on the details of the encounter~(see, e.g.,~\citealt{Perets:2016}).

The signals associated with the compact mergers are likely to appear as head-on
mergers due to the criteria associated with categorizing mergers during
three-body encounters; the exclusion of higher order corrections to Newtonian gravity in
our three-body calculations requires extremely close-encounters due to the relatively small
size of the compact objects involved. 
Despite the uncertainty in the observables produced in three-body collisions,
the rate of occurrence is low enough that our model does not generate a conflict
with present observations.

\subsection{Comparison with observations and previous results}\label{sec:compare}

In our simulations, GCs produce a population of BH-LMXBs with a unique set of characteristic
properties. These properties provide some constraints on the likelihood of a BH-LMXB having
a GC origin. In this section, we identify the key characteristics of BH-LMXBs from GCs
and determine which of the currently known BH-LMXBs are consistent with this population.

As discussed in section~\ref{sec:props} and visible in Figure~\ref{fig:lmxb_mbh},
the spectrum of BH masses in BH-LMXBs from GCs in our simulations is roughly consistent with the
observed population of BH masses. This makes the BH mass a poor candidate for
differentiating between field-formed BH-LMXBs and those with a GC origin.
As a consequence of the age of GCs, the companions are typically
unevolved MS stars, with masses necessarily below the turnoff-mass $m_\mathrm{to} = 0.85 \, M_\odot$.
Additionally, they reside on a tightly confined branch of a temperature-luminosity
diagram (see Figure~\ref{fig:lum_temp}). This provides the first distinctive characteristic
of BH-LMXBs formed in GCs: a companion mass of $m_2 \lesssim 0.85 \, M_\odot$
and a spectral class consistent with late-type K/M stars.
BlackCAT~\citep{Corral-Santana:2016} currently contains 18 observed BH-LMXB systems with
the proper information to compute an estimate of the companion mass.
Of the 18 systems, six BH-LMXBs
have companion masses exceeding the maximum companion mass in our population of BH-LMXBs from GCs.
Two of these six are near the edge of the distribution with with $m_2 \gtrsim 0.9 \, M_\odot$,
while the other four have $m_2 \ge 2.52 \, M_\odot$,
suggesting these are more consistent with a field-formation scenario.

A second property of a BH-LMXB with a GC origin is a characteristically short period.
As shown in Figure~\ref{fig:period}, there is a sharp limit in the distribution 
confining GC-origin BH-LMXBs to periods shorter than $p\sim 6.5 \, \mathrm{h}$.
Of the 27 confirmed BH-LMXBs with measured periods in BlackCAT, 18 have periods with
$p > 7 \, \mathrm{h}$, indicating an unlikely GC origin for an additional set of systems.
Note, however, that these systems are not necessarily distinct from those
ruled unlikely on the basis of companion mass.

Although the GC-origin
BH-LMXBs are more likely to reside at larger values of $|z|$ perpendicular
to the galactic plane (see Figure~\ref{fig:z_hist}),
the overall distribution of the BH-LMXBs from GCs does not provide a strict criterion
for discerning between GC origin and field origin. 
Figure~\ref{fig:z_plot} illustrates
that while the simulated population extends much farther out of the galactic plane than
the observed distribution, there is still a significant population of GC-origin
BH-LMXBs that reside in the plane, overlapping the region where field-formed
binaries are expected to have the highest density.
This makes discerning a potential origin for BH-LMXBs in this region difficult.
Additionally, for the many systems clustered
near the galactic center or those that reside in the plane, the high
density of objects and dust make these systems equally difficult to observe
optically. Although a number of BH-LMXB candidates are detectable in these
regions through X-ray, the detailed properties of these systems remain unknown due
to current optical limitations. 
The spatial distribution of BH-LMXBs from GCs,
in general, makes observations of the population difficult, even
for those out of the plane.
Observation and confirmation of BH-LMXBs rely on a dynamical measurement
of the BH mass through optical spectroscopy, introducing a bias toward
sources at distances $D < 10 \, \mathrm{kpc}$ from the Sun~\citep{Repetto:2015}.
For the population of BH-LMXBs from our model GCs, the MAX and 200 cases both
produce a median distance of $D = 9.7 \, \mathrm{kpc}$, placing roughly half
of the systems beyond the observable range.

\begin{table*}
  \begin{centering}
  \begin{tabular}{l|l|l|l|l|c}
    Name & \hphantom{...} $M_\mathrm{BH} \, [M_\mathrm{\odot}]$ & \hphantom{........} $m_{2} \, [M_\mathrm{\odot}]$ & \hphantom{...........} $p \, [\mathrm{h}]$ & \hphantom{.......} $|z| \, [\mathrm{kpc}]$ & \hphantom{..} References \\
\hline \hline
MAXI J1659-152 & \hphantom{...} $5.8 \hphantom{5} \pm 2.2$ & \hphantom{...} $0.19 \hphantom{7} \pm 0.05$  & $2.414 \hphantom{41} \pm 5\times10^{-3}$ & \hphantom{...} $2.45 \pm 1.05$ & [0,1]\\
SWIFT J1357.2-0933 & $> 8.3$ & $> 0.33$  & $2.8 \hphantom{7841} \pm 3\times10^{-1}$ & $> 1.75$ &  [2,3] \\
SWIFT J1753.5-0127 & $> 7.4 \hphantom{5} \pm 1.2$ & $\ge 0.30 \hphantom{7} \pm 0.03$  & $3.244 \hphantom{41} \pm 1\times10^{-3}$ & \hphantom{...} $1.3 \hphantom{2} \pm 0.4$ & [4-7]\\
XTE J1118+480 & \hphantom{...} $ 7.55 \pm 0.65$ & \hphantom{...} $0.187 \pm 0.083$  & $4.07841 \pm 1\times10^{-5}$ & \hphantom{...} $1.52 \pm0.09$ & \hphantom{.} [8-11]\\
GRO J0422+32 & \hphantom{...} $8.5 \hphantom{5} \pm 6.5$ & \hphantom{...} $0.46 \hphantom{7} \pm 0.31$  & $5.09185 \pm 5\times10^{-6}$ & \hphantom{...} $0.51 \pm 0.06$ & \hphantom{...} [12-15]\\
\hline

  \end{tabular}
  \caption{
    Properties of the five observed systems that are
    consistent with the properties of our simulated population of
    BH-LMXBs with GC origins.
    The columns refer to the primary BH mass $M_\mathrm{BH}$, the
    companion mass $m_2$, the orbital period $p$, and the
    absolute distance perpendicular to the galactic plane $|z|$.
    [0]~\citet{Yamaoka:2012}, 
    [1]~\citet{Kuulkers:2013}, 
    [2]~\citet{Sanchez:2015}, 
    [3]~\citet{Corral-Santana:2013}, 
    [4]~\citet{Shaw:2016}, 
    [5]~\citet{Neustroev:2014}, 
    [6]~\citet{Zurita:2008}, 
    [7]~\citet{CadolleBel:2007}, 
    [8]~\citet{Khargharia:2013}, 
    [9]~\citet{Calvelo:2009}, 
    [10]~\citet{Torres:2004}, 
    [11]~\citet{Gelino:2006}, 
    [12]~\citet{Casares:1995}, 
    [13]~\citet{Beekman:1997}, 
    [14]~\citet{Webb:2000}, 
    [15]~\citet{Gelino:2003} 
    }
  \label{table:obsLMXBs}
  \end{centering}
\end{table*}
Although this model population has characteristics that make observations
of the binary properties difficult, there are some observed systems
that provide a resemblance to those with GC origins.
There are 18 observed and confirmed BH-LMXBs in BlackCAT with measured quantities
that allow for comparison with our simulated population.
Five of the 18 systems have a BH mass, companion mass, and period consistent
with the characteristics of our population of BH-LMXBs from GCs.
These systems are MAXI J1659-152, SWIFT J1357.2-0933, SWIFT J1753.5-0127, XTE J1118+480,
and GRO J0422+32.
In Table~\ref{table:obsLMXBs}, we list the five consistent systems
and the known properties that are compatible with the range of values belonging
to our population of BH-LMXBs from GCs.
While we cannot make any strong claims in regards to the specific origin
of these systems, it is worthwhile to note the similarities of these systems
with respect to the population produced in this study.

The BH-LMXB system XTE J1118+480 is well studied, which provides
some additional parameters worth comparing with our modeled population of
BH-LMXBs from GCs.
In addition to the consistent mass of the companion star in XTE J1118+480,
the spectral type is also aligned with the band of GC-origin companions
in Figure~\ref{fig:lum_temp}.
Although space-velocity measurements of BH-LMXBs are rare,
fortunately there exists a velocity measurement of XTE J1118+480.
In the same heliocentric galactic coordinate system $(U,V,W)$ introduced in
section~\ref{sec:props},~\cite{Mirabel:2001} found a space-velocity for this
system of $(U=-105\pm16,V=-98\pm16,W=-21\pm10) \, \mathrm{km} \, \mathrm{s}^{-1}$.
The large magnitude $v \sim 145 \, \mathrm{km} \, \mathrm{s}^{-1}$ and the large negative $V$ component
are consistent with a high-velocity halo orbit and a lower
than average rotational velocity about the galactic center.
This description is consistent with the velocity distribution of our population
of BH-LMXBs from GCs, which inherit the high-velocity halo-orbits when they
are ejected from the GC.
As a consequence of
the high-velocity halo orbit, which manifests itself as a high computed peculiar velocity,
this system is commonly invoked to support large
natal kicks~(\citealt{Gualandris:2005};~\citealt{Fragos:2009};~\citealt{Repetto:2012};~\citealt{Repetto:2015}).
Confidently identifying an origin for this system could help to shed some
light on the issue.
The relatively low-metallicity environments
of GCs provides an additional constraint on properly categorizing
BH-LMXBs as originating in GCs versus in the field.
Although all of the previous characteristics point to a GC origin, 
perhaps one of the strongest arguments against a GC origin
for this system is the supersolar abundance of elements
in the secondary star found by~\cite{Gonzales:2006}, which 
is consistent with a metal-rich progenitor and makes a GC origin
highly unlikely.
However, there exist a conflicting claim
presented by~\cite{Frontera:2001}, where through
broad-band X-ray spectroscopy, it was concluded that
the companion has a metallicity of $Z/Z_\mathrm{\odot} \sim 0.1$,
consistent with the low metallicities expected of systems at
large $|z|$ or those with a GC origin.
Given that metallicity provides a strong constraint on the origin of a BH-LMXB,
additional observations appear necessary to reduce the uncertainty
of this case.

To our knowledge, there are no known velocity measurements or
metallicity measurements for the four other BH-LMXBs with possible GC origins.
Although an increasing number of BH-LMXB candidates are being discovered in X-rays,
only a few have been confirmed and characterized with detailed optical
follow-up observations. Over time, more data will become available,
better constraining the properties of the galactic BH-LMXB population.
If even a single BH-LMXB could be confidently attributed to a GC
origin this would provide a strong argument in favor of BH retention
in GCs.

\section{Discussion and conclusions}\label{sec:conclusion}
There is growing observational evidence and theoretical support for a
sizable BH population in present-day galactic GCs. These BHs
can acquire low-mass companions through dynamical interactions
within the GC. Those binaries that are ejected from the GC can evolve
into BH-LMXBs and can populate a large region of space above and
below the galactic plane. These binaries could potentially explain
observed BH-LMXBs at large distances from the plane without a need for
large BH birth kicks.

In this study, we have presented a population of Milky Way BH-LMXBs
formed through dynamical interactions in GCs. 
To explore the BH-LMXB population dependence on
BH retention in GCs, we performed simulations for retained BH populations of 20,
200, and 1000 BHs.
The simulated GCs broadly cover the parameter
space and represent a realistic subset of Milky Way GCs.
We generated a large number of binary evolution realizations for each
set of initial GC parameters and number of retained BHs.
This allowed us to derive statistical distributions for the
number of ejected binaries and their relevant properties.
Using the statistics from the GC simulations,
we performed Monte Carlo simulations to obtain
a present day population of BH-LMXBs ejected from GCs.

We find that in the case of minimal BH retention ($N_\mathrm{BH} = 20$)
no observable BH-LMXBs are produced, while the
$N_\mathrm{BH} = 200$ and $N_\mathrm{BH} = 1000$
cases yield $25^{+10}_{-6}$ and $156^{+26}_{-24}$ BH-LMXBs, respectively.
Here, the uncertainties represent the bounds of the 95 per cent confidence interval.
As there is no observable population for minimal BH retention,
this suggests that finding any BH-LMXB of GC origin would imply that GCs
retain sizable BH populations of more than a few tens of BHs.

Aside from the difference in the size of the population, the properties
and distributions of BH-LMXBs are qualitatively similar for the two cases
that produce BH-LMXBs, 200 and MAX.
We find that BH-LMXBs from GCs have velocity distributions
inherited from their host clusters that are consistent with stars
on high-velocity halo orbits.
Additionally, the ejected BH-LMXBs have a spatial distribution
that is also similarly aligned with the GC galactic distribution.
This shared distribution is described by a 
high density in the galactic plane
and near the galactic center, with a significant fraction distributed
well above and below the galactic plane.
The typical binary is located at an absolute distance
of $R=4.5 \, \mathrm{kpc}$ from the galactic core when projected onto the
galactic plane, an absolute distance of $|z|=1.6 \, \mathrm{kpc}$ perpendicular
to the galactic plane, and at a distance of $D=9.74 \, \mathrm{kpc}$ from the Sun.
The presence of a large population of
BH-LMXBs at large distances from the plane is characteristic of BH-LMXBs from GCs,
as field formed BH-LMXBs must be subject to large kicks
in order to access this region.
The average present-day BH-LMXB ejected from a GC is composed of a
$8.25 \, M_\mathrm{\odot}$ BH and a $0.22 \, M_\mathrm{\odot}$
K/M late-type MS star below the turnoff-mass, with a characteristically short
orbital period of $p=0.186 \, \mathrm{h}$.
These properties and their associated distributions are
key observable characteristics of this predicted population
of BH-LMXBs formed in GCs.

Comparing our BH-LMXB systems with the ensemble of observed BH-LMXBs,
we find that five of these are candidates for having a GC origin.
There are a total of 27 confirmed
BH-LMXBs, but just 18 of these have sufficient observations for
comparing measured properties against our results.
The five systems that are compatible with our simulated population of BH-LMXBs
from GCs are SWIFT J1357.2-0933, SWIFT J1753.5-0127, XTE J1118+480, and GRO J0422+32.
XTE J1118+480 is one of the rare systems with a measured space velocity
and it is atypically large for a system formed in the galactic disk,
with $v \sim 145 \, \mathrm{km} \, \mathrm{s}^{-1}$.
This system is commonly discussed in the context of formation kicks, since a high-velocity kick is
required to explain the large distance from the galactic plane, $|z| \sim 1.52 \, \mathrm{kpc}$,
under the assumption that it originated in the plane.
However, if XTE J1118+480 comes from a GC, which produces BH-LMXBs at a median
distance of $|z| \sim 1.6 \, \mathrm{kpc}$ from the plane,
then its position and velocity are a natural consequence of the GC origin
and do not require a large BH birth kick.

\raggedbottom

Future observations of the remaining four system velocities would provide
an important additional piece of evidence in each of these cases. 
Additionally, the companion stars in BH-LMXBs from GCs should have the same low
metallicity as is typical for GCs. This emphasizes the need for reliable
metallicity measurements of the companion metallicity,
which could help to support or reject a GC origin scenario.
The strength in this measurement relies on the distinctly
low-metallicity environments of GCs compared to the disk environment.
The metallicity of the companion in XTE J1118+480 has been measured
by~\cite{Frontera:2001} and~\cite{Gonzales:2006}. However,
the two measurements disagree, with the former finding sub and the latter
finding super solar metallicity. 
Additional observations may be necessary to settle the discussion
for XTE J1118+480.
Future observations will be needed to more reliably determine or
rule out the potential GC origin of the candidate BH-LMXBs. On the basis
of our GC simulations, we reaffirm that if one or multiple can be shown
to come from a GC, then GCs retain sizable BH populations.

An additional result from our simulations is a prediction for the BH-BH merger
rate as function of the GC BH population.
The expected rate of mergers due to all GCs for our maximum retention case, $N_\mathrm{BH} = 1000$,
is $4.81$ $\mathrm{Gpc}^{-3} \, \mathrm{yr}^{-1}$, while in the case of minimal retention,
$N_\mathrm{BH} = 20$,
the rate is as low as $3.95\times 10^{-2}$ $\mathrm{Gpc}^{-3} \, \mathrm{yr}^{-1}$.
This rate represents an average over the cluster lifetimes and assumes a spatial
density of GCs throughout the universe of $\rho_\mathrm{GC} = 0.77 \,\mathrm{Mpc}^{-3}$.
Our maximum retention rate is consistent with previous estimates of the GC merger rate
contribution and is compatible with the recent observations by aLIGO.
Although our model produces rates in good agreement with previous studies,
our simulations result in a larger than average fraction of mergers
occurring in-cluster, as opposed to post-ejection. We attribute the discrepancy
to the increased interaction between the BHs and the lower mass stars as a
consequence of our cluster BH distribution.
The BH-BH binaries
that merge in-cluster are a consequence of the large eccentricities,
acquired through dynamical formation, leading to significantly
shortened orbital decay times.
The dynamically formed BH-BH binaries that merge in-cluster are formed with an
average eccentricity of $e \sim 0.96$. At the time of merger in the aLIGO
band, the residual eccentricities are small and in the range
$10^{-6} \lesssim e \lesssim 10^{-2}$.
However, we find that when passing through the LISA band
years before merger, they still have eccentricities in the range
$10^{-2} \lesssim e \lesssim 1$.
Models in which the BHs are confined to a subcluster at the core of GCs
produce mergers with substantially smaller eccentricities.
As the merger formation channels are sufficiently different for a BH subcluster
model, LISA might be able to help distinguish how a population of retained BHs
is distributed in GCs by observing the distribution of eccentricities.

The present study provides new insights into the population and
properties of BH-LMXBs of GC origin. However, there are a number of
important limitations that should be kept in mind when interpreting
our results.
While there is mounting evidence to support that present-day GCs are
BH retaining, how GCs are able to retain a significant population
of BHs and how those BHs are distributed is still uncertain.
Our choice of distributing
the BHs throughout the cluster is motivated by preserving the
observed structural properties of each modeled GC in the presence
of a large BH population. However, this spreading leads to
an increase in interaction between the BHs and the lower-mass stars,
which is typically a rare occurrence if the BHs remain clustered in the core. 
If GCs are able to retain a significant population of BHs that
remain centrally clustered, formation of BH-NC binaries will likely be suppressed.
The reduced formation of BH-NC binaries would
significantly reduce the number of ejected BH-NCs, directly diminishing
the number of BH-LMXBs from GCs.
Future studies regarding the impact of
the BH distribution within BH-retaining GCs are necessary to fully understand
the consequences of this limitation.
Furthermore, the results presented here rely on the outcomes of many
independent realizations.
Since we perform each simulation independently in a static cluster background,
we are neglecting the change in the BH population and its impact on the
cluster as single BHs and BH binaries are ejected over the cluster lifetime.
Additionally,
we do not account for binary-binary interactions, which have the potential to
disrupt existing binaries or possibly aid in ejecting them.
Models which account for these limitations are necessary to better understand
the impact of ignoring these processes.
While \textit{N}-body simulations and Monte Carlo based models can resolve
some of these issues, the computational expense remains a limiting factor
in performing many realizations. However, as the computational techniques
and resources continue to improve, it will soon be possible to produce many
high-accuracy GC simulations that address these limitations.

\section*{Acknowledgments}
The authors thank Sterl Phinney, Steinn Sigurdsson, and Saul Teukolsky
for valuable discussions.
This work is partially supported by the Sherman Fairchild Foundation and
by NSF under award No. CAREER PHY-1151197. The simulations were carried
out on NSF/NCSA Blue Waters under PRAC award no. ACI-1440083 and on the
Caltech cluster Zwicky, supported by the Sherman Fairchild Foundation
and NSF award No. PHY-0960291.


\bsp    
\label{lastpage}
\end{document}